\documentclass{jpp}
\usepackage{graphicx}

\usepackage[utf8]{inputenc}
\usepackage[T1]{fontenc}
\usepackage{amsmath}
\usepackage{amssymb}

\usepackage{xcolor}
\usepackage{caption}
\usepackage{subcaption}
\usepackage[most]{tcolorbox}

\shorttitle{Guidelines for authors}
\shortauthor{D. Recchiuti, L. Franci, L. Matteini, E. Papini, R. Battiston and M. Piersanti}

\title{Evolution of ion distribution functions \\in ionospheric plasmas \\ perturbed by Alfv\'en waves}

\author{D. Recchiuti\aff{1,2}
  \corresp{\email{dario.recchiuti@unitn.it}},
  L. Franci\aff{3,2}, L. Matteini\aff{4}, E. Papini\aff{2,3}, R. Battiston\aff{1}
 \and M.  Piersanti\aff{5,2}}

\affiliation{\aff{1}Department of Physics, University of Trento,
Trento, 38122, Italy
\aff{2}National Institute of Astrophysics-IAPS, Rome, 00133, Italy
\aff{3}School of Engineering, Physics and Mathematics, Northumbria University, Newcastle Upon Tyne, NE1 8ST, UK
\aff{4}Department of Physics, Imperial College London, London, SW7 2AZ, UK
\aff{5}Department of physical and chemical sciences, University of L'Aquila, 67100, Italy}

\begin{document}

\maketitle

\begin{abstract}

This study investigates ion kinetic effects during the parametric decay instability (PDI) of parallel-propagating Alfv\'en waves under plasma conditions characteristic of the Earth's ionosphere. By using a series of hybrid particle-in-cell simulations, we examine the evolution of ion velocity distribution functions (VDFs) in ultra-low-beta plasmas. Our numerical campaign systematically explores the dependence on key parameters (plasma beta, pump-wave amplitude and polarization, and ion composition). To emphasize the role of kinetic effects, we choose to trigger the PDI with a dispersive mother wave with wavelength comparable to the ion characteristic inertial length. Our results reveal pronounced nonthermal VDF modifications, including parallel heating and the formation of secondary ion beams,  linked to the nonlinear evolution of parametric decay instability. By varying the plasma beta and the pump-wave amplitude, we identify a critical regime where rapid and complete broadening of the velocity distribution function is observed, triggering bidirectional ion acceleration. Notably, simulations modeling realistic ionospheric conditions demonstrate that even low-amplitude Alfv\'enic perturbations can induce significant VDF spreading and ion beam generation, with hydrogen ions exhibiting stronger effects than oxygen. These nonthermal microscopic processes offer a plausible mechanism for particle precipitation in space weather events. This work represents the first comprehensive study with hybrid simulations of PDI-driven ion kinetics in ultra-low-beta plasmas, providing quantitative estimates for the time delay between electromagnetic wave impact and ion VDF modification and new insights into wave-particle interactions that may contribute to ion acceleration, precipitation processes and space plasma dynamics.

\end{abstract}

\section{Introduction}\label{sec:intro}

Alfv\'en waves \citep{alfven1942existence} are an ubiquitous feature of magnetized plasmas and constitute a fundamental component in a wide range of phenomena across space and laboratory environments. They are frequently observed in space plasmas \citep{belcher1971large,gershman2017wave} and serve as a mean through which magnetized plasmas transmit internal information concerning dynamic changes in currents and magnetic fields \citep{gekelman1999review}.

The propagation of finite-amplitude Alfv\'en waves is usually characterized by a class of wave-wave interactions referred to as parametric instabilities \citep{hollweg1994beat}. In these nonlinear processes, a ``pump'' (or mother) wave couples with a compressive acoustic-like perturbation and other electromagnetic fluctuations.  This interaction leads to various parametric instabilities, the specific manifestation of which depends on the plasma characteristics. One of the most well-known of these processes is the Parametric Decay Instability \citep[PDI,][]{goldstein1978instability} characterized by the excitation of a compressive wave that possesses a wave vector ($k_s$) larger than that of the mother wave ($k_m$). During this interaction, energy is progressively transferred from the pump wave to the growing acoustic unstable wave and to a daughter (or reflected) Alfv\'en wave with $|k_r| < |k_m|$. This three-wave interaction satisfies the condition of momentum conservation along the propagation direction, $k_r = k_m - k_s$, implying that the daughter wave always propagates in opposite direction with respect to the mother wave \citep{malara1996parametric,del2001parametric}.

PDI is believed to play a crucial role in numerous space plasma environments where Alfv\'en waves are typically observed \citep[e.g.][]{belcher1971large}. Primarily, it offers a natural mechanism to generate reflected Alfv\'en waves, as observed in situ in the solar wind \citep{bavassano2000evolution}. The presence of counter-propagating waves is essential to trigger non-linear turbulent interactions. In this context, PDI has been proposed as a key mechanism for developing a turbulent cascade in the solar wind \citep{chandran2018parametric,malara2022parametric}, which subsequently provides a possible mechanism for heating and accelerating the plasma \citep{perez2013direct}. Furthermore, PDI produces compressible modes that can contribute directly to plasma heating through various mechanisms. These include the formation of shocks \citep{del2001parametric} or, when kinetic effects are taken into account, processes such as particle trapping and phase-space mixing \citep{araneda2008proton,matteini2010kinetics}, as well as proton energization at steepened fronts \citep{gonzalez2020role,gonzalez2021proton}.

Both theoretical models and numerical simulations indicate that PDI should contribute significantly to the dynamics of the solar wind within the inner heliosphere, particularly in near-Sun regions where $\beta \ll 1$ \citep{tenerani2013parametric,reville2018parametric,shoda2019three,verdini2019turbulent,hahn2022evidence}. This picture is consistent with the increase in the PDI growth rate as plasma beta decreases \citep{goldstein1978instability}.

Given its dependence on the plasma beta, the influence of PDI on the plasma dynamics is expected to be even greater in space plasmas characterized by a much lower $\beta$ than the solar wind, such as the Earth's ionosphere. However, to date, the effects of PDI in such environments remain  unexplored. The main objective of this work is then to examine ion-kinetic effects during the parametric decay of parallel-propagating Alfv\'en waves under background plasma conditions consistent with the Earth's ionosphere. We specifically focus on studying deformations of the ion VDF, including local field-aligned acceleration and the potential generation of secondary beams and higher energy tails.

This investigation is motivated by the extensive observation of transverse electromagnetic waves propagating along field lines in Earth's ionosphere, often corresponding to 
whistler waves \citep{chaston2014observations,shen2021conjugate, recchiuti2025automatic}. Furthermore, ElectroMagnetic Ion Cyclotron (EMIC) waves have been demonstrated to be associated with proton precipitation \citep{tian2022effects,d2024case}. Motivated by the need to understand the underlying kinetic processes governing these phenomena, a PDI triggered by a dispersive mother electromagnetic wave with wavelength comparable to the ion characteristic inertial length is here studied, so to emphasize the role of kinetic effects.


To gain a comprehensive understanding of PDI in a realistic ionospheric environment, we conducted a methodical simulation campaign. First, we performed numerical simulations of the PDI in a more well-known plasma regime, used as a baseline, and then we progressively tuned the simulation parameters to reach the aimed low-beta and low-wave-amplitude regime representative of the Earth's ionosphere. Such a multi-step approach has been crucial to disentangle the complex interplay of ambient parameters influencing the  evolution of the instability and its significant impact on the ion VDF. The study reached its final goal with simulations employing realistic parameters of the top-side ionosphere, both in terms of the wave-amplitude and of the plasma background and composition, allowing for a detailed investigation of the effects of perturbations induced by Alfv\'en waves of different amplitudes.

The work is structured as follows: Section \ref{sec:setup} describes the numerical model and the simulation set-up; Section \ref{sec:results} details the results from the numerical simulations; finally, Section \ref{sec:conclusion}  provides the conclusions and discusses potential applications to space plasmas.

\section{Model and simulation setup}\label{sec:setup}

Our investigation employs a Hybrid Particle-In-Cell (HPIC) code. Hybrid models differentiate the treatment of various plasma components, representing some species of the plasma as particles and the rest as a fluid \citep{winske1985hybrid}. Given that our focus is strictly on ion-scale interactions, it is sufficient to model the ion VDFs explicitly. In this regime, electron kinetic effects are considered negligible, so that electrons can be modelled using a fluid approximation. Specifically, the model treats ions as macro-particles, which are statistically-representative portions of the particle distribution function in phase space, while electrons are treated as a massless, charge-neutralizing fluid.


\subsection{The CAMELIA code}

For this study, we employed the HPIC code CAMELIA \citep[Current Advance Method Et cycLIc leApfrog,][]{hellinger2003hybrid,hellinger2003hybridB,Franci_al_2018b} based on the CAM-CL code by \citep {matthews1994current}. Thanks to its inherent capability for straightforwardly treating multiple ion species, CAMELIA is well suited to reproduce the ionospheric multi-species plasma dynamics. The code  has consistently proved to be able to accurately reproduce observed ion properties in space plasmas, including preferential ion heating \citep{hellinger2005alfven}, kinetic instabilities \citep{matteini2006parallel,hellinger2008oblique}, and kinetic plasma turbulence \citep{franci2015high,franci2015solar,franci2018three,Franci_al_2020}. It has also been successfully used for exploring the plasma parameter space to probe conditions that are representative of different regions of the inner heliosphere, from the near-Sun solar wind 
\citep{franci2020arxiv,franci2022anisotropic} to the Earth's magnetosheath \citep{Franci_al_2020,Franci_2022universe}
with a specific focus on the role of the plasma beta \citep{franci2016plasma}, a key parameter of this study.




In the code, the spatial scales are normalized with respect to the ion inertial length $d_i$, which defines the scale at which ions and electrons decouple, defined as $d_i = c/\omega_i$, where $\omega_i = ( n_i q_i^2/m_i\epsilon_0)^{1/2}$ represents the plasma frequency of the ion species $i$. Here, $n_i$, $q_i$ and $m_i$ are the number density, charge and mass of the ion species $i$, respectively, and $\epsilon_0$ is the permittivity of free space. Correspondingly, wavenumbers are expressed in units of $d_i^{-1}$. 
Units of time are normalized to the inverse ion cyclotron frequency $\Omega_i^{-1}$, defined as $\Omega_i = q_iB_0/m_i$. All the simulations are initialized with a uniform background magnetic field $B_0$ (set to unity in the simulation domain). Velocities are normalized with respect to the the Alfv\'en velocity $v_A = B_0 /(\mu_0 n_i m_i)^{1/2}$. 



Here we summarize the numerical setup of our simulations:

\begin{itemize}

	\item \textbf{1D domain}: to reduce the computational cost of the simulations, we opted for a one-dimensional domain, aligned with the x direction, corresponding to the direction of the background magnetic field. This is justified by the fact that we estimate field-aligned dynamics to be dominant in this low-beta (strong field) environment. It is important to note, however, that all three spatial components of vector field quantities (such as magnetic field and ion bulk velocity) are retained.

    \item \textbf{Spatial resolution}: we set a spatial resolution  $\Delta x=  5 \cdot 10^{-2} d_i$, consistent with previous studies employing the same code \citep[e.g.,][]{matteini2010kinetics,franci2016plasma}. 
	
    \item \textbf{Particles per cell}: to ensure a high accuracy of our results and prevent non-physical artifacts \citep[such as numerical heating,][]{markidis2011energy,horky2017numerical,alves2021numerical}, we employ a minimum value of particles per cell (ppc) of $10^4$. This value is consistent with, and often exceeds, the one used in previous studies using 1D CAMELIA simulations \citep[e.g.,][]{matteini2010kinetics}.
	
	\item \textbf{Perturbing wave}: we perturb the plasma with different small-amplitude dispersive Alfv\'en waves directed along the ambient magnetic field $B_0$ (i.e., in the positive direction of the x-axis). The wave has amplitude $\delta B / B_0$, and wave vector $k_m = k_0 w_m$ ($w_m$ being an integer number), where $k_0 = 2 \pi / L_{\textrm{box}}$ is the wave vector of the largest mode contained in the simulation box, which has a length $L_{\textrm{box}}$. Hereafter, this perturbing wave will be referred to as the ``mother'' or ``pump'' wave. We examine both right (R) and left (L) wave polarizations.
	
	\item \textbf{Plasma beta}: Our investigation probes different values of the plasma beta (i.e., the ratio of the particle thermal pressure to the magnetic pressure) spanning several orders of magnitude for both ions ($\beta_i$) and electrons ($\beta_e$).

\end{itemize}

These parameters ensure an accurate description of the particle distribution function with low numerical noise, hence enabling the investigation of both wave-particle and wave-wave interactions and allowing for the analysis of ion VDF deviations from Maxwellian equilibrium \citep[e.g.][]{matteini2010kinetics}.

\subsection{Ionospheric parameters from IRI model}

The ultimate objective of this study is to model the ionospheric environment in a way that is as much realistic as possible given the small-scale local approach of the model, thereby facilitating future comparisons between model predictions and observational data. Consequently, we establish a reference altitude of 500 km, consistent with the orbital altitudes of several advanced spacecraft missions such as SWARM \citep{olsen2013swarm} and CSES  \citep[China Seismo-Electromagnetic Satellite,][]{shen2018state}. To achieve this objective, we derived ionospheric parameters from the International Reference Ionosphere (IRI) model \citep{bilitza2022international}, an internationally recognized empirical model that provides monthly averaged values of electron density, electron temperature, ion temperature, and ion composition throughout the ionospheric altitude range.

For the IRI model input, we chose equatorial latitudes, an early morning temporal window, and the year 2021 (a period indicative of low solar activity, far from the solar maximum) to establish representative ionospheric conditions at 500 km altitude. The output from the IRI model, detailed in Table \ref{tab:ionoParam}, subsequently served as the comprehensive reference for the ionospheric environment. Table \ref{tab:ionoParam} details the key parameters for each particle species $s$, including their percentage composition in the plasma ($\%_s$), temperature ($T_s$), density ($n_s$), cyclotron frequency ($\Omega_{g,s}$), Alfv\'en speed ($v_{A,s}$),  plasma beta ($\beta_s$), plasma frequency ($\omega_s$), and inertial length ($d_s$). Only electrons, protons ($H^+$), and $O^+$ are included, as they represent the primary constituents of the plasma at this altitude. Indeed, Table \ref{tab:ionoParam} shows that $H^+$ and $O^+$ collectively account for over $98\%$ of the ionospheric ion population.

\begin{table}
	\centering
	\begin{tabular}{cccccccccc}
	\hline
	$s$ & $\%_s$ & $T_s$ (K) & $n_s (m^{-3})$ & $\Omega_{g,s}$ (Hz) & $v_{A,s} (m/s)$ & $\beta_s$ & $\omega_s$ (Hz) & $d_s$ (m) \\
	\hline
	e & - & $1.4\cdot10^3$ & $2.3\cdot10^{11}$ & 8.4$\cdot 10^5$ & $5.8\cdot 10^7$ & $1.3\cdot 10^{-5}$ & $4.3\cdot 10^6$ & $11.1$\\
	\hline
	$H^+$ & 4.2 & $1.1\cdot10^3$ & $9.7\cdot10^9$ & 457 & $6.6\cdot 10^6$ & $4.2\cdot 10^{-7}$ & $2.1\cdot 10^4$ & $2.3\cdot 10^3$\\
	\hline
	$O^+$ & 93.9 & $1.1\cdot10^3$ & $2.2\cdot10^{11}$ & 29 & $3.5\cdot 10^5$ & $9.4\cdot 10^{-6}$ & $2.4\cdot 10^4$ & $2.0\cdot 10^3$\\
	\hline
	\end{tabular}
	\caption{Ionospheric parameters derived from IRI model.}
    \label{tab:ionoParam}
\end{table}

\section{Results}\label{sec:results}

\subsection{Numerical Validation at low beta values}

To validate our numerical framework, we first ran similar simulations to those in \cite{matteini2010kinetics}, before gradually transitioning to our target conditions.
In Table \ref{tab:runABC}, we report the input parameters used in the validation runs (A, B and C).
Specifically, our Run A is similar to Run B from \citet{matteini2010kinetics}, in which the authors observed a PDI and, as a consequence, a velocity beam aligned with the ambient magnetic field. Differently from that work, however, we decided to initialize our simulation with a wave vector ($k_m d_i \simeq 1 $), which aligns with our specific research objectives.
Furthermore, we investigated the impact of varying $k_m$ on the evolution of the VDF,  employing simulation boxes of different lengths in Runs A, B, and C, while keeping $w_m$ the same.

\begin{table}
	\centering
	\begin{tabular}{cccccccccc}
	\hline
	Run & $\beta_e$ & $\beta_i$ & $\delta B/B_0$ & $w_m$ & Pol. & $L_{\textrm{box}}(d_i)$ & $k_0(d_i^{-1})$ & $k_m(d_i^{-1})$ & $ppc$\\
	\hline
	A & $1\cdot10^{-2}$ & $5\cdot10^{-2}$ & $5\cdot10^{-2}$ & $20$ & R & $121.6$ & $0.05$ & $1.03$ & $10^4$\\
	\hline
	B & $1\cdot10^{-2}$ & $5\cdot10^{-2}$ & $5\cdot10^{-2}$ & $20$ & R & $243.2$ & $0.025$ & $0.52$ & $10^4$\\
    \hline
	C & $1\cdot10^{-2}$ & $5\cdot10^{-2}$ & $5\cdot10^{-2}$ & $20$ & R & $60.8$ & $0.1$ & $2.07$ & $10^4$\\
	\hline
	\end{tabular}
	\caption{Run A, B and C parameters.}
    \label{tab:runABC}
\end{table}

We first present the results of Run A, which examines a single-ion-species plasma characterized by $\beta_e = 1\cdot 10^{-2}$ and $\beta_i = 5\cdot 10^{-2}$. The plasma is perturbed by a right-handed polarized Alfv\'en wave with $k_m = 1.03\ d_i^{-1}$ and amplitude $\delta B/B_0 = 5 \cdot 10^{-2}$. The simulation has $L_{\textrm{box}} = 121.6\ d_i$ and $10^4$ ppc. 

\begin{figure}
  \centering
  \includegraphics[width=0.75\linewidth]{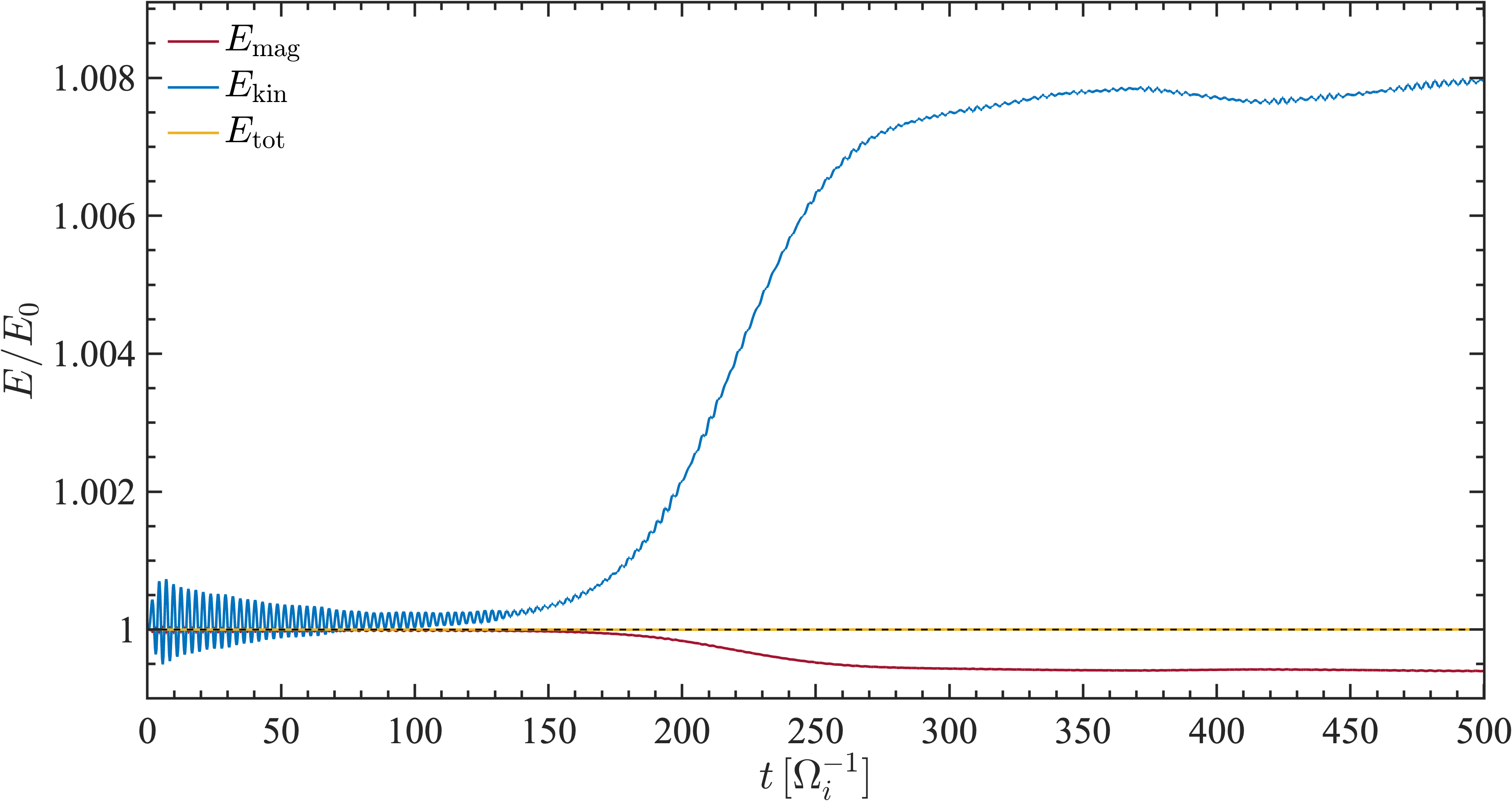}
  \caption{Run A: Time evolution of the energy components normalized to their respective initial values: kinetic energy $E_{\textrm{kin}}/E_{\textrm{kin},0}$ (blue), magnetic energy $E_{\textrm{mag}}/E_{\textrm{mag},0}$ (red), and total energy $E_{\textrm{tot}}/E_{\textrm{tot},0}$ (yellow). An horizontal dashed black line is overplotted for comparison to show a remarkably good energy conservation.}
\label{fig:runAenergie}
\end{figure}

Figure \ref{fig:runAenergie}  shows the temporal evolution of  kinetic energy $E_{\textrm{kin}}$ (blue), magnetic energy $E_{\textrm{mag}}$ (red) and total energy $E_{\textrm{tot}}$ (yellow), normalized to their respective initial values $E_{\textrm{kin},0}$, $E_{\textrm{mag},0}$, and $E_{\textrm{tot},0}$. This plot illustrates the conservation of total energy and the transfer of energy from the magnetic field ($E_{\textrm{mag}}$, decreasing) to the kinetic energy of the ions ($E_{\textrm{kin}}$, increasing). This transfer initiates at $t\approx 100$ and continues until $t\approx 300$ where $E_{\textrm{kin}}$ reaches a plateau and remains approximately constant. During the initial phase ($t\leq100$), the plot exhibits numerical noise attributable to spatial grid discretization. This is a common feature of particle-in-cell simulations \citep{matteini2010kinetics}. To keep such numerical noise low (below $10^{-3}$), a high number of $ppc$ was employed.

To verify that this energy transfer is indeed caused by a PDI, we analyzed the temporal evolution of the energy associated with backward and forward Alfv\'en waves propagation. This is typically investigated through the use of Elsässer variables \citep{elsasser1950hydromagnetic}, which in Alfv\'en units are defined as $z^{\pm} = \delta\mathbf{v} \mp \delta\mathbf{B}$, where $\delta\mathbf{v}$ and $\delta\mathbf{B}$ denote velocity and magnetic field fluctuations, respectively. 
Consistent with \citet{del2001parametric}, we use a notation in which a positive sign indicates propagation in the positive $x$-direction. 
Moreover, here $z^{\pm}$ variables have been adjusted to take into account dispersive effects in the propagation of the pump wave, using the cold plasma dispersion relation for parallel waves \citep[see e.g.][]{boyd2003physics}. 
Following the same approach of \citet{del2001parametric} and \citet{matteini2010kinetics}, we investigated the evolution of the spatially-averaged energies 
\begin{equation}
E^{\pm} = <\frac{1}{2}\left| z^{\pm} \right|^2 >,
\end{equation}
and the normalized cross-helicity
\begin{equation}
\sigma = \frac{E^+ - E^-}{E^+ + E^-},    
\end{equation}
which is a measure of the prevailing mode.

\begin{figure}
  \centering
\includegraphics[width=0.75\linewidth]{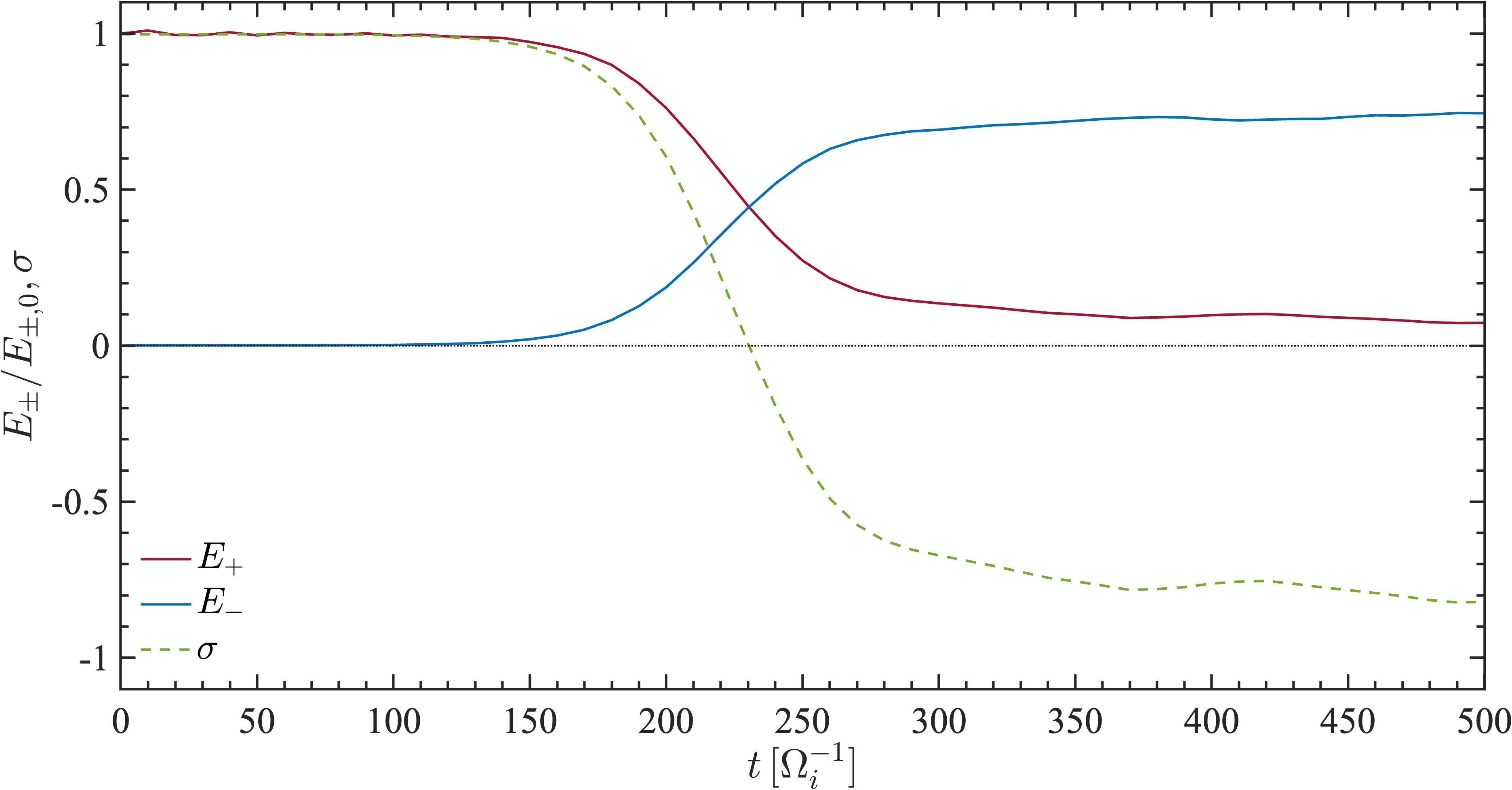}
  \caption{Run A: Time evolution of the spatially-averaged Elsässer energies normalized to their initial values, $E_+/E_{+,0}$ (solid red), $E_-/E_{-,0}$ (solid blue) and of normalized cross-helicity $\sigma$ (dashed green). A dashed black line is included to indicate the zero reference.}
\label{fig:runAEpEm}
\end{figure}

Figure \ref{fig:runAEpEm} shows the evolution of these quantities normalized to their initial values as a function of time: $E^+/E^-(0)$ (solid red), $E^-/E^-(0)$ (solid blue), and $\sigma$ (dashed green). Initially, for pure Alfv\'en waves propagating in the positive $x$ direction, $E^+=\sigma=1$. As the mother wave is damped, $E^+$ decreases and $E^-$ (initially zero) increases,  revealing the development of a backward propagating daughter wave. Correspondingly, $\sigma$ crosses zero when the backward and forward propagating perturbations attain equivalent energy, transitioning to negative values as $E^-$ exceeds $E^+$.

\begin{figure}
\centering
\includegraphics[width=0.49\textwidth]{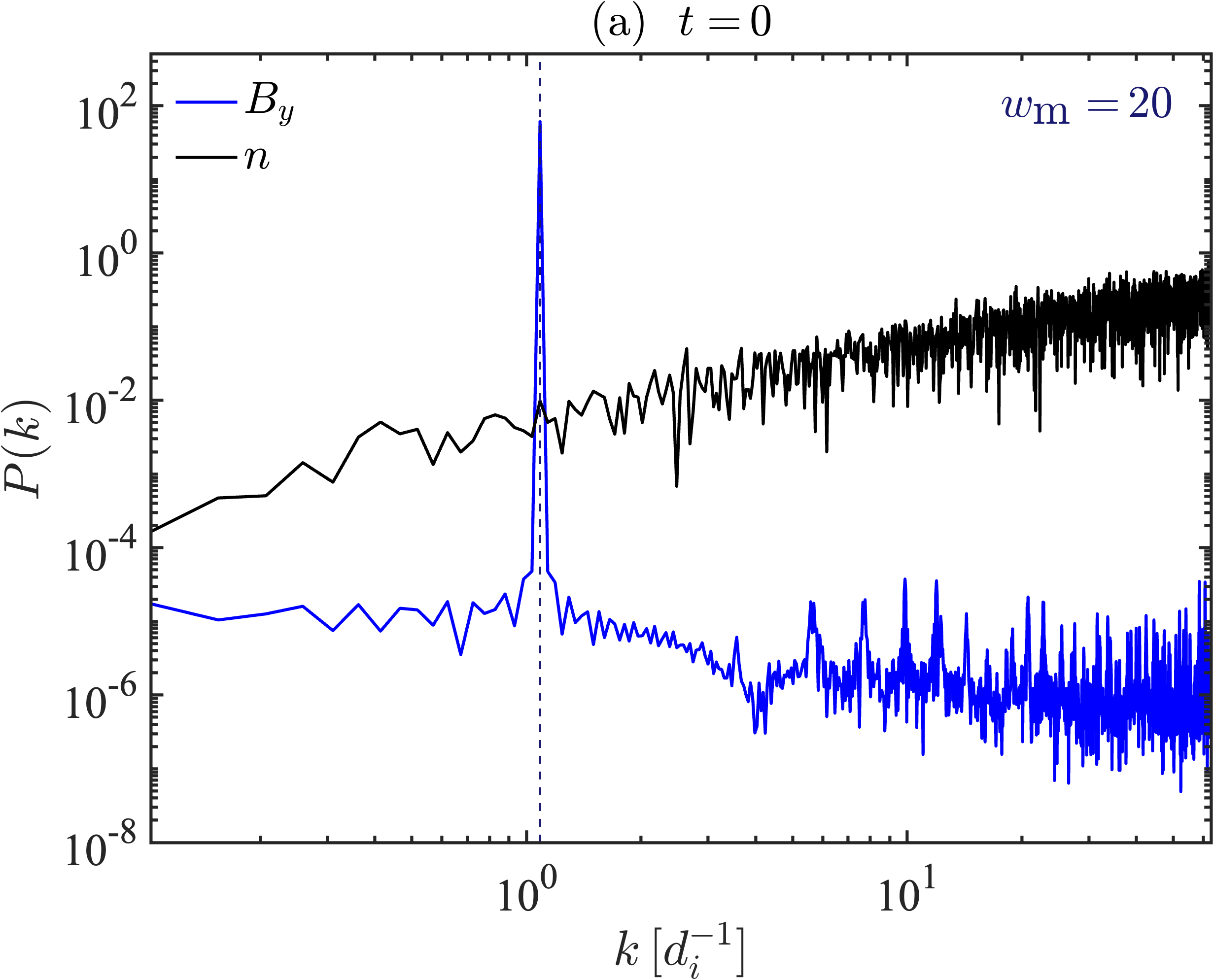}
\includegraphics[width=0.49\textwidth]{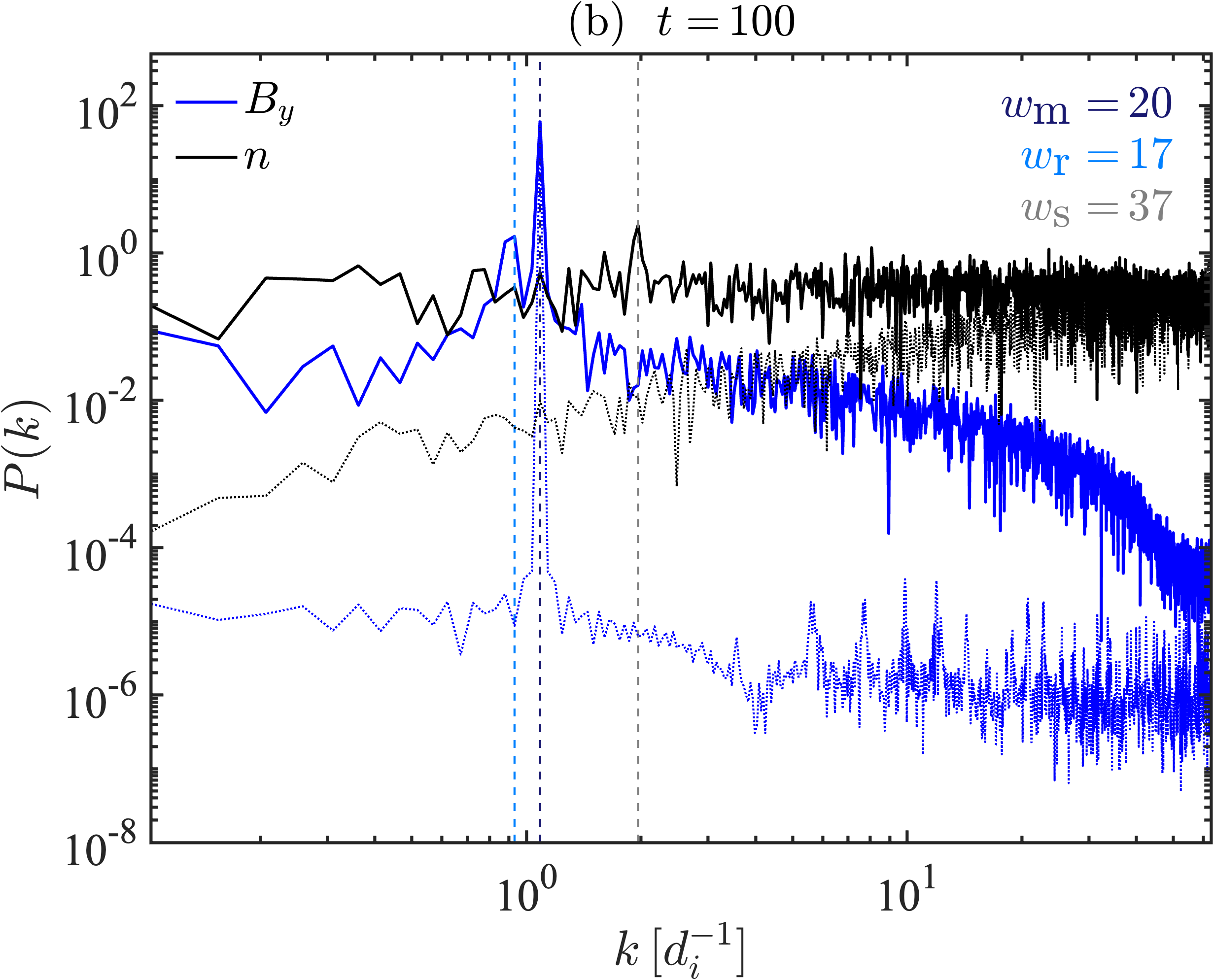}\\
\includegraphics[width=0.49\textwidth]{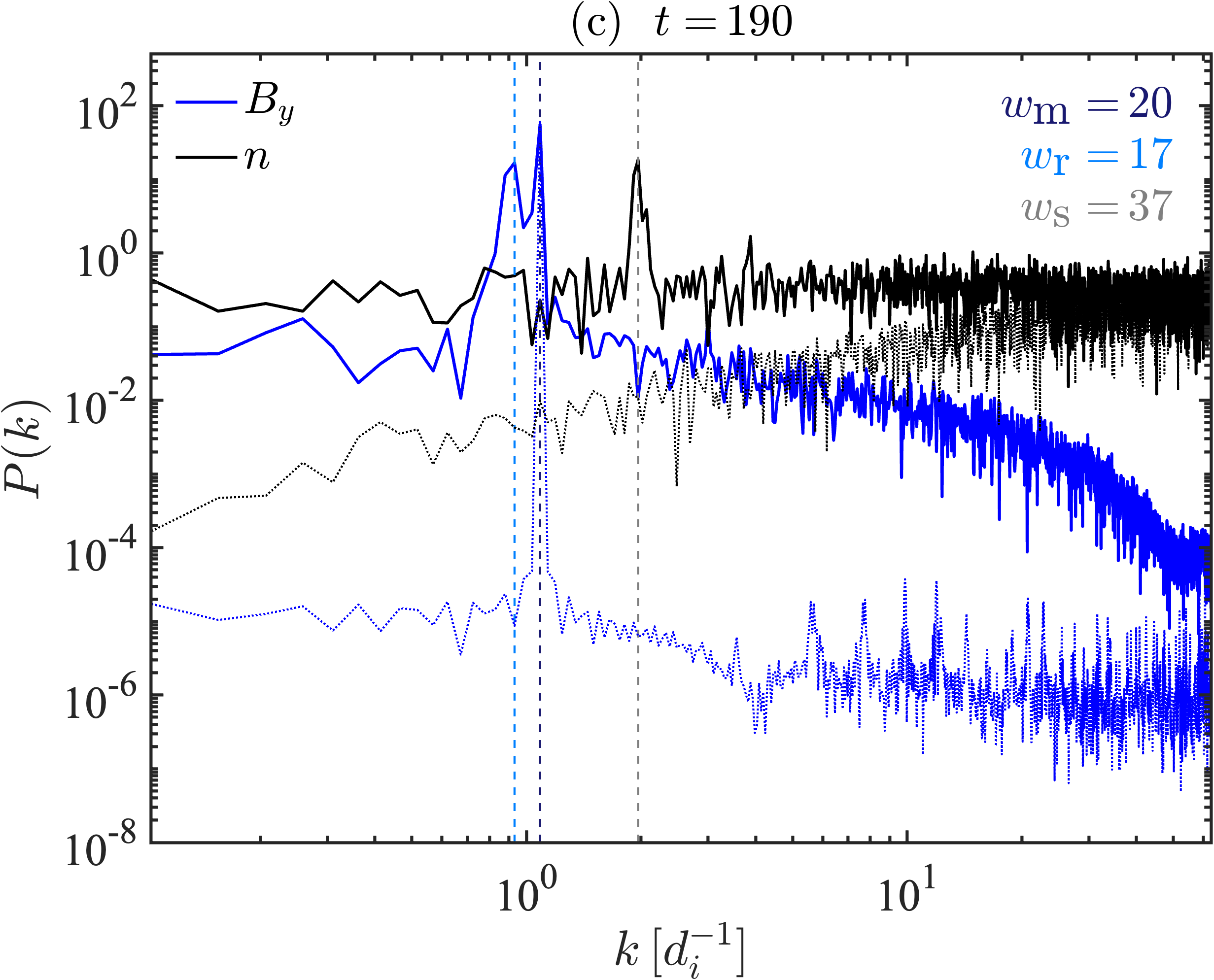}
\includegraphics[width=0.49\textwidth]{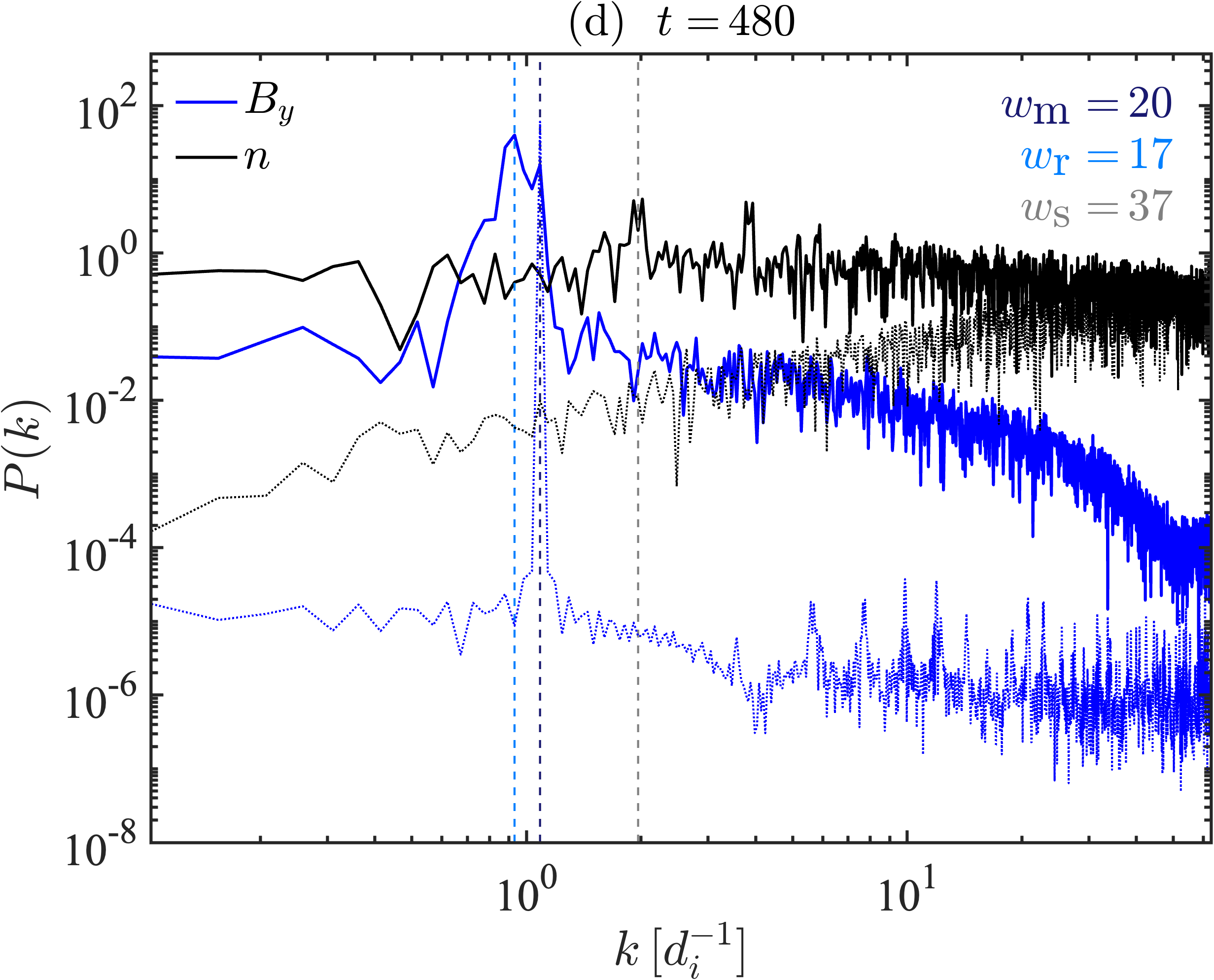}
\caption{Run A: Power spectrum of the $y$ component of the magnetic field (blue) and ion density (black). Values at $t = 0$ are overlaid as thinner dashed lines of the same color. Vertical dashed lines highlight significant peaks corresponding to mother wave (blue), daughter wave (cyan) and acoustic wave (black). The respective wavenumbers ($w_\mathrm{m}$, $w_\mathrm{r}$ and $w_\mathrm{s}$) are indicated in the upper right corner with the same colors.
} 
\label{fig:runA1Dspectrum}
\end{figure}

The outcome of this parametric decay can be elucidated by examining the power spectrum $P(k)$ of the $y$ component of magnetic field (blue) and of the density (black) at different times. Figure \ref{fig:runA1Dspectrum} shows $P(k)$ at $t=0$ (Panel a), $t=100$ (Panel b),  $t=190$ (Panel c), and $t=480$ (Panel d). To facilitate a clearer visualization of the temporal evolution, the panels for $t > 0$ also include the corresponding values at $t = 0$, indicated by thinner dashed lines of the same color. These also provide a reference for the level of numerical noise in the density spectrum, due to the finite number of ppc, which corresponds to its initial spectrum. The mother wave perturbing the system has wave number $w_m=20$, which corresponds to the magnetic energy peak at $k_m = 1.03$ (highlighted by the thin blue vertical line). At time $t=0$ no other signatures are present in the spectra. The decay generates a higher frequency compressive acoustic‐like wave, which appears as a peak on the density (black line) at $k_s = 1.91$ ($w_s = 37$) in Panels b (evidenced by the vertical black dashed line). Simultaneously, a lower frequency backward Alfv\'en wave emerges, corresponding to a second peak in the magnetic field fluctuations at $k_r = 0.88$  ($w_r = 17$, vertical dashed cyan line). As time progresses, this second peak's amplitude exceeds the initial one, as clearly depicted in Panel d. These observations are consistent with the resonant condition for wave numbers in parametric decay: $k_r = k_m - k_s$.

\begin{figure}
\centering
\begin{subfigure}{0.6\textwidth}
    \includegraphics[width=\textwidth]{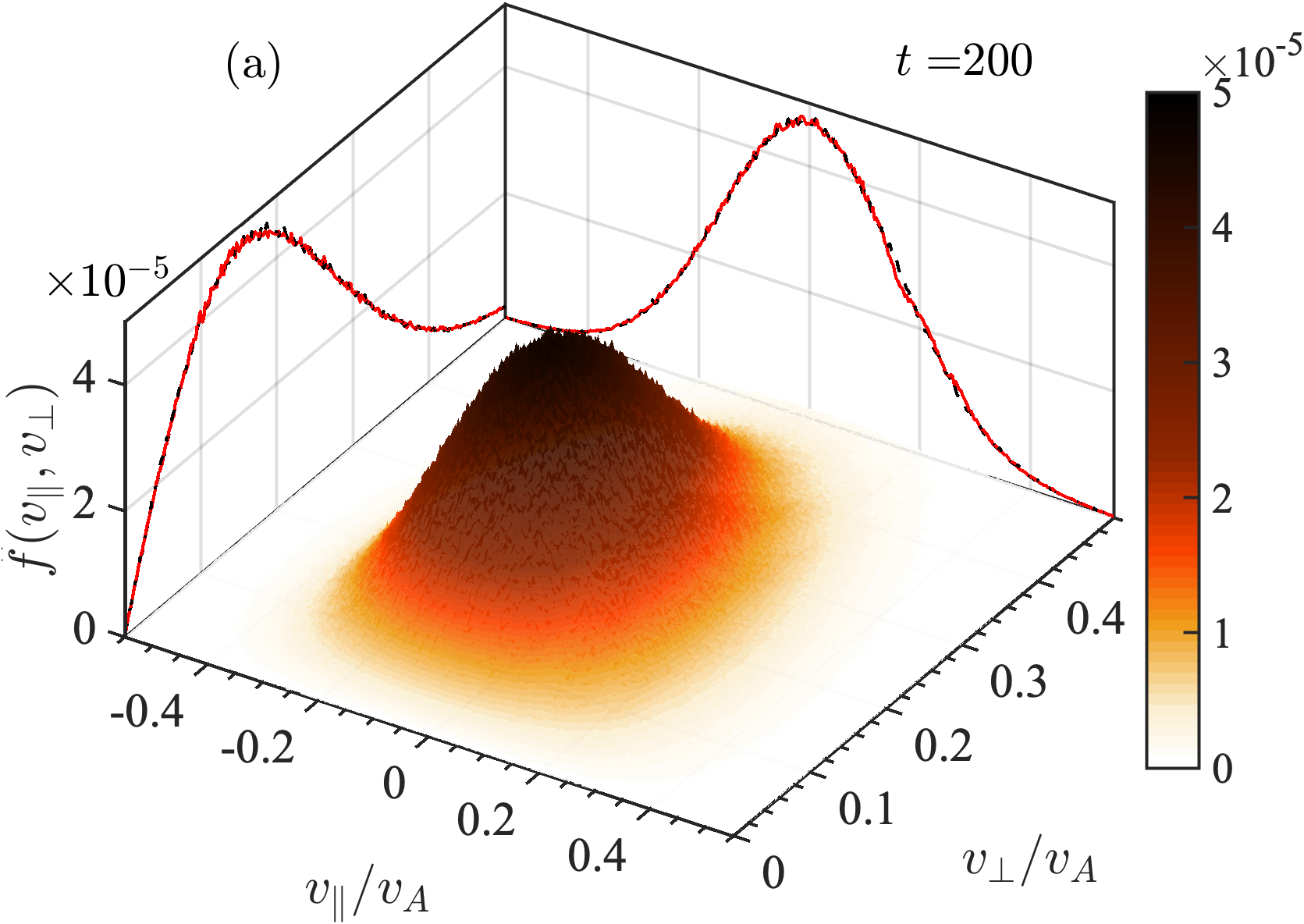}
    \label{fig:runAvparT200}
\end{subfigure}
\hfill
\begin{subfigure}{0.6\textwidth}
    \includegraphics[width=\textwidth]{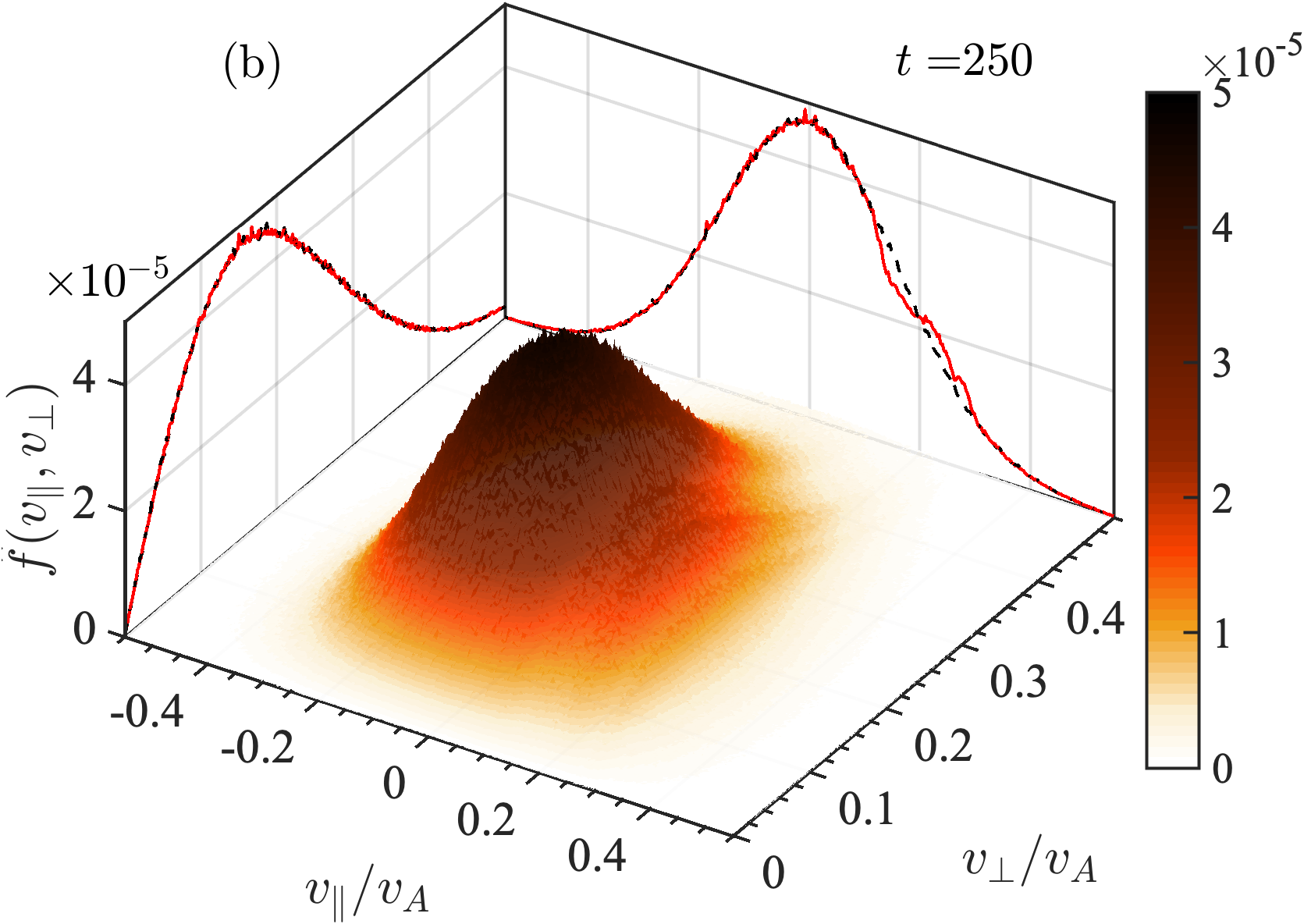}
    \label{fig:runAvparT250}
\end{subfigure}
\hfill
\begin{subfigure}{0.6\textwidth}
    \includegraphics[width=\textwidth]{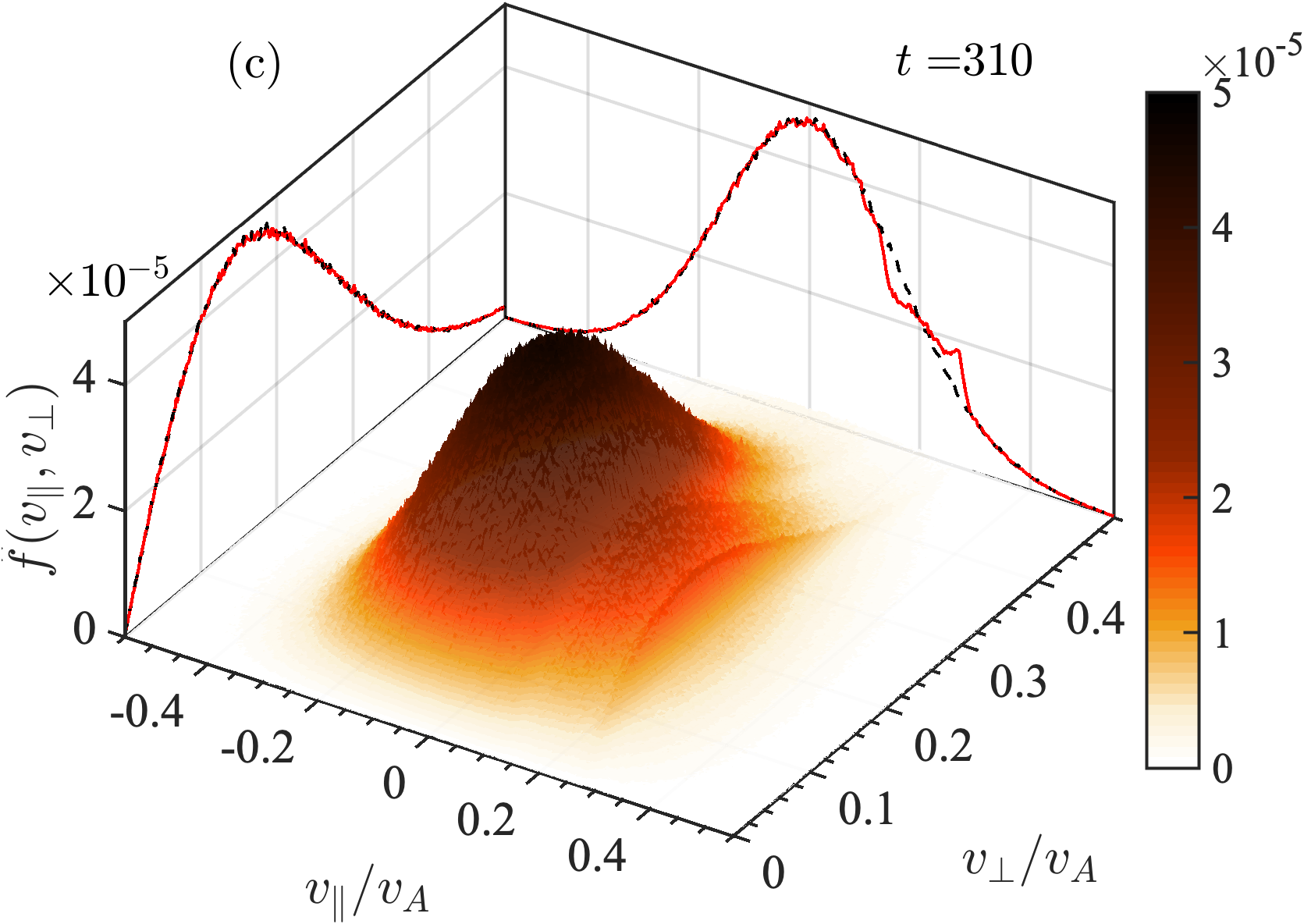}
    \label{fig:runAvparT310}
\end{subfigure}
        
\caption{Run A: Ion VDF displayed as a 3D color scale. Red lines represent VDF projections on the parallel and perpendicular directions. Black dashed lines denote the corresponding projections of the initial distribution (t=0).
} 
\label{fig:runAvdf}
\end{figure}

A comparative analysis of Panels c and d reveals that, subsequent to the linear growth phase of the acoustic mode (which persists until approximately $t \simeq 190$), the instability enters a saturation phase, marked by a saturation in the density peak's growth. Nevertheless, nonlinear wave interactions continue to occur during this post-saturation regime \citep{matteini2010kinetics}. The evolution of the ion VDF clearly illustrates this phenomenon.
To optimally visualize the ion VDF ($f(v_\parallel,|v_\perp|)$), a combined plotting technique is employed: the distribution itself is rendered in a three-dimensional color scale, while its projections on the parallel and perpendicular directions are plotted as red lines.
Figure \ref{fig:runAvdf} presents significant snapshots of the ion VDF evolution: t=200 (Panel a), t=250 (Panel b), and t=310 (Panel c). The black dashed lines represent the projections of the initial distribution (t=0).


Saturation phase begins around $t \approx 300$. This is attributed to particle trapping, a well-established mechanism for suppressing wave growth in collisionless plasmas \citep{matteini2010kinetics}. Within this kinetic regime, this process yields two significant consequences. First, wave-wave interactions from parametric decay are modified relative to fluid-based predictions due to kinetic effects \citep{inhester1990drift,vasquez1995simulation,nariyuki2006kinetically,araneda2007collisionless}. Second,  trapping and its associated wave-particle interactions, which originate from the saturation phase, can substantially influence ion dynamics. A direct consequence of trapping is the acceleration of particles that resonate with the wave, leading to the formation of a faster ion population. While prior work by \citep{terasawa1986decay,vasquez1995simulation} recognized ion heating in the parallel direction due to proton trapping during the parametric decay of a monochromatic Alfv\'en wave, they described it solely as a parallel temperature increase, attributed to a broadening of the distribution function. In contrast, Figure \ref{fig:runAvdf} clearly demonstrates actual ion acceleration, resulting in a forward-propagating ion beam. This corresponds to the appearance of a secondary peak at $v_{\parallel}/v_A \approx 0.3$, aligning with the results obtained by \citep{matteini2010kinetics} and confirming the validity of our approach.

Because the instability growth rate depends on the mother wave characteristics, decreasing the wave vector ($k_m$) of the pump wave, as implemented in Run B (not shown),  results in a slower decay.
Indeed, the energy transfer from the magnetic field to the kinetic energy of the ions in Run B starts around  $t\approx 600$, with the cross-helicity reaching zero at $t\approx 750$.  Beyond this temporal delay, the effects on the ion velocity distribution function (not shown) are consistent with those observed in Run A, manifesting as the emergence of a velocity beam at $v_{\parallel}/v_A \approx 0.3$.

\begin{figure}
  \centering
  \includegraphics[width=0.75\linewidth]{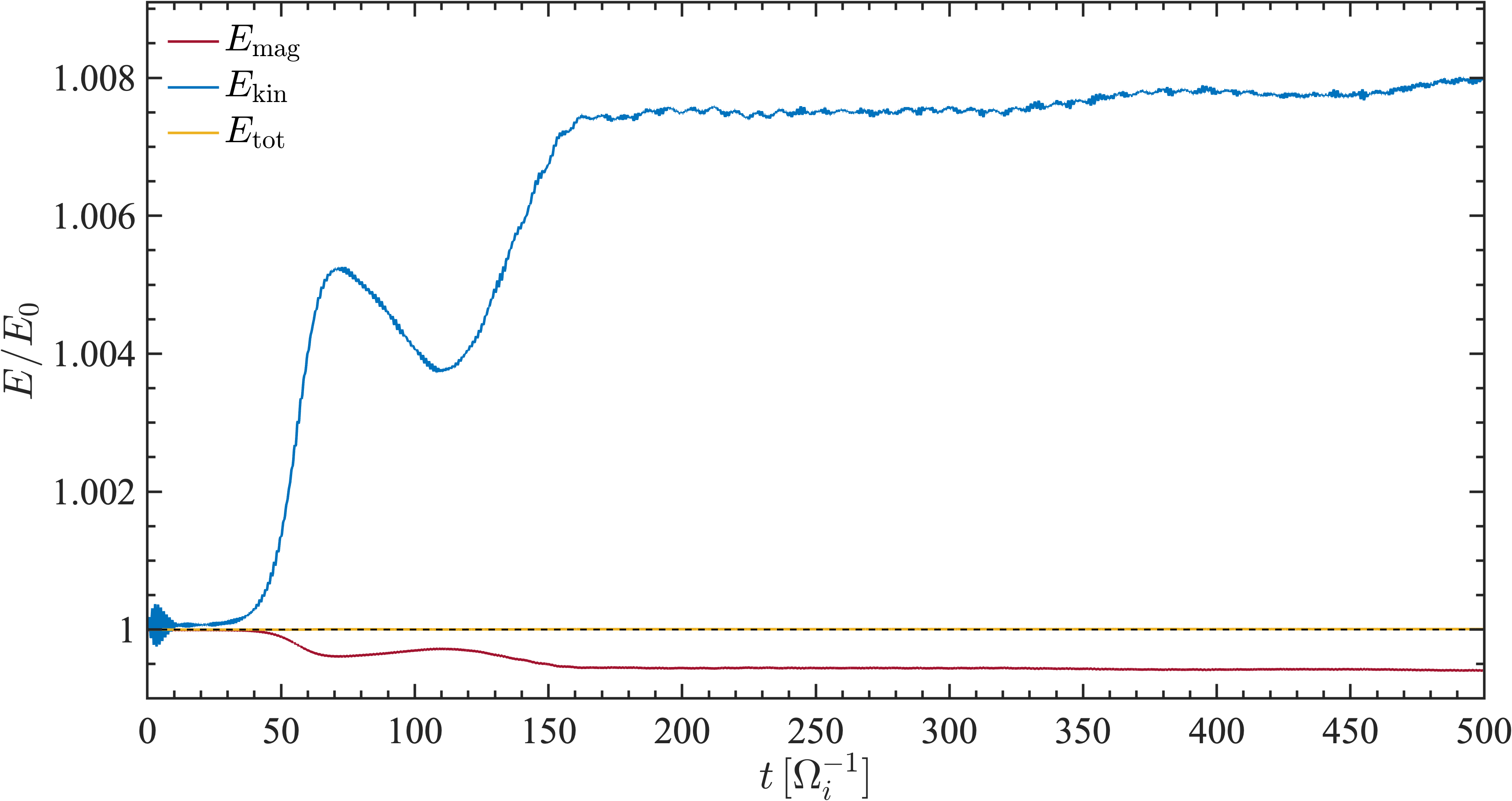}
  \caption{Run C: Time evolution of $E_{\textrm{kin}}/E_{\textrm{kin},0}$ (blue), $E_{\textrm{mag}}/E_{\textrm{mag},0}$ (red), and  $E_{\textrm{tot}}/E_{\textrm{tot},0}$ (yellow). An horizontal dashed black line is overplotted  to show energy conservation.}
\label{fig:runCenergie}
\end{figure}

Consistent with this, a doubled wave vector for the pump wave ($k_m = 2.07$), as in Run C, yields a faster decay. As illustrated in Figure \ref{fig:runCenergie}, which shows the temporal evolution of $E_{\textrm{kin}}/E_{\textrm{kin},0}$, $E_{\textrm{mag}}/E_{\textrm{mag},0}$ and $E_{\textrm{tot}}/E_{\textrm{tot},0}$, a first decrease in magnetic energy accompanied by a corresponding increase in ion kinetic energy is observed at $t \approx 80$. Notably, a second decay phase emerges at $t \approx 140$.

\begin{figure}
  \centering
\includegraphics[width=0.75\linewidth]{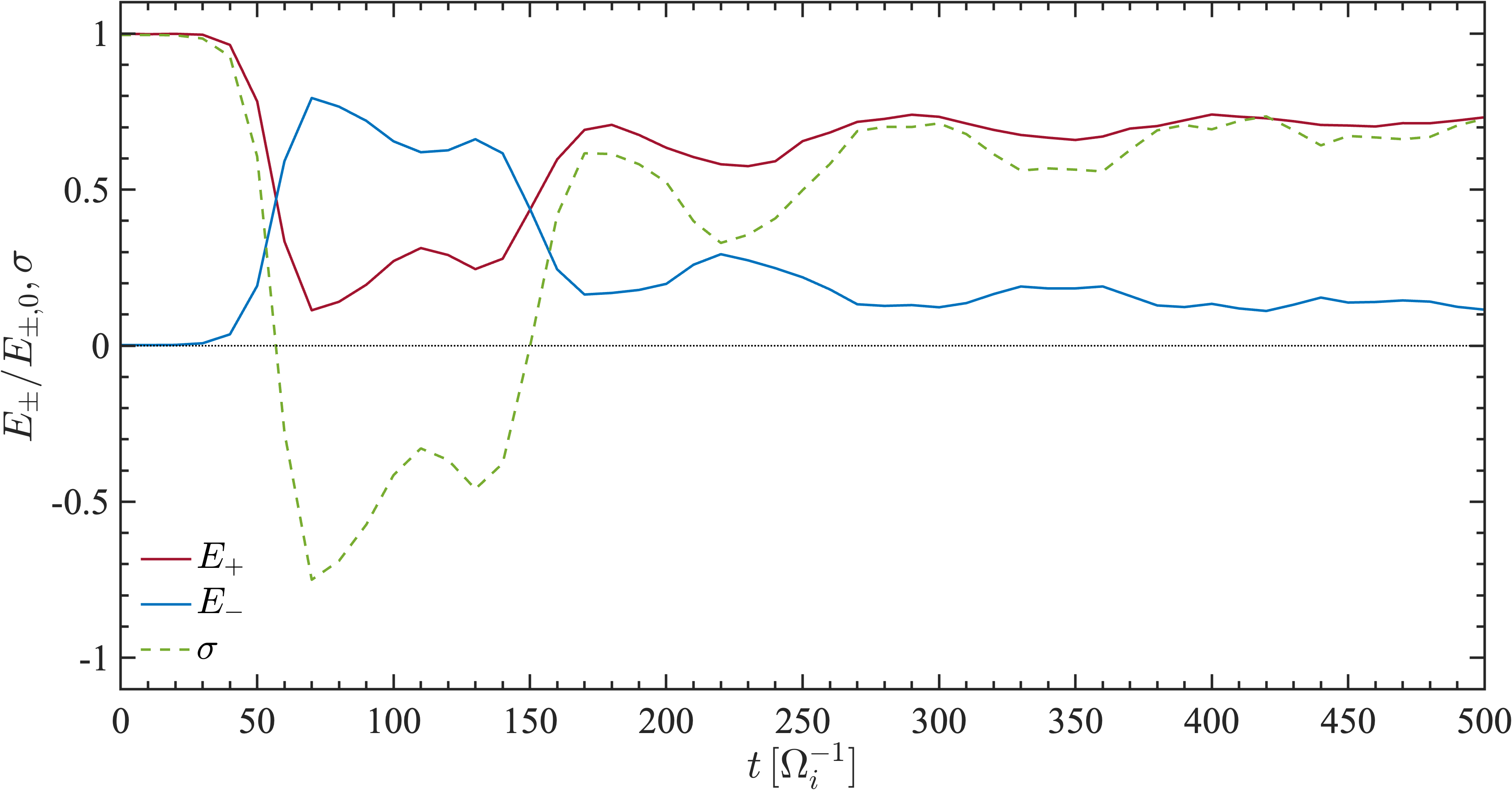}
  \caption{Run C: Time evolution of $E_+/E_{+,0}$ (solid red), $E_-/E_{-,0}$ (solid blue) and  $\sigma$ (dashed green). A dashed black line is included to indicate the zero reference.}
\label{fig:runCEpEm}
\end{figure}

The double decay is further evidenced by the evolution of $E^+$, $E^-$, and $\sigma$, shown in Figure \ref{fig:runCEpEm}. Specifically, the cross-helicity intersects zero twice, at $t \approx 80$ and $t \approx 140$, signifying equal energy in the backward and forward propagating perturbations at these two distinct times. Concurrently, $E^+$ and $E^-$ exhibit equal values and interchange dominance at these same times.

\begin{figure}
\centering
\begin{subfigure}{0.48\textwidth}
    \includegraphics[width=\textwidth]{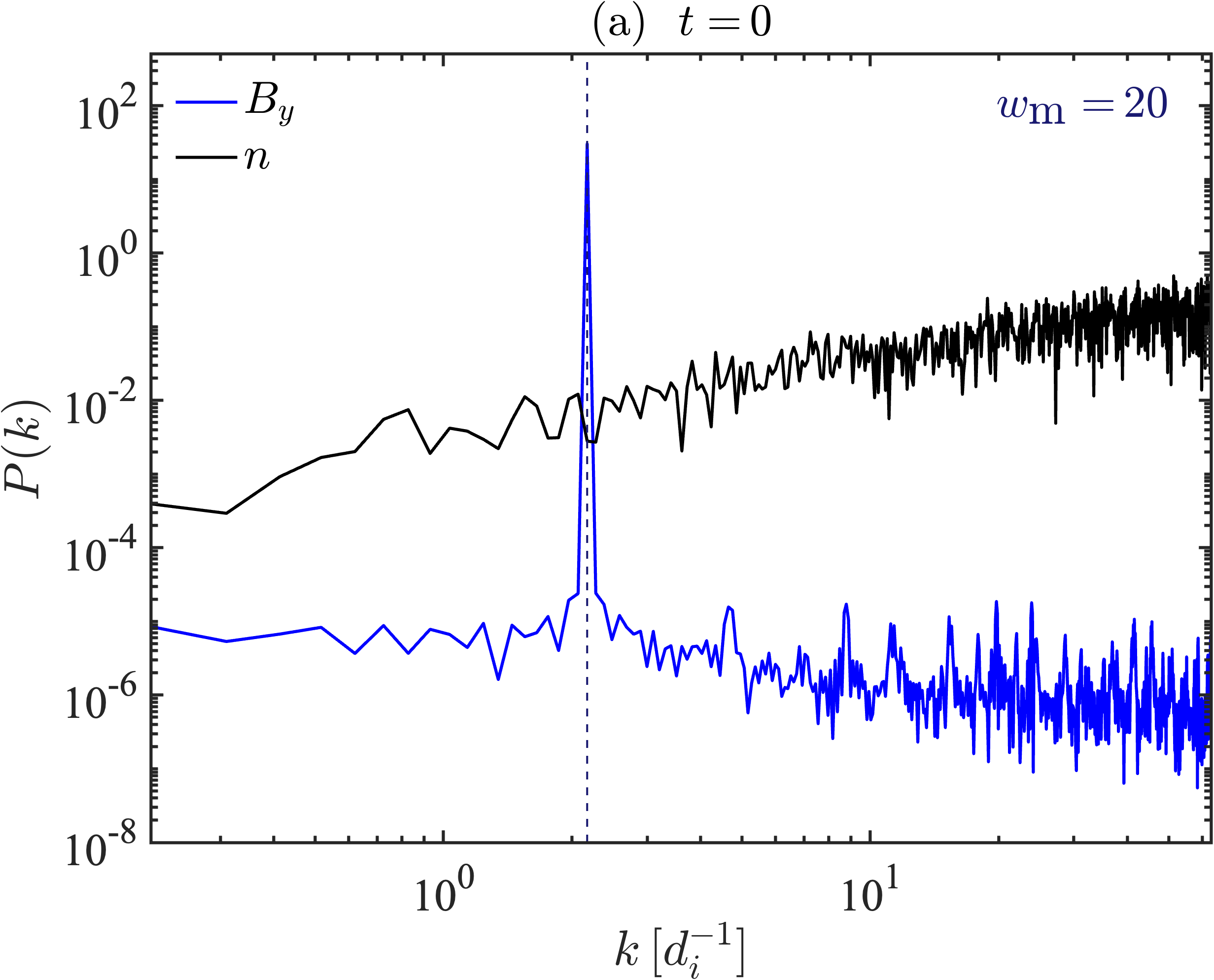}
    \label{fig:runB1DspectrumT0}
\end{subfigure}
\hfill
\begin{subfigure}{0.48\textwidth}
    \includegraphics[width=\textwidth]{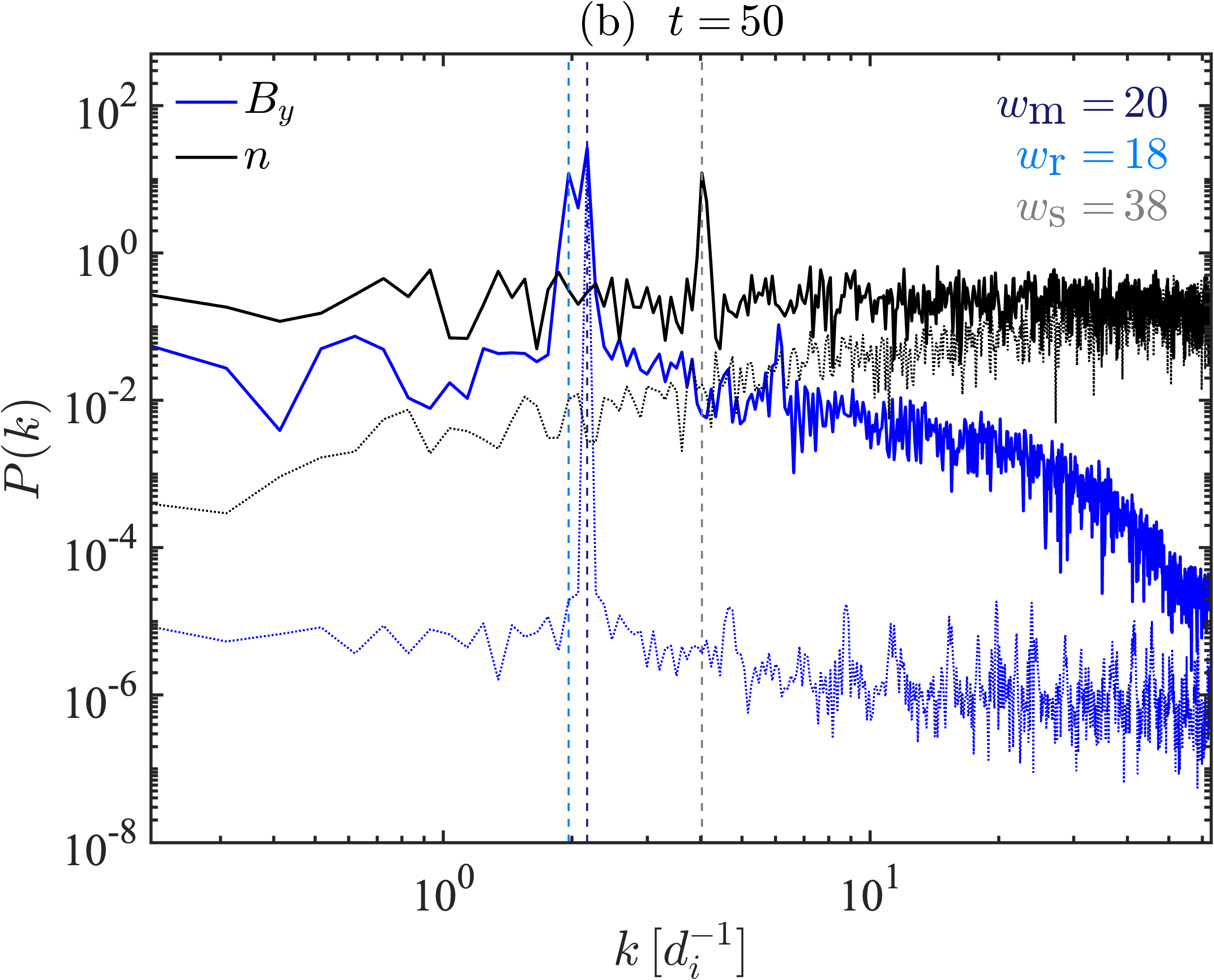}
    \label{fig:runB1DspectrumT50}
\end{subfigure}
\hfill
\begin{subfigure}{0.48\textwidth}
    \includegraphics[width=\textwidth]{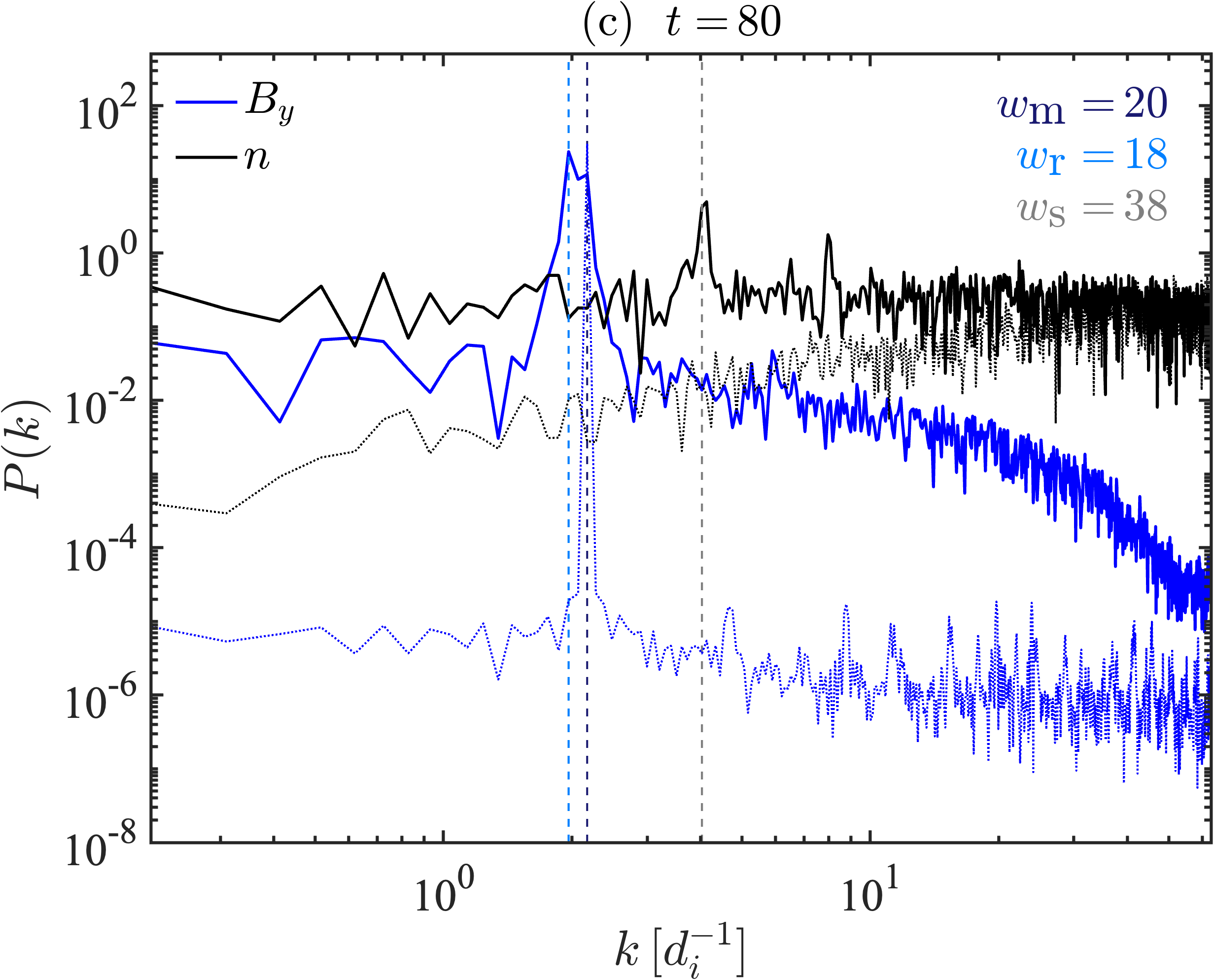}   \label{fig:runB1DspectrumT80}
    \end{subfigure}
    \hfill
\begin{subfigure}{0.48\textwidth}
    \includegraphics[width=\textwidth]{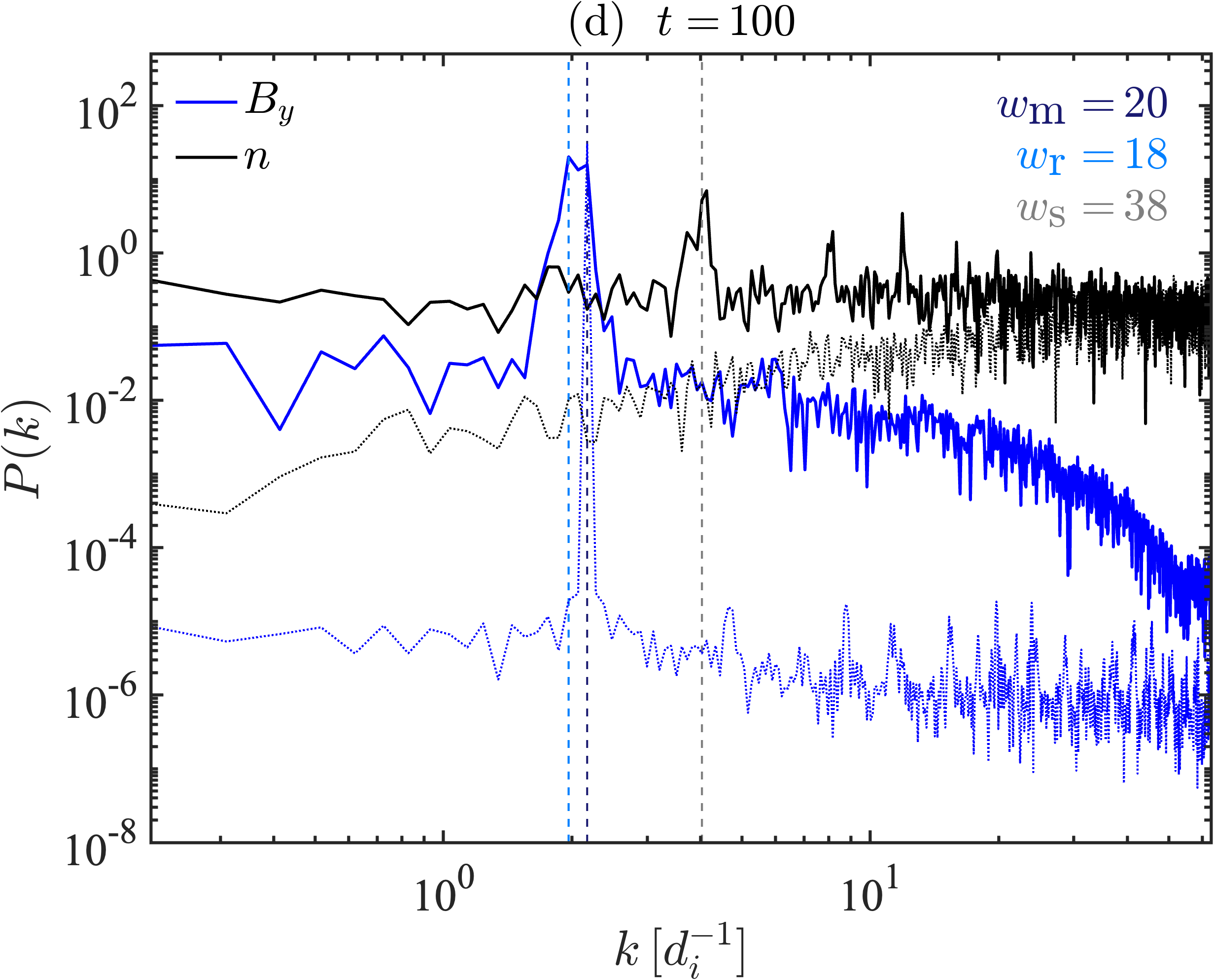}    \label{fig:runB1DspectrumT100}
\end{subfigure}
\hfill
\begin{subfigure}{0.49\textwidth}
    \includegraphics[width=\textwidth]{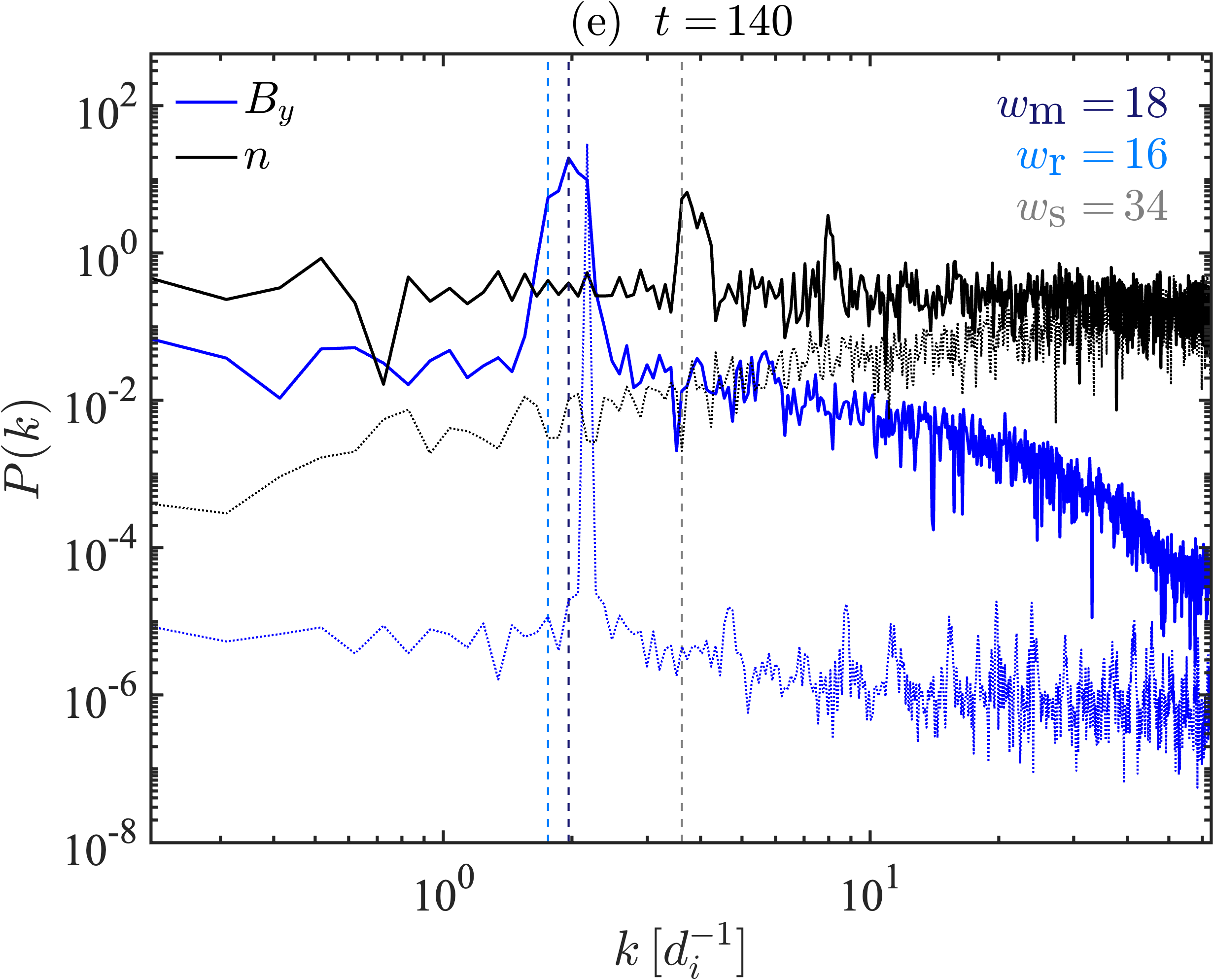}    \label{fig:runB1DspectrumT140}
\end{subfigure}
\hfill
\begin{subfigure}{0.49\textwidth}
    \includegraphics[width=\textwidth]{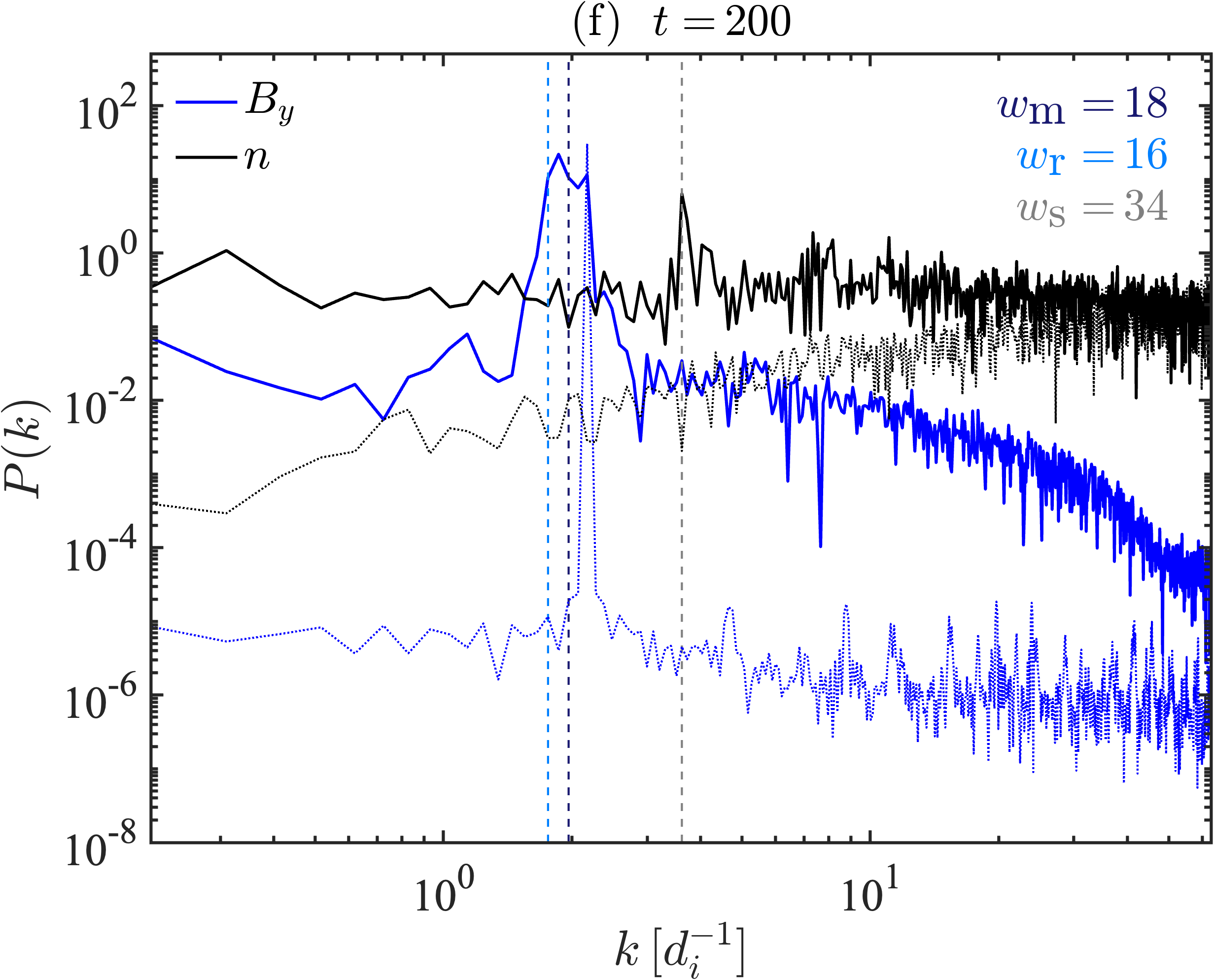}    \label{fig:runB1DspectrumT200}
\end{subfigure}
        
\caption{Run C: Power spectra of the $y$ component of the magnetic field (blue) and density (black) at various time. Values at $t = 0$ are overlaid as thinner dashed lines of the same color. Vertical lines highlight significant peaks.
} 
\label{fig:runB1Dspectrum}
\end{figure}

The double decay process is effectively investigated through the temporal evolution of the power spectrum, presented in Figure \ref{fig:runB1Dspectrum}. At the initial stage (t = 0, Panel a),  the spectrum prominently displays only the mother wave peak at $k = 2.07$. Subsequently, a rapid decay process generates a higher frequency compressive wave, identified by the density peak (black line) at $k_s = 3.93$ ($w_s = 38$), which becomes prominent at $t = 50$ (Panel b), simultaneously with the manifestation of a backward Alfv\'en wave as a small peak in the blue line at $k = 1.86$ ($w_r = 18$). The amplitude of this backward daughter wave undergoes rapid growth, exceeding the original mother wave peak by t = 80 (Panel c). This rapid evolution culminates in the appearance of a secondary density peak at $k = 3.52$ ($w_r=34$) by t = 100 (Panel d). This newly formed peak quickly increases in magnitude, exceeding the initial density peak at $t = 140$ (Panel e). At the same time, a third magnetic field peak, representing a granddaughter wave, emerges at an even lower wavelength $k= 1.66$ ($w_r=16$,  vertical blue dashed-dotted line), corresponding to a second decay process: what was previously the daughter wave ($w_r=18$) decays through the coupling with the new compressive wave ($w_s = 34$), such that the resonant condition for wave numbers is satisfied for the second decay. The granddaughter wave rapidly increases and becomes the most prominent feature in the magnetic spectrum, as shown in Panel f (t = 200).

        

\begin{figure}
\centering
\begin{subfigure}{0.6\textwidth}
    \includegraphics[width=\textwidth]{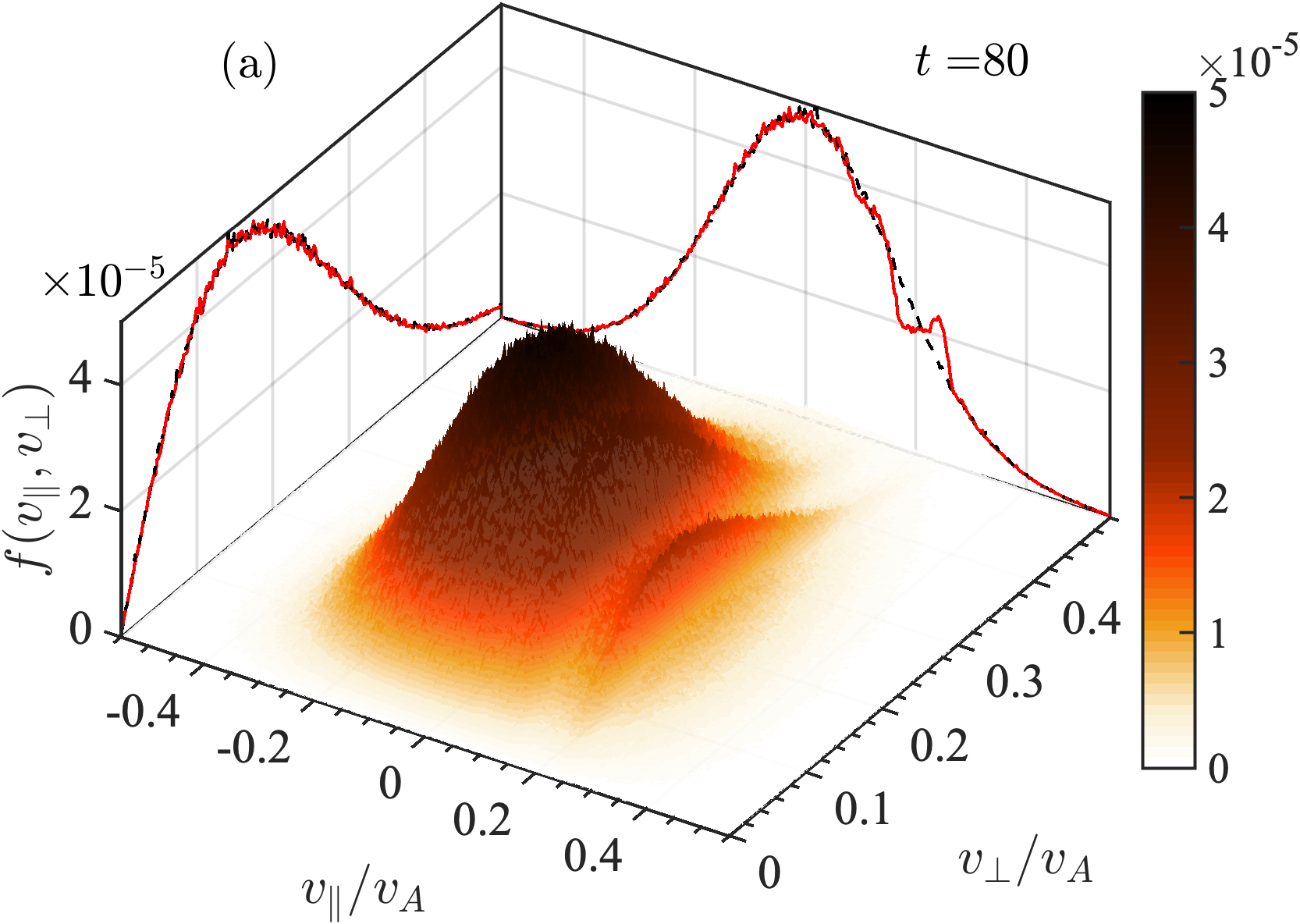}
    \label{fig:runCvparT80}
\end{subfigure}
\hfill
\begin{subfigure}{0.6\textwidth}
    \includegraphics[width=\textwidth]{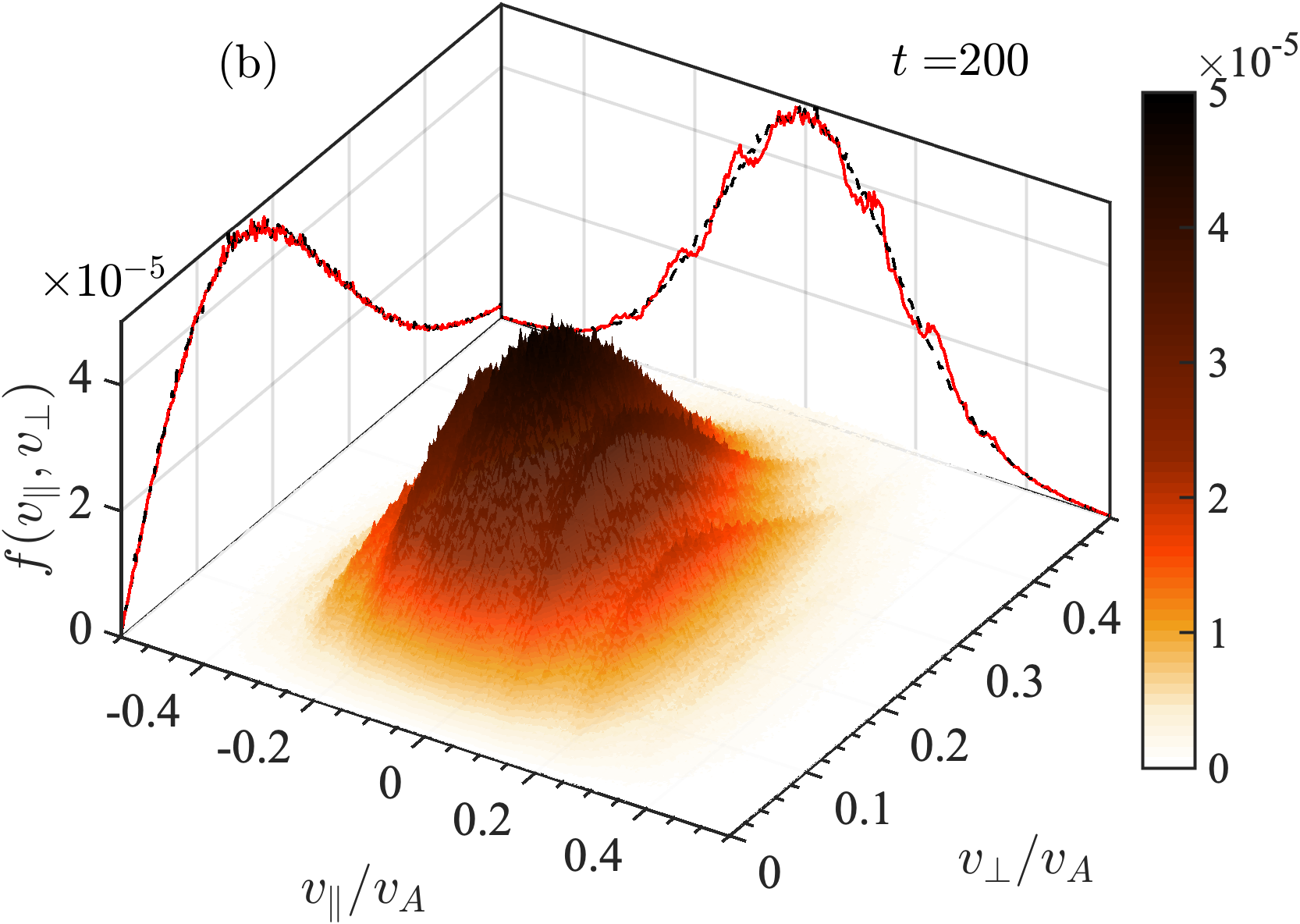}
    \label{fig:runCvparT200}
\end{subfigure}
        
\caption{Run C: Ion VDF displayed as a 3D color scale. Red lines represent VDF projections on the parallel and perpendicular directions. Black dashed lines denote the corresponding projections of the initial distribution (t=0).
} 
\label{fig:runCvdf}
\end{figure}

The emergence of both daughter and granddaughter waves has been previously documented \citep[e.g.][]{kojima1989nonlinear,del2001parametric}, with an associated outcome of ion acceleration and heating \citep{umeda2018decay}. Notably, their influence on the ion VDF is the generation of ion beams propagating in both positive and negative directions along the x-axis. Indeed, as illustrated by the ion VDF (Figure \ref{fig:runCvdf}), at t=80 (Panel a), after the daughter wave's development,  distinct faster ion populations emerge, corresponding to peaks at $v_{\parallel}/v_A \approx 0.1$ and $v_{\parallel}/v_A \approx 0.25$. Subsequently, following the second decay and the emergence of the granddaughter wave, the VDF exhibits a symmetric shape, with additional ion beams appearing at $v_{\parallel}/v_A \approx -0.1$ and $v_{\parallel}/v_A \approx -0.25$ (Panel b, t= 200).

\subsection{Lower beta: toward realistic ionospheric regimes}

\begin{figure}
\centering
\begin{subfigure}{0.49\textwidth}
    \includegraphics[width=\textwidth]{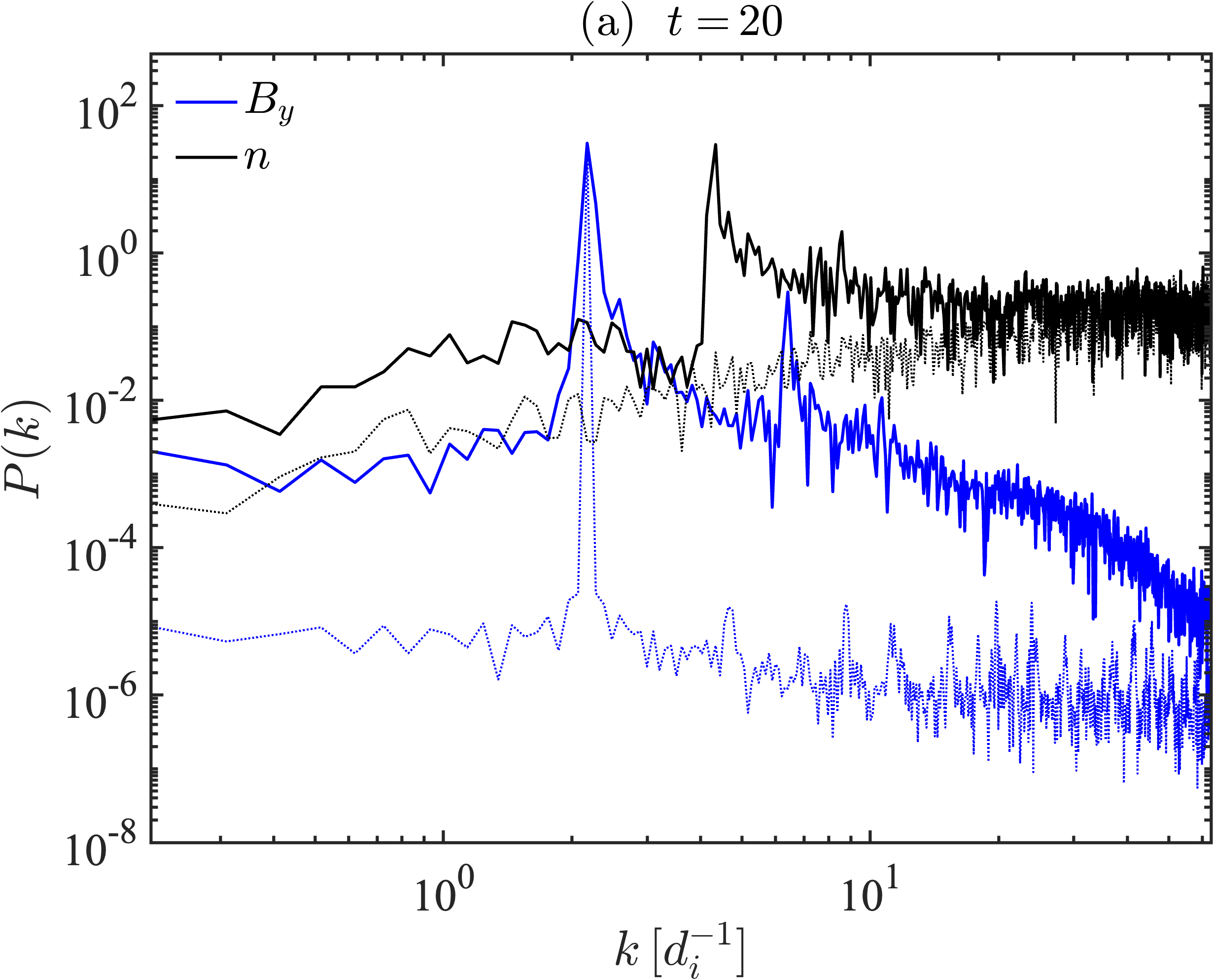}
    \label{fig:runE1DspectrumT20}
\end{subfigure}
\hfill
\begin{subfigure}{0.49\textwidth}
    \includegraphics[width=\textwidth]{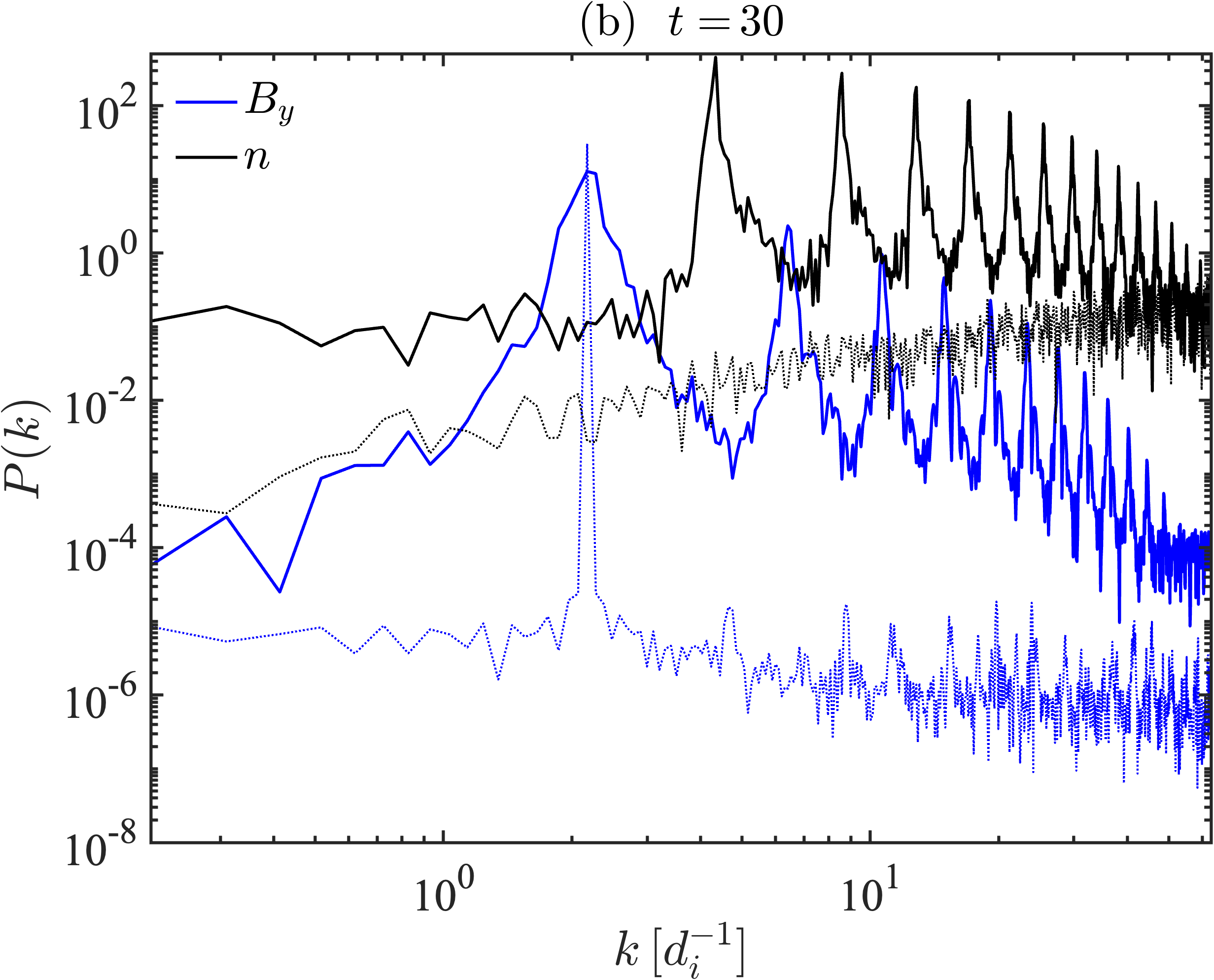}
    \label{fig:runE1DspectrumT30}
\end{subfigure}
\hfill
\begin{subfigure}{0.49\textwidth}
    \includegraphics[width=\textwidth]{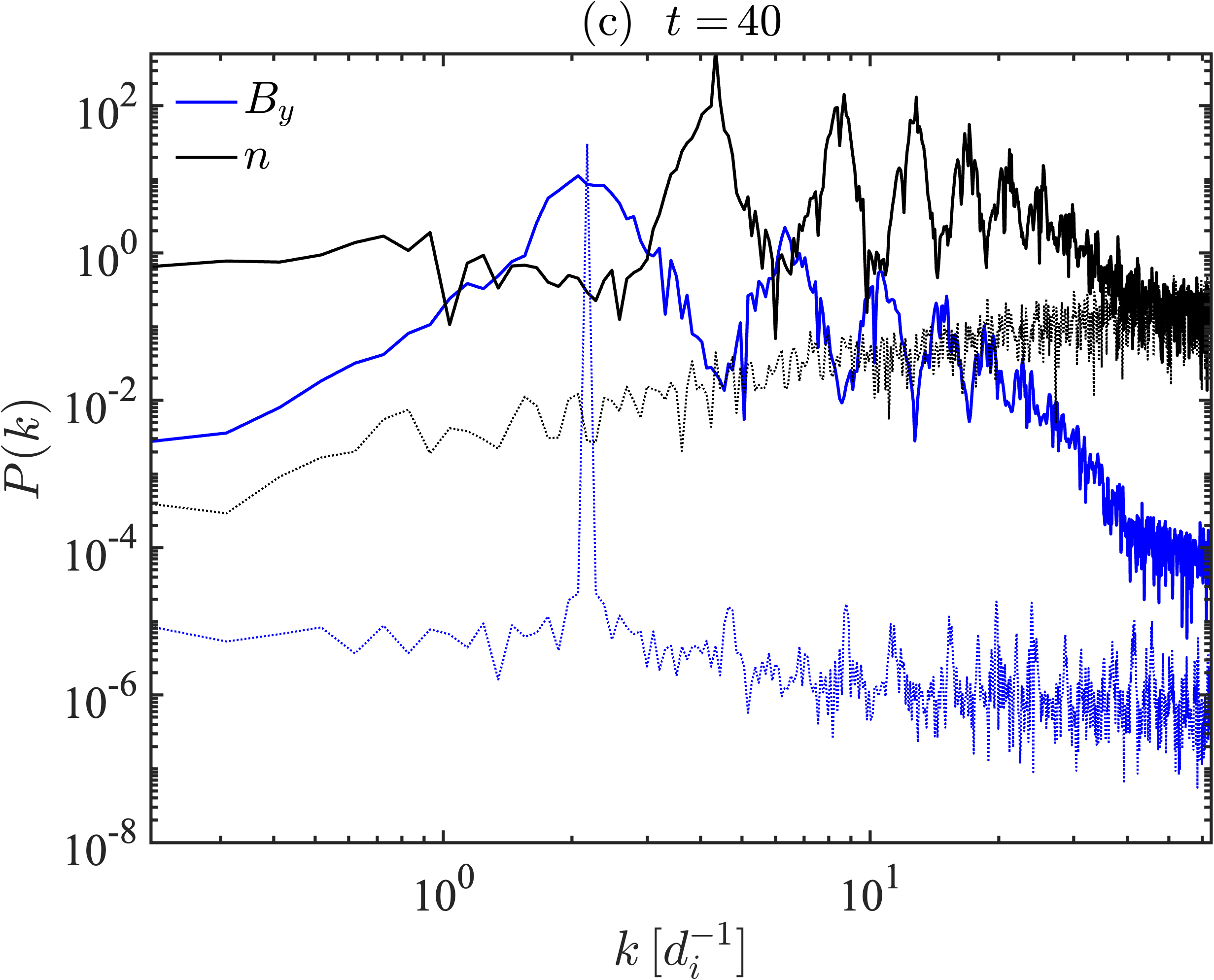}
    \label{fig:runE1DspectrumT50}
    \end{subfigure}
    \hfill
\begin{subfigure}{0.49\textwidth}
    \includegraphics[width=\textwidth]{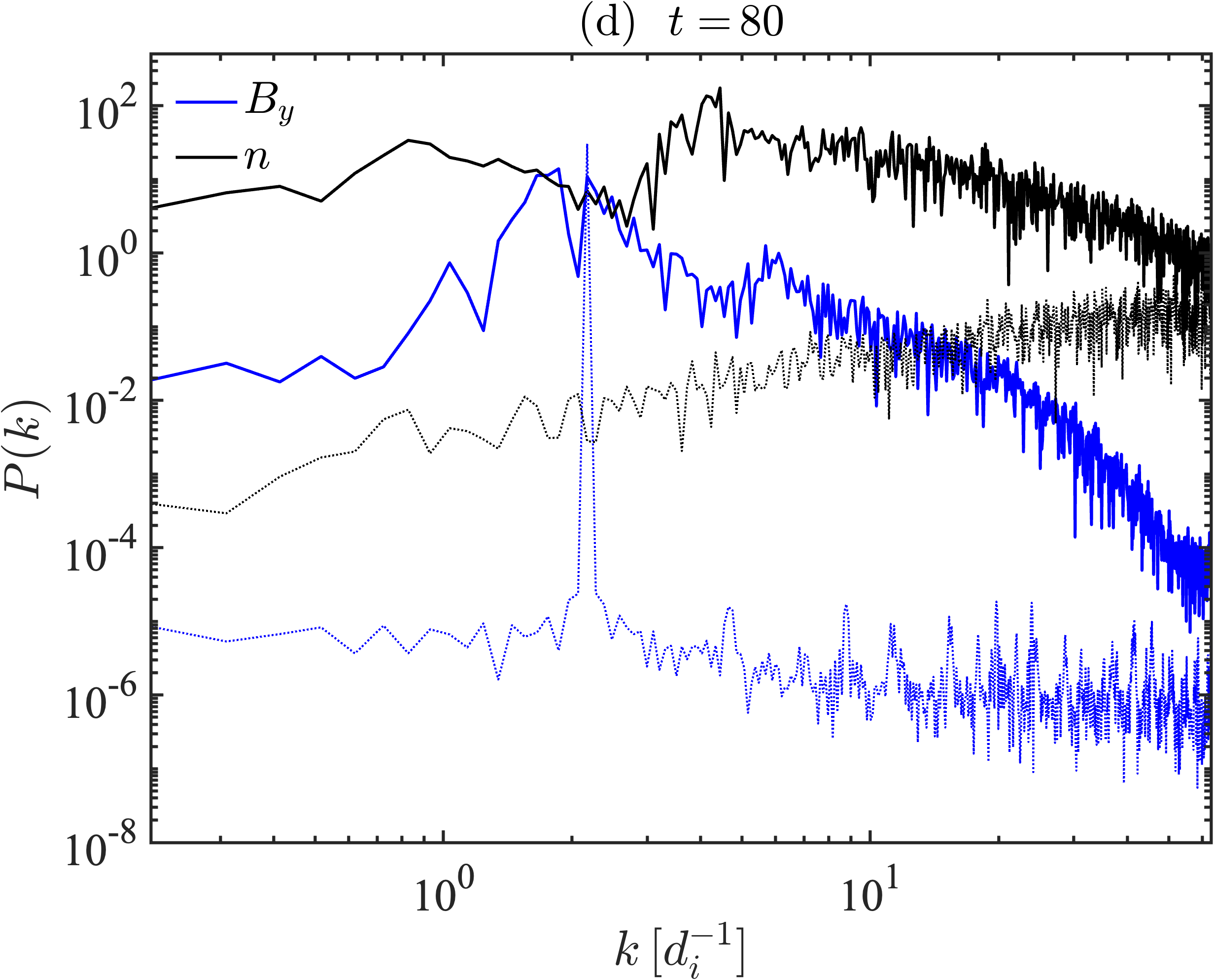}
    \label{fig:runE1DspectrumT80}
\end{subfigure}
        
\caption{Run E: Power spectra of the $y$ component of the magnetic field (blue) and density (black) at different times. Power spectra at $t = 0$ are overlaid as thinner dashed lines of the same color.
} 
\label{fig:runE1Dspectrum}
\end{figure}
\begin{figure}
\centering
\begin{subfigure}{0.49\textwidth}
    \includegraphics[width=\textwidth]{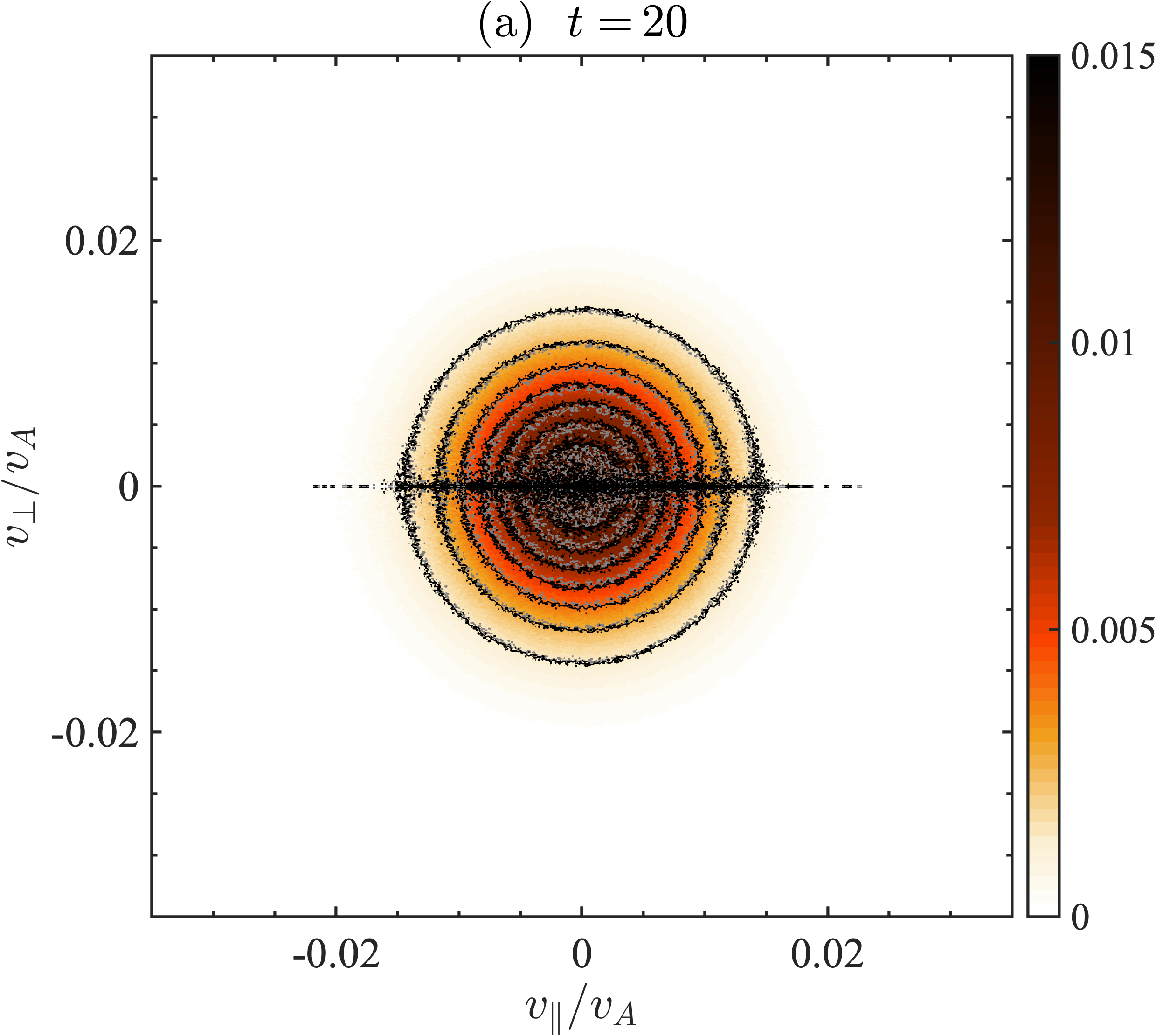}
    \label{fig:runEvdfT20}
\end{subfigure}
\hfill
\begin{subfigure}{0.49\textwidth}
    \includegraphics[width=\textwidth]{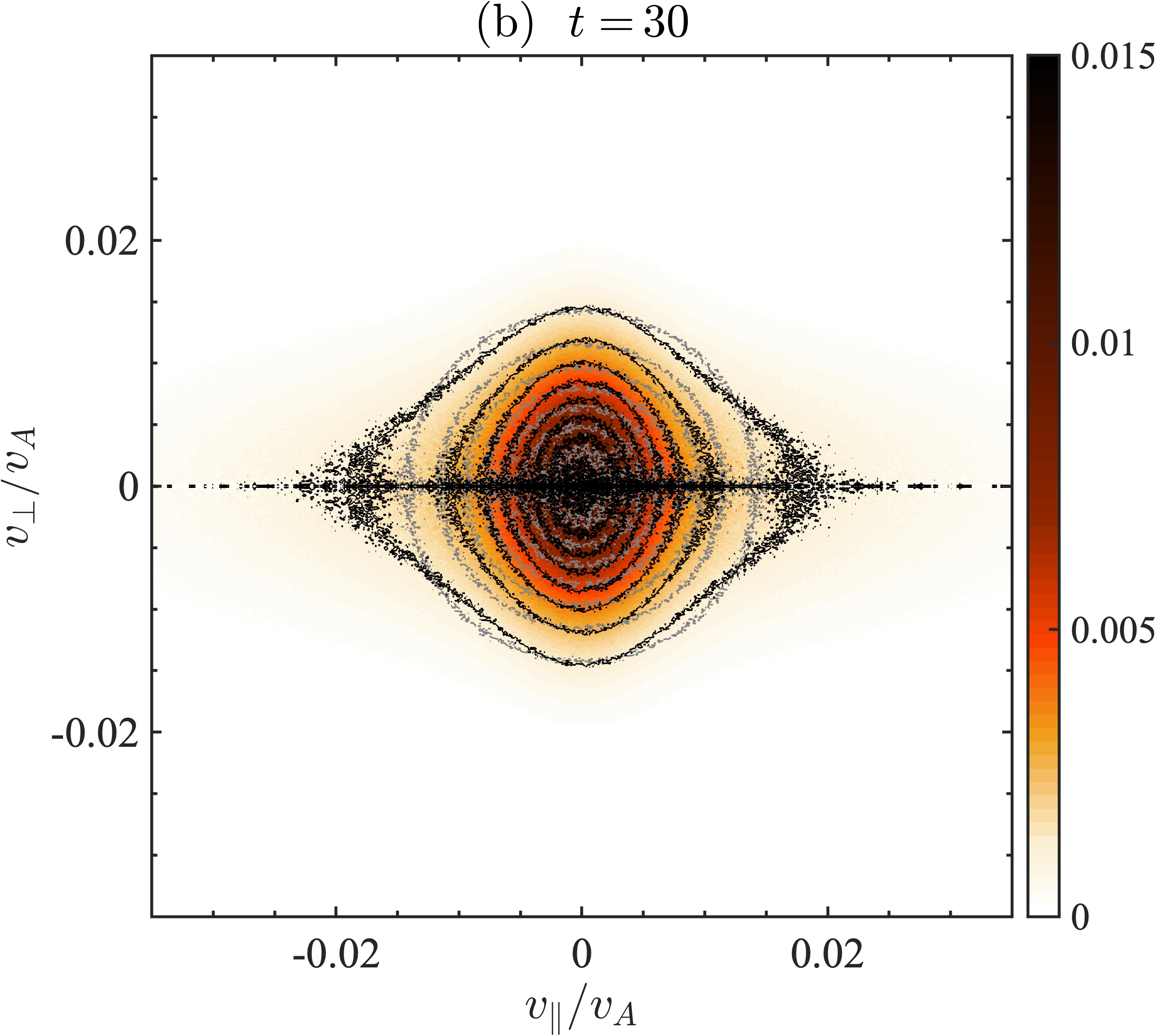}
    \label{fig:runEvdfT30}
\end{subfigure}
\hfill
\begin{subfigure}{0.49\textwidth}
    \includegraphics[width=\textwidth]{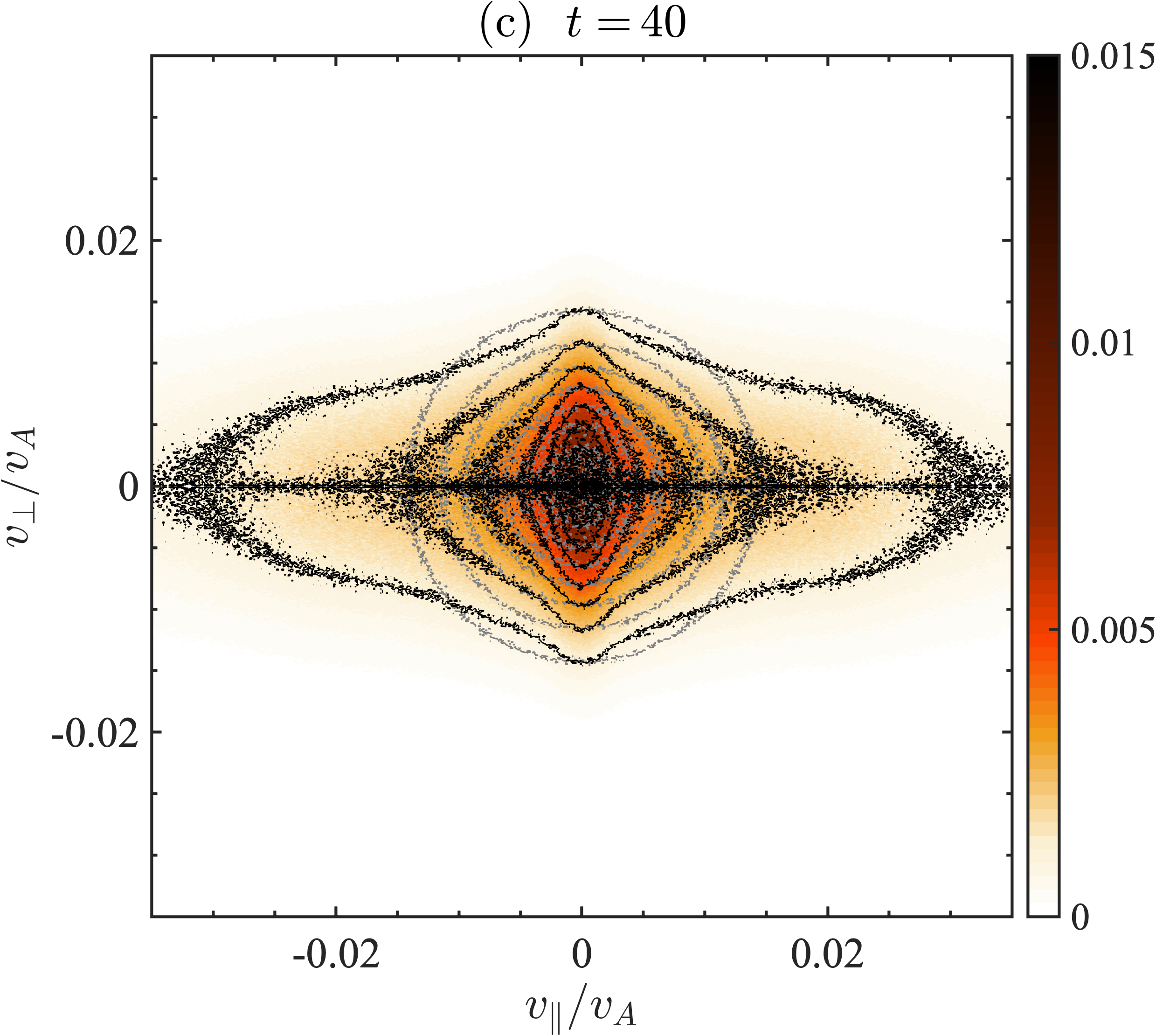}
    \label{fig:runEvdfT40}
    \end{subfigure}
    \hfill
\begin{subfigure}{0.49\textwidth}
    \includegraphics[width=\textwidth]{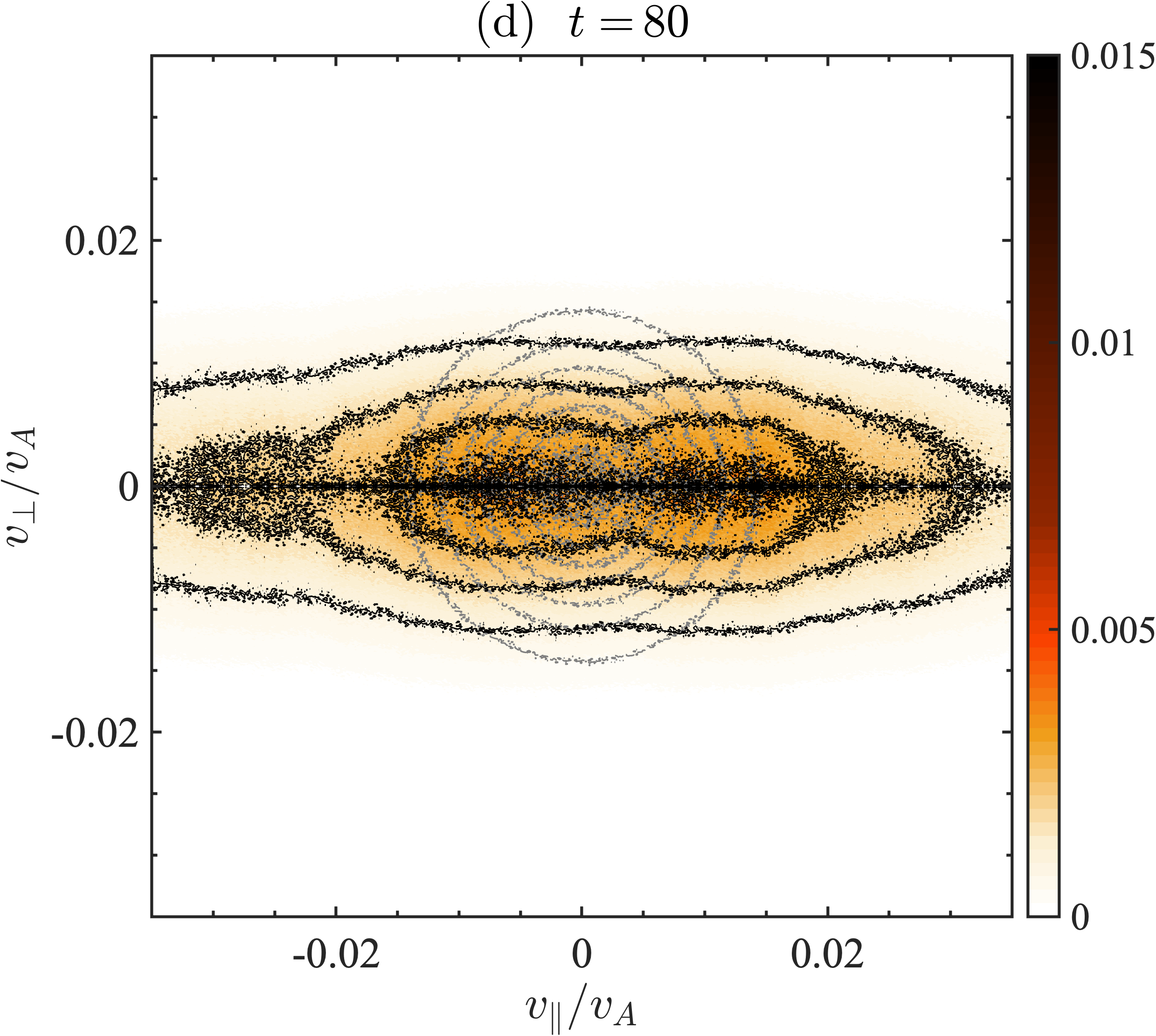}
    \label{fig:runEvdfT80}
\end{subfigure}
        
\caption{Run E: Gyro-averaged VDF in the $v_\parallel$-$v_\perp$ plane. Gray thin points represent the initial configuration (t=0).
} 
\label{fig:runEvdf}
\end{figure}

As indicated in Table \ref{tab:ionoParam}, the Earth's ionospheric plasma environment is characterized by significantly lower plasma beta values compared to those considered in Runs A, B, or C. To investigate the effects of PDI on lower-beta plasmas, we performed additional simulations (Run D and Run E) by progressively reducing both ion and electron beta values, with all other parameters maintained at the values established in Run C (see Table \ref{tab:runDE}). Reducing the plasma beta, as in Run D ($\beta_i = B_e = 10^{-3}$) or in Run E ($\beta_i = B_e = 10^{-4}$), shows the expected dependence of the PDI on the plasma beta. Indeed, the PDI growth rate exhibits a faster evolution, leading to a more rapid transfer of energy from the magnetic field to ion kinetic energy and a quicker, more pronounced modification of the VDF. Figures \ref{fig:runE1Dspectrum} and \ref{fig:runEvdf}, presenting respectively the power spectrum and the gyro-averaged 
VDF at significant times for Run E, illustrate this effect. 

\begin{table}
	\centering
	\begin{tabular}{cccccccccc}
	\hline
	Run & $\beta_e$ & $\beta_i$ & $\delta B/B_0$ & $w_m$ & Pol. & $L_{\textrm{box}}(d_i)$ & $k_0(d_i^{-1})$ & $k_m(d_i^{-1})$ & $ppc$\\
	\hline
	D & $1\cdot10^{-3}$ & $1\cdot10^{-3}$ & $5\cdot10^{-2}$ & $20$ & R & $60.8$ & $0.1$ & $2.07$ & $10^4$\\
	\hline
	E & $1\cdot10^{-4}$ & $1\cdot10^{-4}$ & $5\cdot10^{-2}$ & $20$ & R & $60.8$ & $0.1$ & $2.07$ & $10^4$\\
	\hline
	E2 & $1\cdot10^{-4}$ & $1\cdot10^{-4}$ & $5\cdot10^{-2}$ & $20$ & R & $60.8$ & $0.1$ & $2.07$ & $8\cdot 10^4$\\
	\hline
	\end{tabular}
	\caption{Run D, E and E2 parameters.}
    \label{tab:runDE}
\end{table}

As evident from Figure \ref{fig:runE1Dspectrum}, a density peak (black) emerges at $k = 3.93$ as early as t=20 (Panel a). By t = 30 (Panel b), the density spectrum exhibits multiple peaks, appearing as harmonics of the initial peak. These density peaks undergo rapid growth and subsequent decay, showing reduced amplitude and broadening by t = 40 (Panel c) and complete disappearance by t = 80 (Panel d). In the corresponding magnetic field spectrum (blue line), a secondary harmonic begins to emerge at t=20. By t=30, the mother wave's magnetic field peak at $k = 2.07$ shows a reduction in amplitude. Concurrently, the magnetic field spectrum displays a pattern similar to that of the density, characterized by multiple evenly spaced peaks intermediate to the density peaks. Similar to their density counterparts, these magnetic field harmonics rapidly diminish and vanish. Notably, no clear secondary peak at lower $k$ corresponding to a daughter wave is observed.

\begin{figure}
\centering
\includegraphics[width=0.49\textwidth]{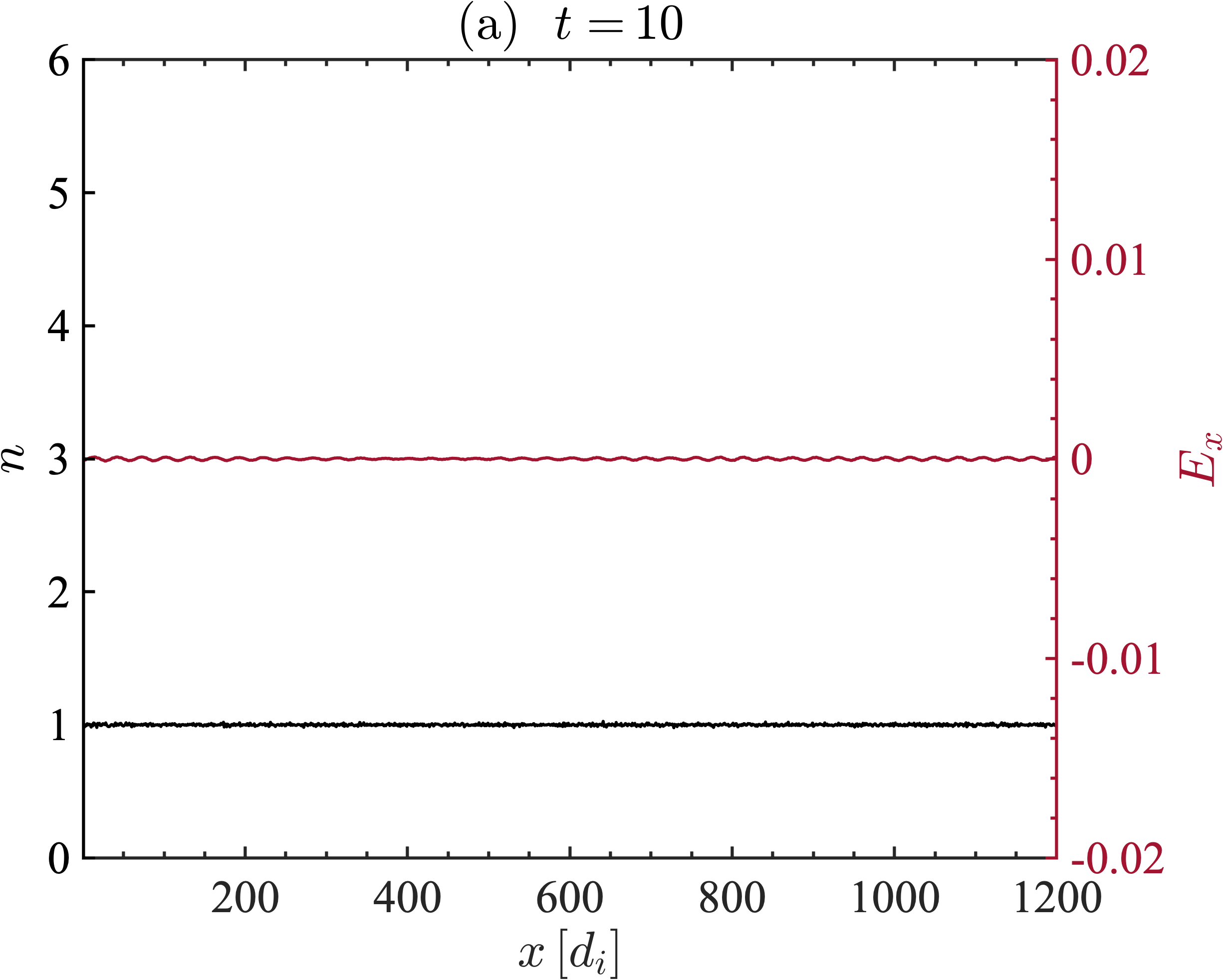}
\includegraphics[width=0.49\textwidth]{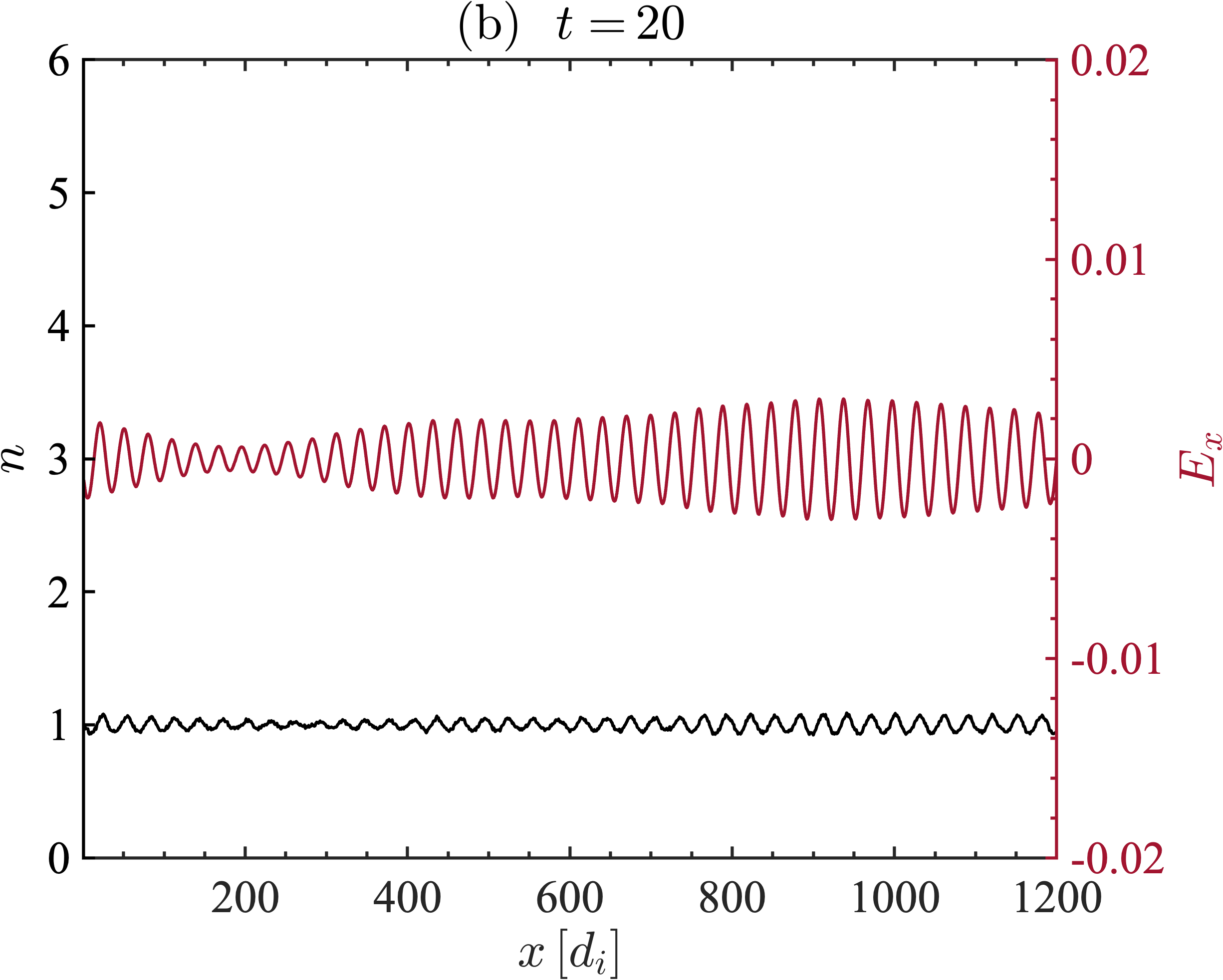}\\
\includegraphics[width=0.49\textwidth]{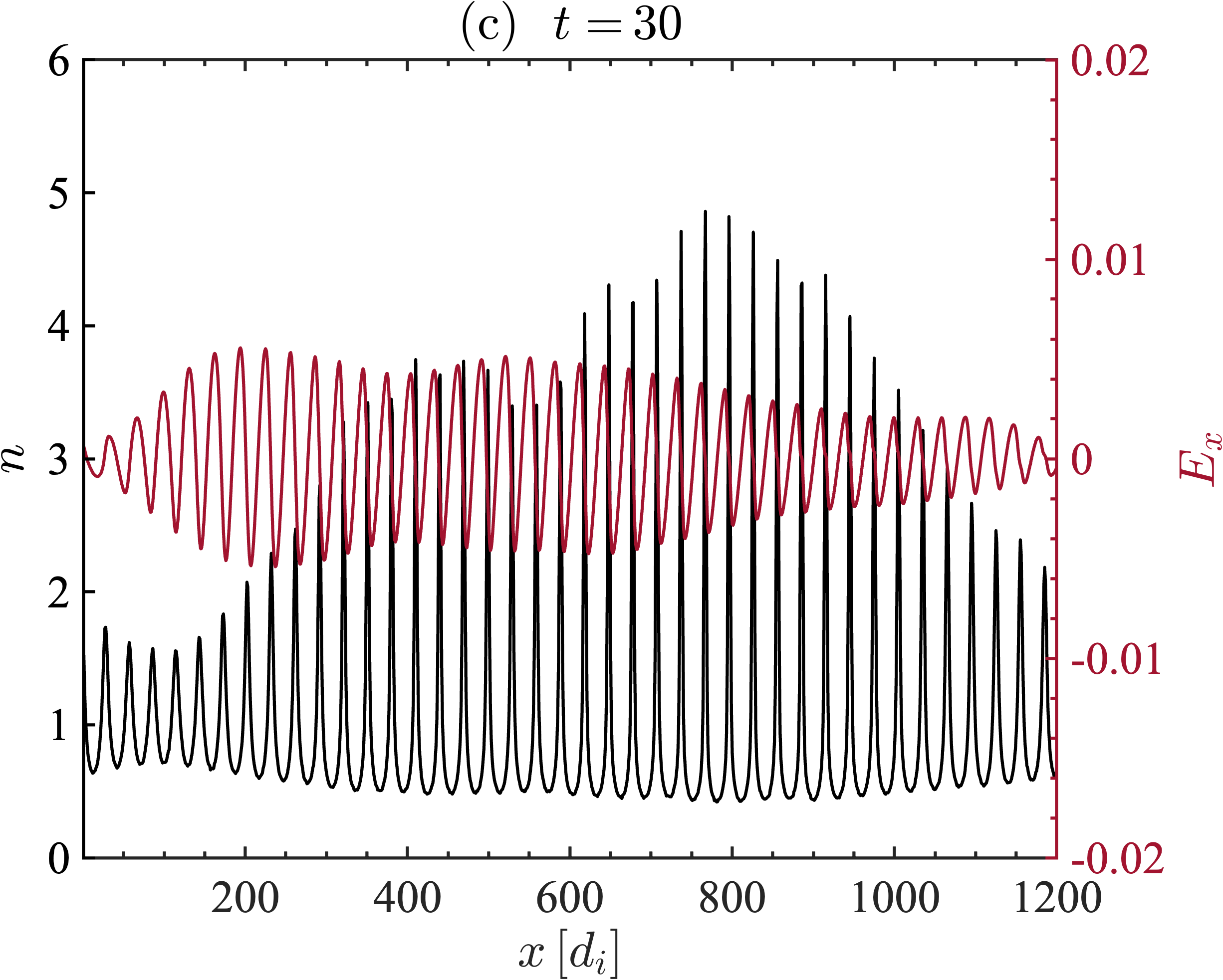}
\includegraphics[width=0.49\textwidth]{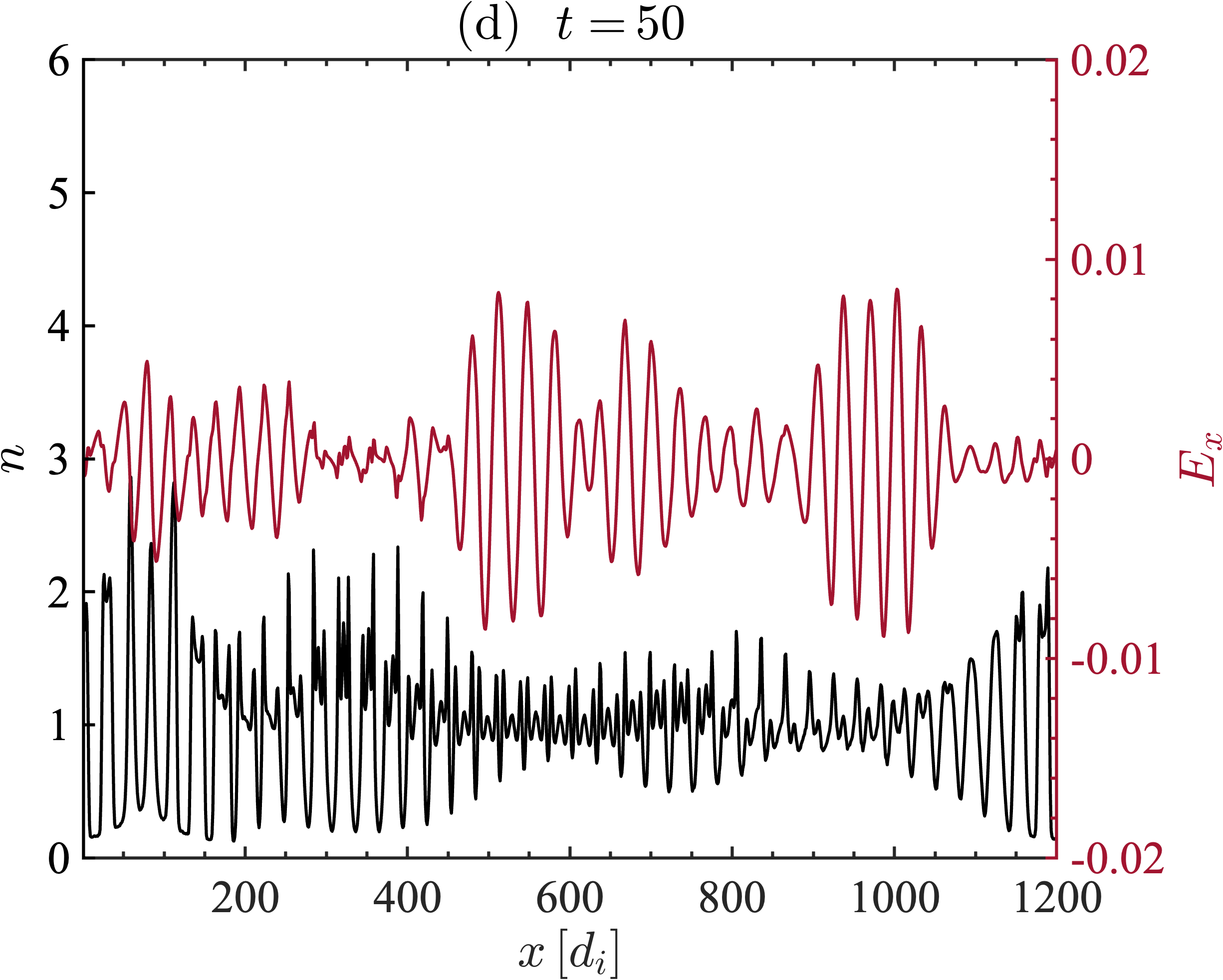}
\caption{Run E: Density (black) and parallel electric field ($E_x$, red) profiles along the simulation box at significant simulation times.
} 
\label{fig:runEdensityExProfile}
\end{figure}

The observed rapid growth and disappearance of these large peaks in the spectrum, which correspond to the harmonics of the first density peak, can be elucidated by examining the temporal evolution of the density profiles along the simulation box, depicted in Figure \ref{fig:runEdensityExProfile} with a black line. As shown in Panel b, at t=20, when the first density peak on the spectrum has developed, only minor fluctuations around the baseline value ($n=1$) are observed. Subsequently, the density fluctuations undergo rapid increase. By t=30 (Panel c), when the spectrum is characterized by the presence of several large-amplitude prominent density peaks, the density profile manifests numerous steep fronts. These fronts exhibit a rapid reduction in amplitude, as discernible in Panel d (t=50).

The generated density fronts, in turn, induce parallel electric field ($E_x$) steep fronts, as depicted by the red line in Figure \ref{fig:runEdensityExProfile}. These $E_x$ fronts represent a viable mechanism for proton acceleration and heating \citep{gonzalez2021proton}, thereby explaining the expansion of the VDF in both positive and negative parallel directions shown in Figure \ref{fig:runEvdf}, where we present the gyro-averaged VDF $f(v_\parallel,v_\perp)$ in the $v_\parallel$-$v_\perp$ plane. 
The rapid broadening of the VDF evident in Figure \ref{fig:runEvdf} directly implies proton acceleration and heating. Although qualitatively analogous behavior has been described in prior literature  \citep[e.g.][]{matteini2010kinetics,nariyuki2014ion,gonzalez2021proton}, our study's exceptionally low beta values ($\beta_e=\beta_i = 10^{-4}$) distinguish it from these previous works, leading to a new VDF modification, not observed therein. 

Nevertheless, the observed effect is robust and not attributable to numerical artifacts. This has been verified through an identical simulation that uses an increased number of particles per cell (Run E2, $ppc = 8 \cdot 10^4$), which yielded consistent results (not shown).

\begin{figure}
  \centering
\includegraphics[width=0.75\linewidth]{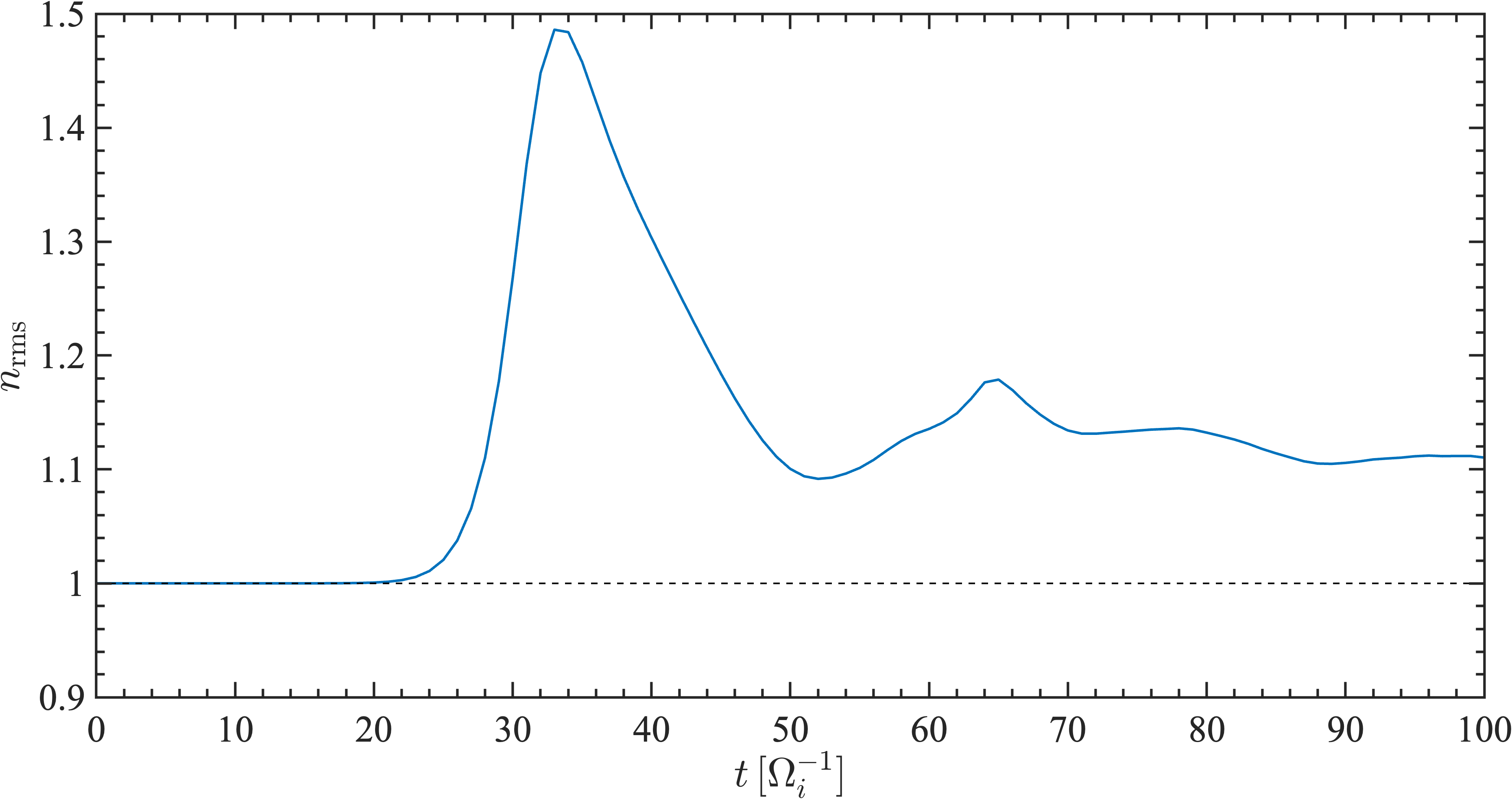}
  \caption{Run E: Temporal evolution
of RMS of density fluctuations. A dashed black line is included to indicate the reference value of unity.}
\label{fig:runErmsDensity}
\end{figure}

Instead, this is a physical effect and is related to the modulation of the total magnetic field profile that are induced by the pump wave during its coupling with density variations. The resulting modulation of the magnetic pressure ($\propto B^2$), triggers an analogous modulation of the plasma pressure, following approximately pressure balance. 
Indeed, the root mean square (RMS) value of the density fluctuations, $n_{\mathrm{rms}}$, shown in Figure \ref{fig:runErmsDensity}, confirms the rapid generation of density variations necessary to equilibrate the magnetic pressure. These begin to increase after t=20, reaching a maximum at approximately t=36. At this point, the VDF already exhibits significant spreading in the parallel direction compared to its initial configuration (see Figure \ref{fig:runEvdf}). Consequently, due to ion heating, we observe a substantial increase in the parallel temperature $T_\|$, leading to a stronger contribution of the temperature to the thermal pressure. At this point, high density variations are no longer required to maintain pressure equilibrium with the magnetic field variations. Therefore, density fluctuations decrease to a minimum around $t=50$ and subsequently display a more gentle oscillatory behavior, mirroring a similar oscillatory pattern of the ion VDF after $t=50$.

\begin{figure}
\centering
\includegraphics[width=0.7\textwidth]{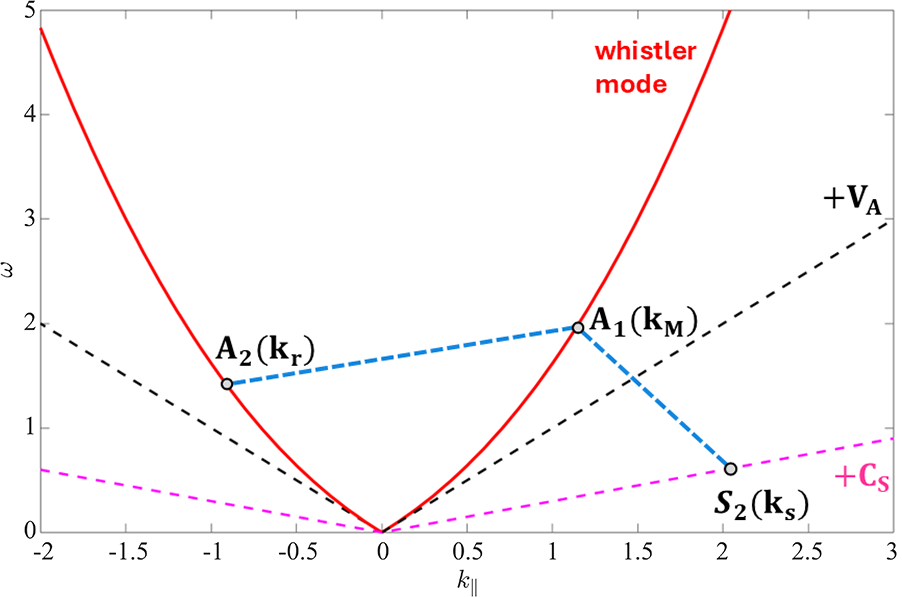}\\
\includegraphics[width=0.7\textwidth]{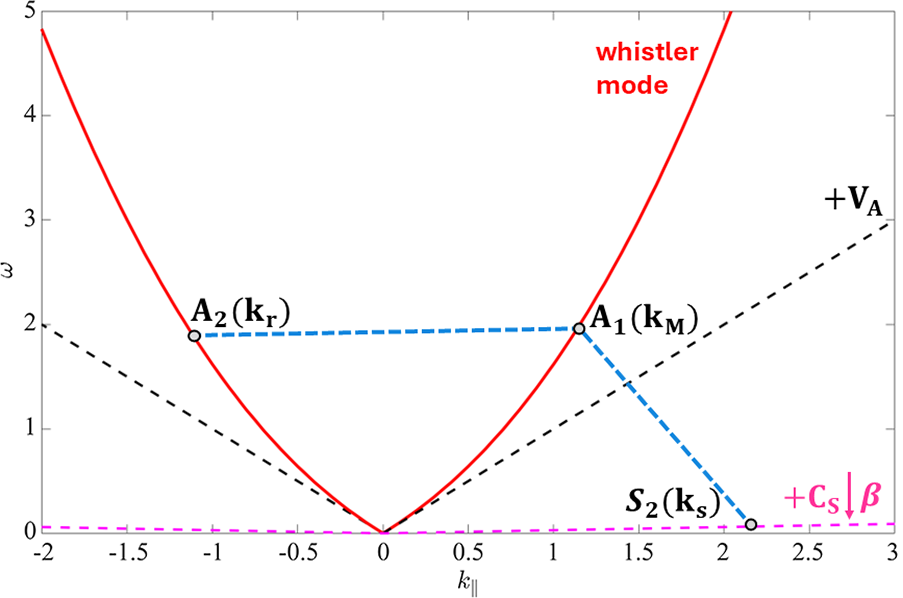}
\caption{Three-wave couplings of the decay instability in the frequency–wavenumber domain, where the wavevector is parallel to the mean magnetic field, for $\beta > 10^{-3}$ (top panel) and for $\beta \lesssim 10^{-3}$  (bottom panel). The lines for $\omega/k_\parallel = V_A$ are dashed black, while the ones for $\omega/k_\parallel = C_S$ are dashed magenta. The whistler mode branch is represented as solid red lines. 
} 
\label{fig:parametrica}
\end{figure}

The absence of a distinct peak corresponding to the daughter wave in the magnetic power spectrum (Figure \ref{fig:runE1Dspectrum}) can be elucidated by considering the specific behavior of the PDI at very low beta, using the frequency-wavevector domain ($k,\omega$). As detailed e.g. in \citet{comicsel2019multi} and illustrated in Figure \ref{fig:parametrica}, wave–wave coupling in the decay instability forms a parallelogram in the frequency–wavenumber domain, ensuring the conservation of both energy and momentum corresponding to $\omega_m=\omega_r+\omega_s$ and $k_m=k_s+k_r$, respectively,  throughout the wave decay process. The pump Alfv\'en wave (denoted by $A_1$ in Figure \ref{fig:parametrica}), propagating parallel to the mean magnetic field with wavenumber $k_m$, decays into a backward (anti-parallel) propagating Alfv\'en mode ($A_2$, wavenumber $k_r$) and a forward-propagating sound wave ($S_2$, wavenumber $k_s$). However, given that ${V_A}/{C_S} \propto 1/{\sqrt{\beta}}$, a lower plasma beta value results in a greater difference between the slopes of the $\omega/k = V_A$ and $\omega/k = C_S$ lines, being $C_S \ll V_A$. As shown in the bottom panel, for ultra-low-beta values ($\beta \lesssim 10^{-3}$), the configuration of the parallelogram in the frequency-wavenumber domain yields a reflected daughter wave wavenumber ($k_r$) that is nearly equal to the mother wave wavenumber ($|k_r| \simeq |k_m|$).

While not explicitly shown, the results from Run D, an intermediate scenario between Run C and Run E, further confirm the observations from Run E. Specifically, the ion VDF exhibits broad expansion in both positive and negative parallel directions, though it is characterized by a slightly slower temporal evolution and a less pronounced modification with respect to $t=0$ compared to Run E.

\subsection{Realistic perturbing wave}

An increased mother wave amplitude is known to accelerate the decay process and enhance the magnitude of density fluctuations \citep{matteini2010kinetics}. However, the mother wave amplitude of $\delta B/B_0 = 5\cdot 10^{-2}$ used in Runs A-E is  unrealistically high within an ionospheric environment. Considering the geomagnetic field's typical magnitude of $10^4$ nT \citep{campbell2003introduction}, such perturbations would correspond to several hundreds or thousands of nT, making them particularly rare \citep{le2011c}. Consequently, Run F was performed as a modified replication of Run E2, wherein the pump wave's amplitude was reduced by a factor of 10 ($\delta B/B_0 = 5\cdot 10^{-3}$, refer to Table \ref{tab:runDE}). This not only enabled us to model a more representative ionospheric perturbation, but also allowed us to investigate the kinetic behaviour of the PDI in a regime characterized by a $\delta B/B_0$ notably smaller than that commonly found in existing literature, as \citet{matteini2010kinetics} consider $\delta B/B_0$ from 0.05 to 0.5; \citet{comicsel2019multi} use  $\delta B/B_0=$ 0.2; \citet{gonzalez2023particle} employ $\delta B/B_0=$ 0.5).

Furthermore, the presence of a perfectly monochromatic, delocalized,  wave is atypical in the ionosphere, which is instead characterized by a complex electromagnetic environment with signatures at various frequencies and waves with rapid frequency variations such as whistlers, at specific locations. To model a more realistic ionospheric situation, Runs G and H therefore consider a mini-spectrum of perturbing waves rather than a single monochromatic wave. Run G was designed as a replication of Run F,  but in this case, the plasma is perturbed with a modulated wave-packet of Alfv\'en waves with wavenumbers ranging from $w_{min}=19$ to $w_{max}=21$, corresponding to wave vectors from $k_{min}=1.96\ d_i^{-1}$ to $k_{max}=2.17\ d_i^{-1}$.
To investigate the role of polarization within this parameter range, Run H is a replication of Run G but with an inverted wave polarization (left-handed instead of right-handed). A summary of the parameters for these runs is provided in Table \ref{tab:runFGH}.

\begin{table}
	\centering
	\begin{tabular}{cccccccccc}
	\hline
	Run & $\beta_e$ & $\beta_i$ & $\delta B/B_0$ & Pol. & $L_{\textrm{box}}(d_i)$ & $k_0(d_i^{-1})$ & $k_{min}(d_i^{-1})$ & $k_{max}(d_i^{-1})$ & $ppc$\\
	\hline
	F & $1\cdot10^{-4}$ & $1\cdot10^{-4}$ & $5\cdot10^{-3}$ & R & $60.8$ & $0.1$ & $2.07$ & - & $8\cdot 10^4$\\
	\hline
	G & $1\cdot10^{-4}$ & $1\cdot10^{-4}$ & $5\cdot10^{-3}$ & R & $60.8$ & $0.1$ & $1.96$ & 2.17 & $1\cdot 10^4$\\
	\hline
    H & $1\cdot10^{-4}$ & $1\cdot10^{-4}$ & $5\cdot10^{-3}$ & L & $60.8$ & $0.1$ & $1.96$ & 2.17 & $1\cdot 10^4$\\
	\hline
	\end{tabular}
	\caption{Run F, G and H parameters.}
    \label{tab:runFGH}
\end{table}

Results from Run F (not explicitly shown) confirm that a smaller amplitude pump wave leads to a slower decay and a lower level of density fluctuations. Specifically, the maximum density fluctuation ($\delta n = \mathrm{max}(n) -1$)
reached a value of  $\delta n \simeq 1.2$, compared to $\delta n \simeq 6$ for run E. The density peak in the spectrum emerges only at approximately $t=60$, and the numerous harmonic features present in Figure \ref{fig:runE1Dspectrum} are absent (as well as a clear daughter wave peak). This, in turn, implies a slower evolution and less pronounced modification of the VDF.
In this case
the VDF displays only a minor deviation from its initial state by $t=200$. 
Only by $t = 400$ 
the VDF exhibit a plateau around $v_\parallel/v_A = 0$ and small velocity beams at $v_\parallel/v_A \simeq \pm 5\cdot 10^{-3}$.

        


Similar modifications to the VDF are also evident in Run G, in which the plasma is perturbed by a wave-packet, as summarized in table \ref{tab:runFGH}. In this case, the modifications are of a slightly greater magnitude. A plausible explanation for this phenomenon involves ponderomotive forces compelling the plasma towards the static approximation ($n \propto B^2$), as described by \cite{spangler1982properties,spangler1989kinetic}. Additionally, the breakdown of this approximation due to $B^2$ modulation \citep{machida1987simulation,nariyuki2006remarks} and modulational instability of wave packets \citep{machida1987simulation,vasquez1993strongly,velli1999propagation,buti2000hybrid} may also contribute. Furthermore, \cite{nariyuki2007consequences} explained how an alternative way for the mother wave energy dissipation can be provided by the modulational instability driven by incoherent modes. Finally, \cite{khachatryan2005effect} showed that the energy gain of particles interacting with a varying-frequency (chirped) electromagnetic pulse increases with both the pulse amplitude and the chirp strength.

Run H exhibits the same VDF modifications as Run G and F,  but with a more pronounced effect and a faster evolution. 
Indeed, in the case of a left-handed polarized wave-packet, the perpendicular magnetic field fluctuations ($B_\perp$) shrink and grow in amplitude, as observed in \cite{velli1999propagation,buti2000hybrid, matteini2010kinetics}. Conversely, for a right-handed polarized spectrum, wave packets undergo gradual dispersion at advanced times, as also reported by \cite{matteini2010kinetics}. The results from Runs G and H confirm the behavior reported in these works. Specifically, $B_\perp$ fluctuations reach a value of approximately $3\cdot 10^{-2}$ for Run H (left-handed polarization) by $t=500$, whereas a value of $\delta B_\perp \simeq 1\cdot 10^{-2}$ is observed for Run G (right-handed polarization) at the same time.

\subsection{Realistic ionospheric environment}

As shown in Table \ref{tab:ionoParam}, the plasma at 500 km altitude is primarily composed of $O^+$ ions ($\simeq 94\%$). However, a non-negligible concentration of $H^+$ ions ($\simeq 4.2\%$) also  exists among other minor ion populations. To more accurately simulate ionospheric conditions, subsequent simulations (Runs I, J, J2 and K) modeled a two-species plasma composed of heavier $O^+$ ion  ($95\%$ number density) and 16 times lighter $H^+$ ions ($5\%$ number density). Run I was conducted as a replication of Run H, but with a two-species plasma. Building upon this, subsequent simulations were performed to more closely model realistic wave-particle interactions in the ionosphere. Specifically, for Run J, J2 and K we adjusted the plasma beta to match values derived from the IRI model. The final simulation, Run K, explored the effects of an even smaller amplitude spectrum of perturbing waves, with the goal of modeling a typical interaction in an ionospheric environment.

For Runs I, J, J2 and K, the perturbing power spectrum has wavenumbers from $w_{min}=19$ to $w_{max}=21$ and left-handed polarization,  consistent with Run H. Table \ref{tab:runIJK} summarizes the parameters for these runs, including the beta values for all plasma components: electrons ($\beta_e$), oxygen ions ($\beta_O$), and hydrogen ions ($\beta_H$). Hereafter, $d_i$ will denote the inertial length of the dominant species ($O^+$).

\begin{table}
	\centering
	\begin{tabular}{ccccccccc}
	\hline
	Run & $\delta B/B_0$ & $\beta_e$ & $\beta_O$ & $\beta_H$ & $L_{\textrm{box}}(d_i)$ & $n_O$ & $n_H$ & $ppc$\\
	\hline
	I & $5\cdot10^-3$ & $1\cdot10^{-4}$ & $1\cdot10^{-4}$ & $4\cdot10^{-6}$ & 60.8 & 0.95 & 0.05 &$1\cdot 10^4$\\
	\hline
	J & $5\cdot10^-3$ & $1\cdot10^{-5}$ & $1\cdot10^{-5}$ & $4\cdot10^{-7}$ & 60.8 & 0.95 & 0.05 &$1\cdot 10^4$\\
		\hline
	J2 & $5\cdot10^-3$ & $1\cdot10^{-5}$ & $1\cdot10^{-5}$ & $4\cdot10^{-7}$ & 60.8 & 0.95 & 0.05 &$5\cdot 10^5$\\
	\hline
	K & $2\cdot10^-3$ & $1\cdot10^{-5}$ & $1\cdot10^{-5}$ & $4\cdot10^{-7}$ & 60.8 & 0.95 & 0.05 &$1\cdot 10^4$\\
	\hline
	\end{tabular}
	\caption{Run I, J, J2 and K parameters.}
    \label{tab:runIJK}
\end{table}

Results from Run I (not explicitly shown) confirm the behavior observed in Run H, with the VDF exhibiting a central plateau around $v_\parallel/v_A = 0$ and the appearance of ion velocity beams. However, the VDF of $H^+$ ions undergoes more pronounced modifications and develops more evident ion beams. These beams occur at $v_\parallel/v_A \simeq \pm 5\cdot 10^{-2}$ for $H^+$ ions, and at $v_\parallel/v_A \simeq \pm 1\cdot 10^{-2}$ for $O^+$ ions. 



Analysis of Run J (not explicitly shown) reveals a modification of the ion VDF consistent with the observations from Runs D and E. Specifically, the VDFs for both $O^+$ and $H^+$ ions demonstrate a pronounced and rapid broadening along both the positive and negative parallel directions. To rule out possible numerical artifacts arising from insufficient resolution, the results of Run J were corroborated by an identical simulation (Run J2) employing an increased number of particles ($ppc = 5 \cdot 10^5$), which yielded consistent results (not shown). The consistent outcomes from this validation run confirm the physical robustness of the observed phenomena.

To reconcile the consistent behavior exhibited by the VDF across Runs D, E, and J, we introduce the derived parameter $\beta^* = \frac{\beta_{\textrm{tot}}}{(\delta B /B)^2}$, defined as the ratio between the total plasma beta ($\beta_{\textrm{tot}} = \beta_i + \beta_e$) and the energy density associated to the magnetic field fluctuations. The values of $\beta^*$ for our simulation campaign are presented in Table \ref{tab:betaStar}. Runs D, E, and J operate in the regime where $\beta^* < 1$. To better interpret this result, it is instructive to examine its reciprocal $1/\beta^* $. This parameter expresses the ratio between the magnetic field fluctuation energy and the background thermal pressure ($1/\beta^* \propto \frac{\delta B^2}{nk_B T}$, where $k_B$ is the Boltzmann constant), serving as a key indicator of the compressive nature of the plasma response. In Runs D, E and J, $1/\beta^* $ is greater than unity, implying that the magnetic field fluctuation energy is greater than the background thermal pressure. The super-thermal magnetic fluctuations lead to a modulation in the enhancement of the parallel electric field, which accelerates the protons strongly, leading to the picture of Figure \ref{fig:runEvdf}, with a very wide VDF. The same effects have been observed in \citet[][Run E]{matteini2010kinetics}. Notably, the calculated $1/\beta^*$ of 2.3 for this reference simulation is consistent with the empirical threshold derived from our results.

\begin{table}
    \centering
    \label{tab:betaStar}
    \begin{tabular}{cccccc}
        \hline
        Run & $\mathbf{\beta_{\text{tot}}}$ & $\delta B/B_0$ & $\mathbf{\beta^*}$ & $\mathbf{1/\beta^*}$ \\
        \hline
        A & $6\cdot 10^{-2}$ & $5\cdot 10^{-2}$ & $24$ & $4.17 \cdot 10^{-2}$ \\
        \hline
        B & $6\cdot 10^{-2}$ & $5\cdot 10^{-2}$ & $24$ & $4.17 \cdot 10^{-2}$ \\
        \hline
        C & $6\cdot 10^{-2}$ & $5\cdot 10^{-2}$ & $24$ & $4.17 \cdot 10^{-2}$ \\
        \hline
        D & $2 \cdot 10^{-3}$ & $5\cdot 10^{-2}$ & $0.8$ & $1.25$ \\
        \hline
        E & $2. \cdot 10^{-4}$ & $5\cdot 10^{-2}$ & $8\cdot 10^{-2}$ & $12.5$ \\
        \hline
        F & $2.0 \cdot 10^{-4}$ & $5\cdot 10^{-3}$ & $8$ & $0.125$ \\
        \hline
        G & $2.0 \cdot 10^{-4}$ & $5\cdot 10^{-3}$ & $8$ & $0.125$ \\
        \hline
        H & $2.0 \cdot 10^{-4}$ & $5\cdot 10^{-3}$ & $8$ & $0.125$ \\
        \hline
        I & $2.0 \cdot 10^{-4}$ & $5\cdot 10^{-3}$ & $8$ & $0.125$ \\
        \hline
        J & $2 \cdot 10^{-5}$ & $5\cdot 10^{-3}$ & $0.8$ & $1.25$ \\
        \hline
        K & $2 \cdot 10^{-5}$ & $2\cdot 10^{-3}$ & $5$ & $0.2$ \\
        \hline
    \end{tabular}
    \caption{$\beta^*$ values for the simulation campaign.}
    \label{tab:betaStar}
\end{table}


Our final simulation (Run K), was dedicated to investigate the effects of a more realistic, smaller amplitude perturbing wave, in an ionospheric-like background. The amplitudes considered in Runs F-I were still significantly higher than those typically observed in ionospheric environments. For context, observational data from satellite missions such as SWARM, orbiting at altitudes of approximately 500 km, show that the ambient magnetic field at equatorial latitudes is of the order of 30000 nT \citep{leger2015flight,fratter2016swarm}. Therefore, $\delta B/B_0 = 5 \cdot 10^3$ implies a perturbation of approximately 150 nT. While perturbations of this magnitude can occur during geomagnetic storms \citep{yang2021cses}, typical magnetic field perturbations during geomagnetically quiet periods are of the order of few tens of nT \citep{yagova2023geomagnetic}. Consequently, Run K was conducted as a replication of Run J but with $\delta B/B_0 = 2 \cdot 10^{-3}$, simulating an ionospheric perturbation of few tens of nT, compatible with quiet or moderately disturbed periods.

\begin{figure}
\centering
\begin{subfigure}{0.48\textwidth}
    \includegraphics[width=\textwidth]{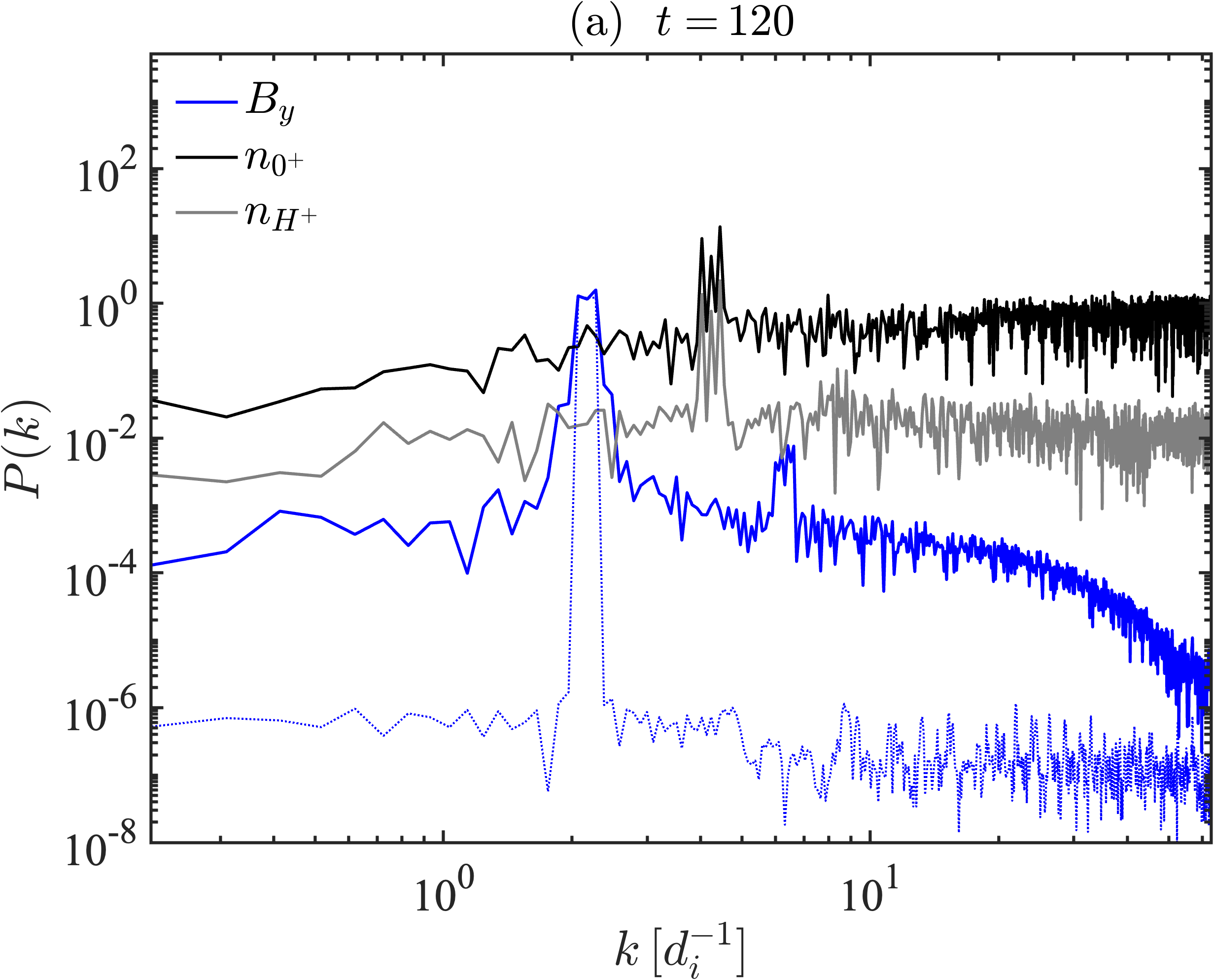}
    \label{fig:runK1DspectrumT120}
\end{subfigure}
\hfill
\begin{subfigure}{0.48\textwidth}
    \includegraphics[width=\textwidth]{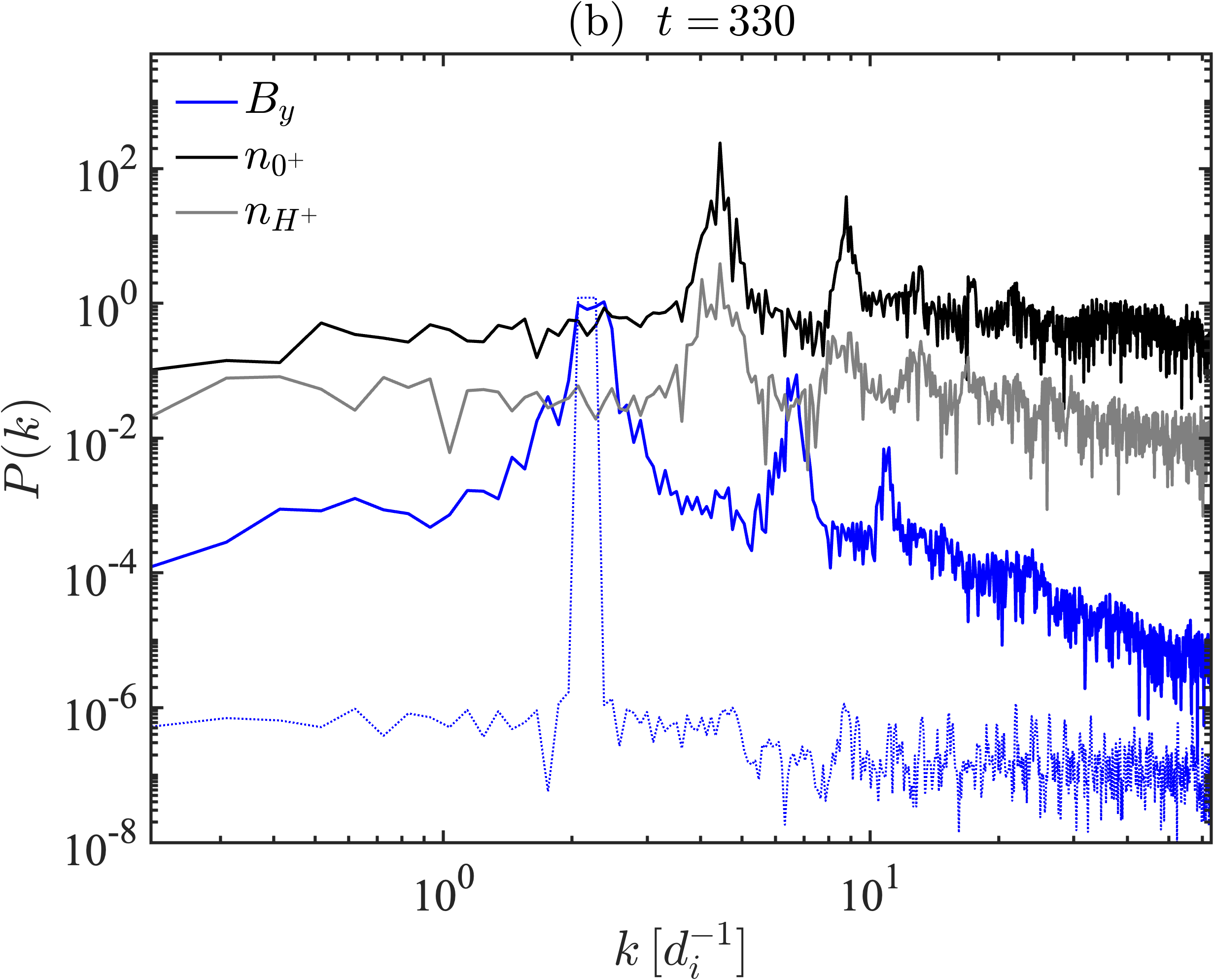}
    \label{fig:runK1DspectrumT330}
\end{subfigure}
\hfill
\begin{subfigure}{0.48\textwidth}
    \includegraphics[width=\textwidth]{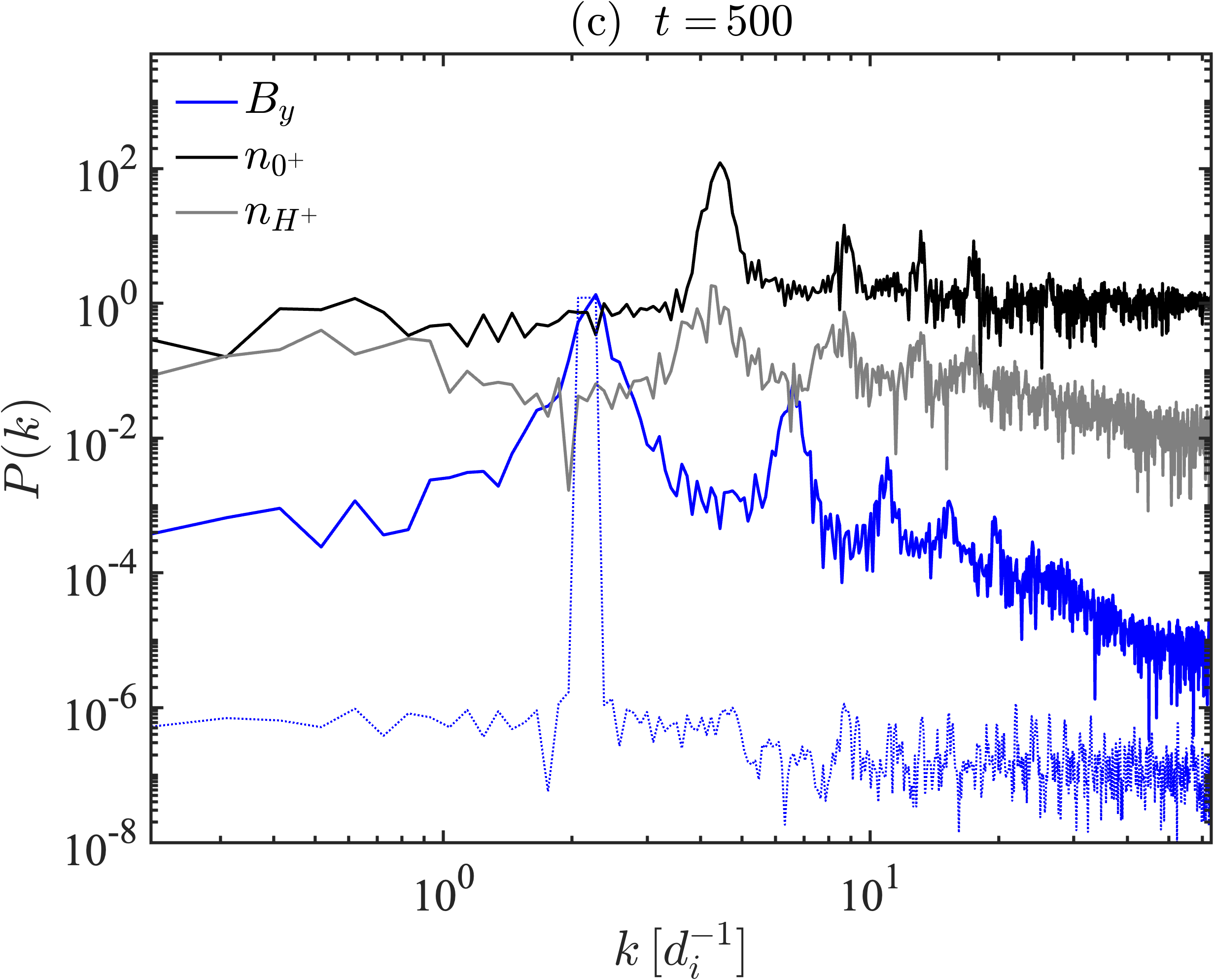}
    \label{fig:runK1DspectrumT500}
\end{subfigure}
\hfill
\begin{subfigure}{0.48\textwidth}
    \includegraphics[width=\textwidth]{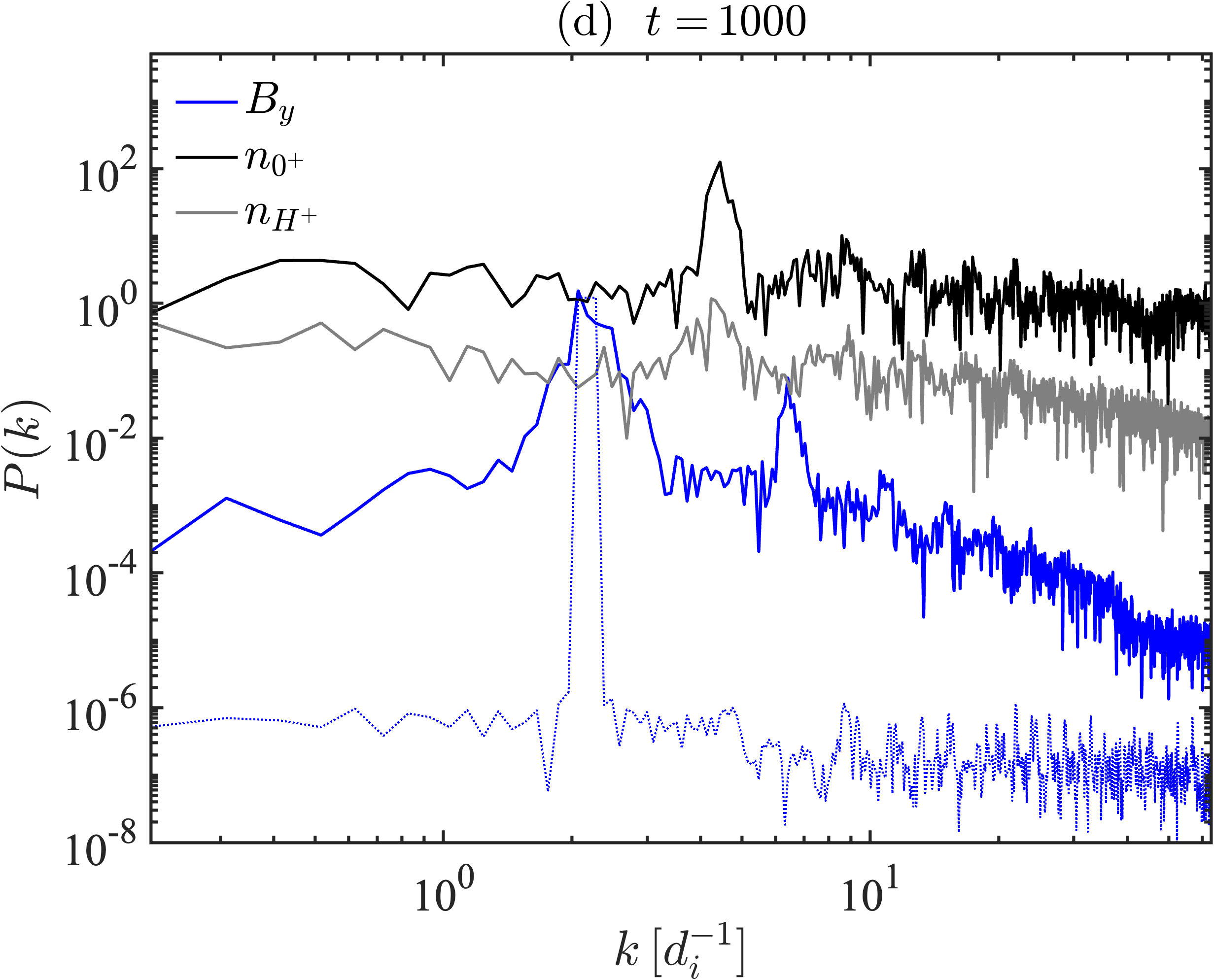}
    \label{fig:runK1DspectrumT1000}
\end{subfigure}
        
\caption{Run K: Power spectra of the $y$ component of the magnetic field (blue), $O^+$ density (black), and $H^+$ density (grey). Power spectrum of $B_y$ at $t = 0$ are overlaid as thinner blue dashed lines.
} 
\label{fig:runK1Dspectrum}
\end{figure}

Figure \ref{fig:runK1Dspectrum} shows significant snapshots of the temporal evolution of the power spectrum of the $y$ component of magnetic field (blue), $O^+$ density (black) and $H^+$ density (grey). A spectral peak on both ion densities emerges at $k\simeq$ 4 around $t=120$ (Panel a). This primary mode undergoes nonlinear development, producing distinct harmonics that collectively peak in amplitude around $t=330$ (Panel b). The system then stabilizes, entering a new quasi-equilibrium phase, characterized by only minor spectral variations at longer times (see Panels c and d, representing the power spectrum at $t=500$ and $t=1000$ respectively).

        

\begin{figure}      
\centering
\includegraphics[height=0.35\textwidth]{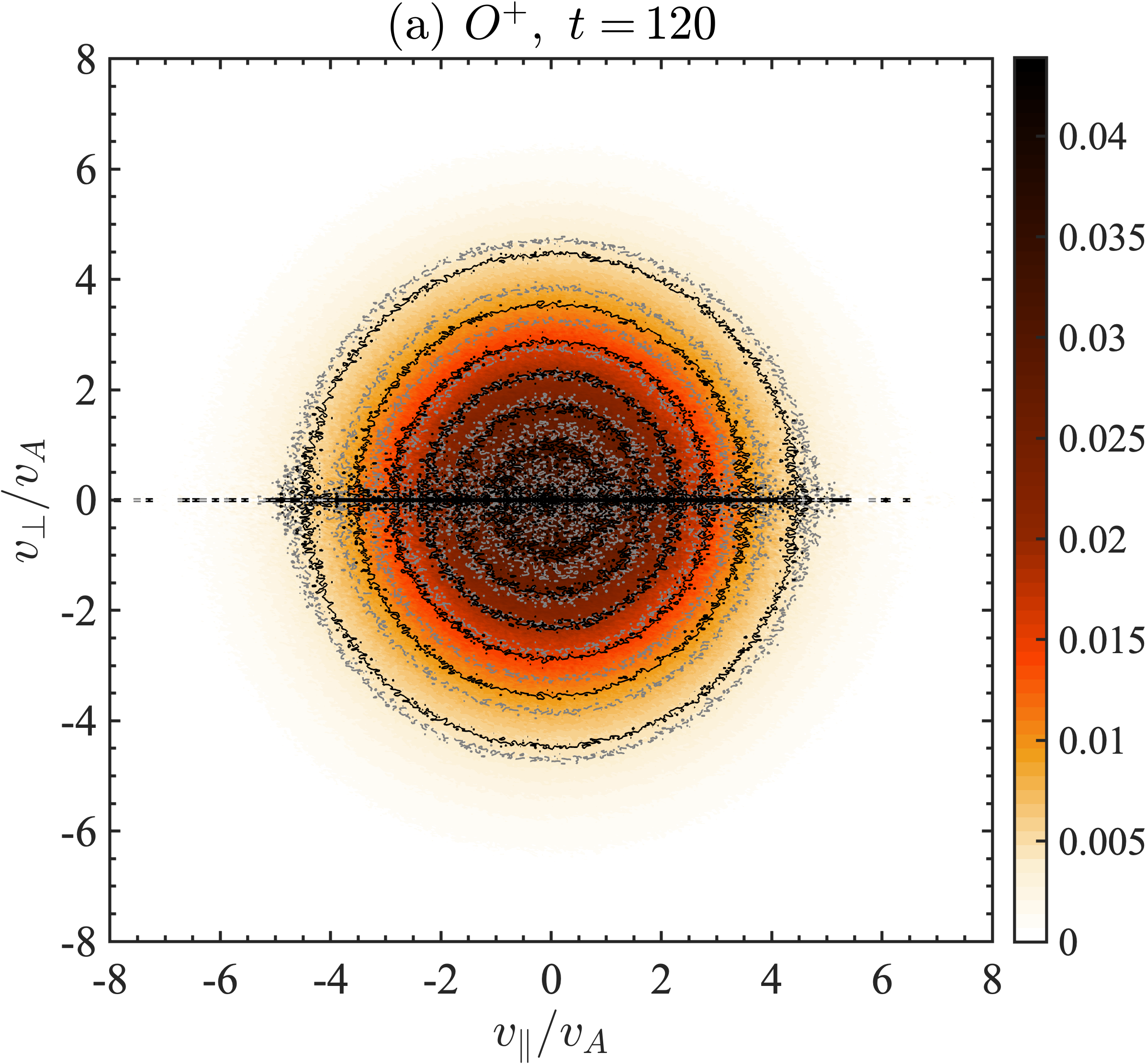}
\includegraphics[height=0.35\textwidth]{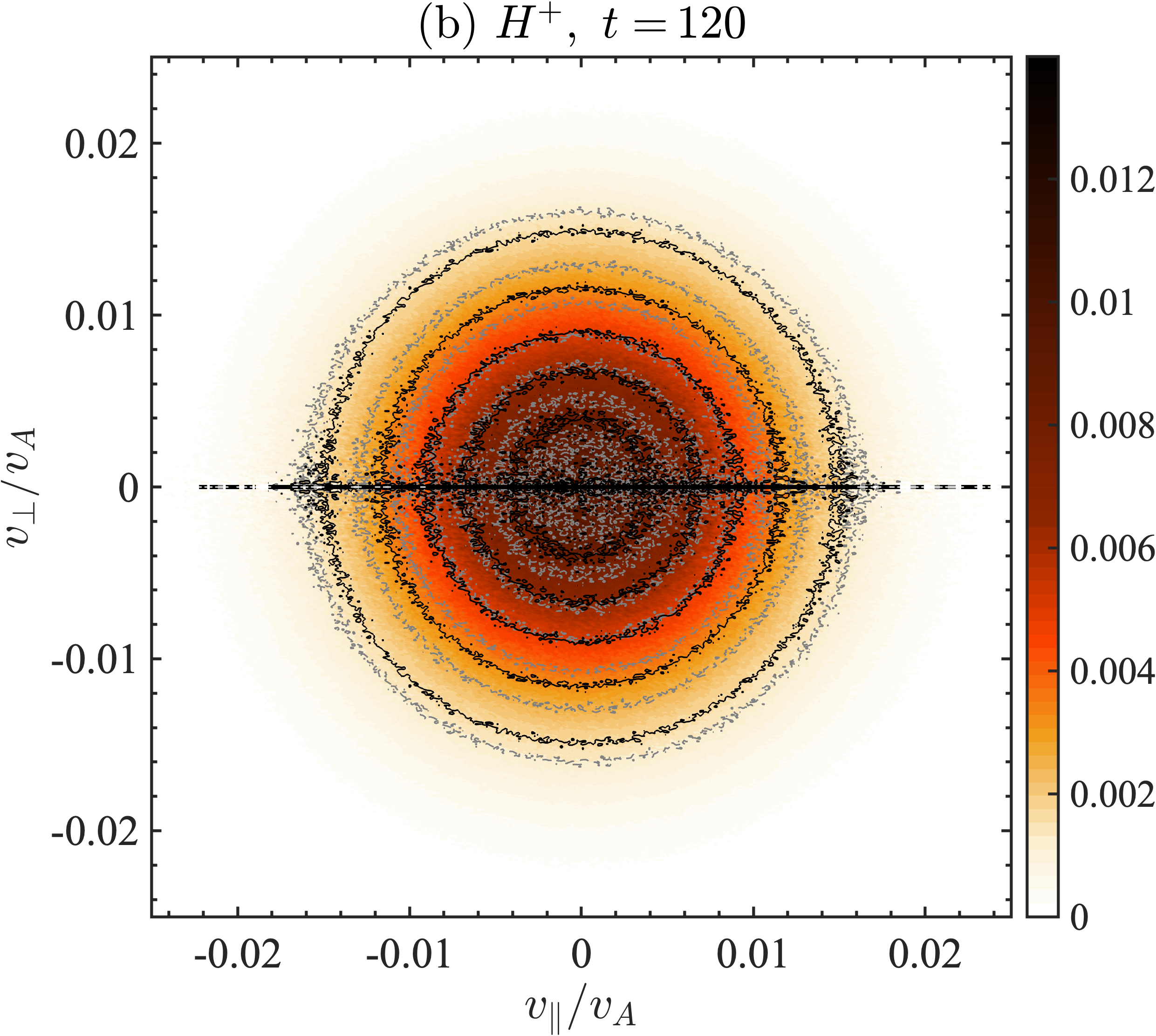}\\
 \includegraphics[height=0.35\textwidth]{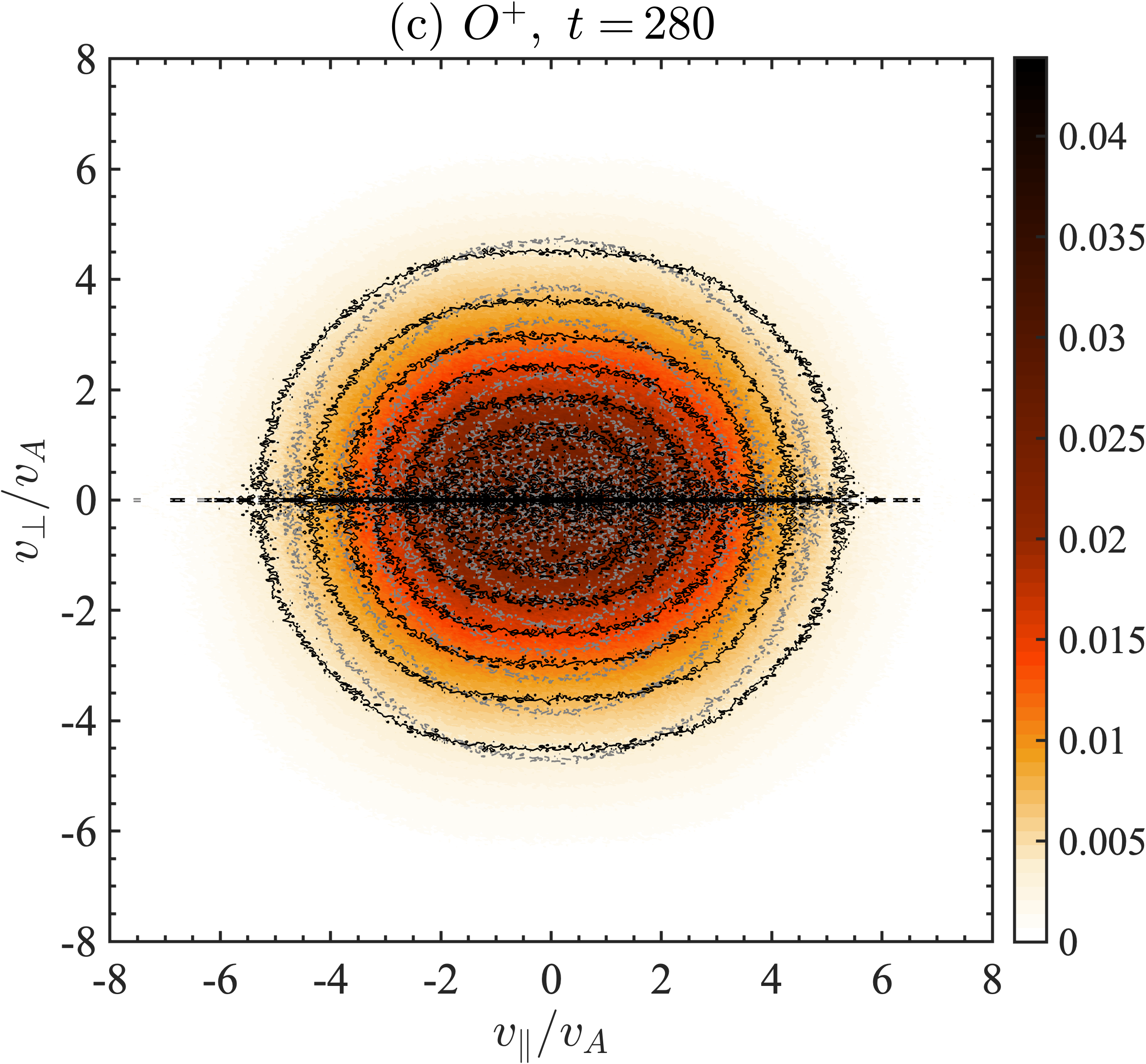}
\includegraphics[height=0.35\textwidth]{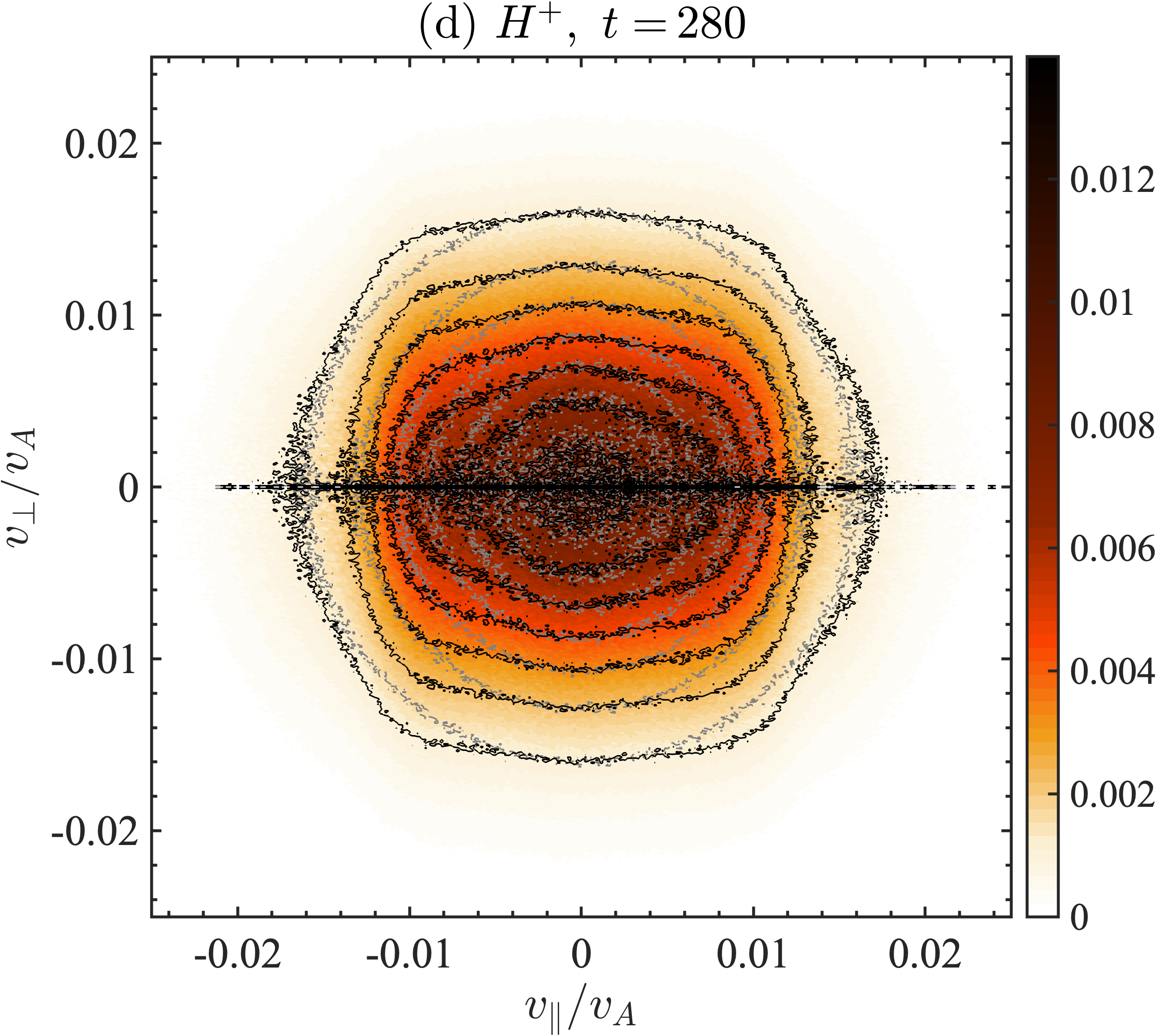}\\
\includegraphics[height=0.35\textwidth]{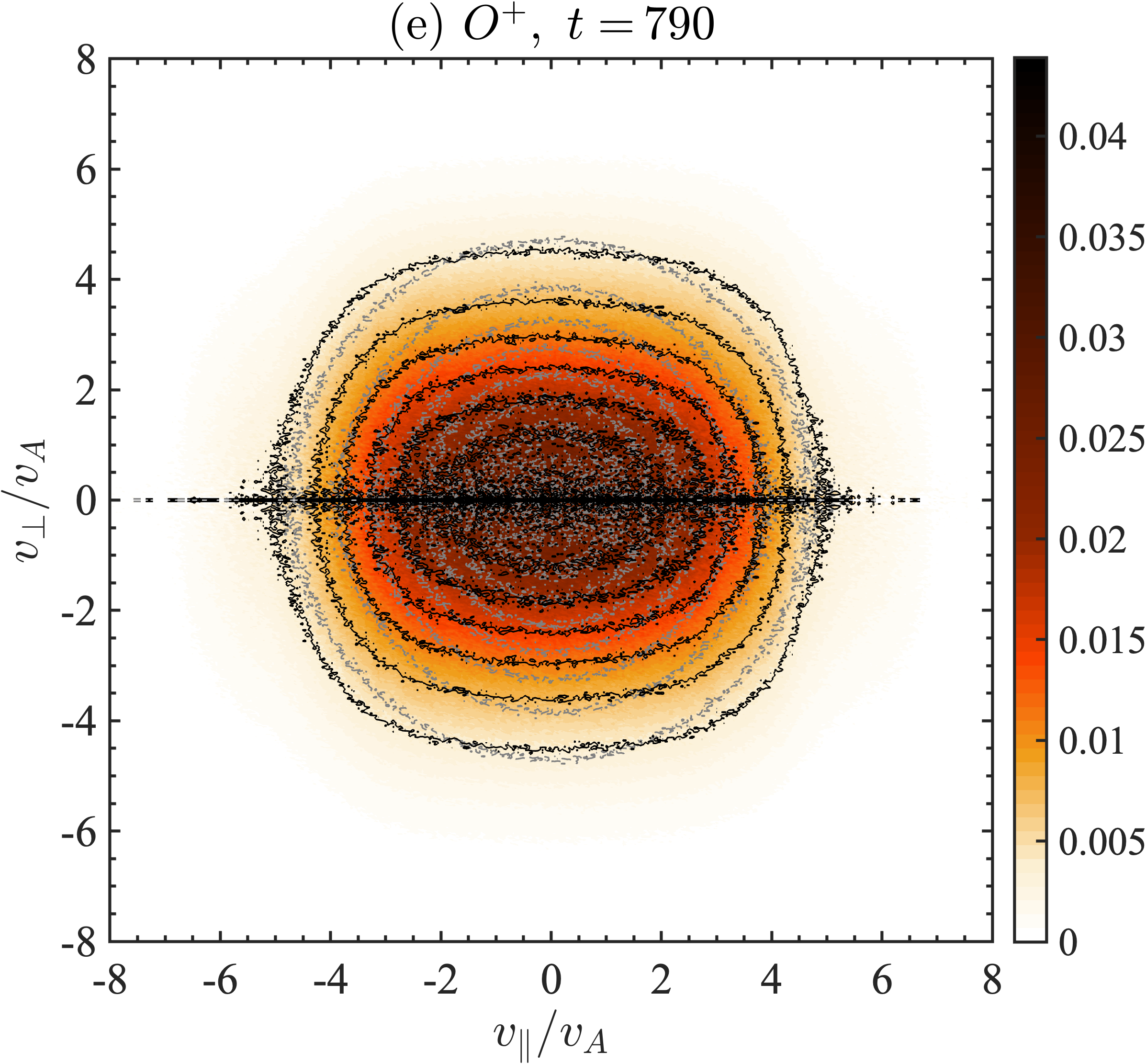}
\includegraphics[height=0.35\textwidth]{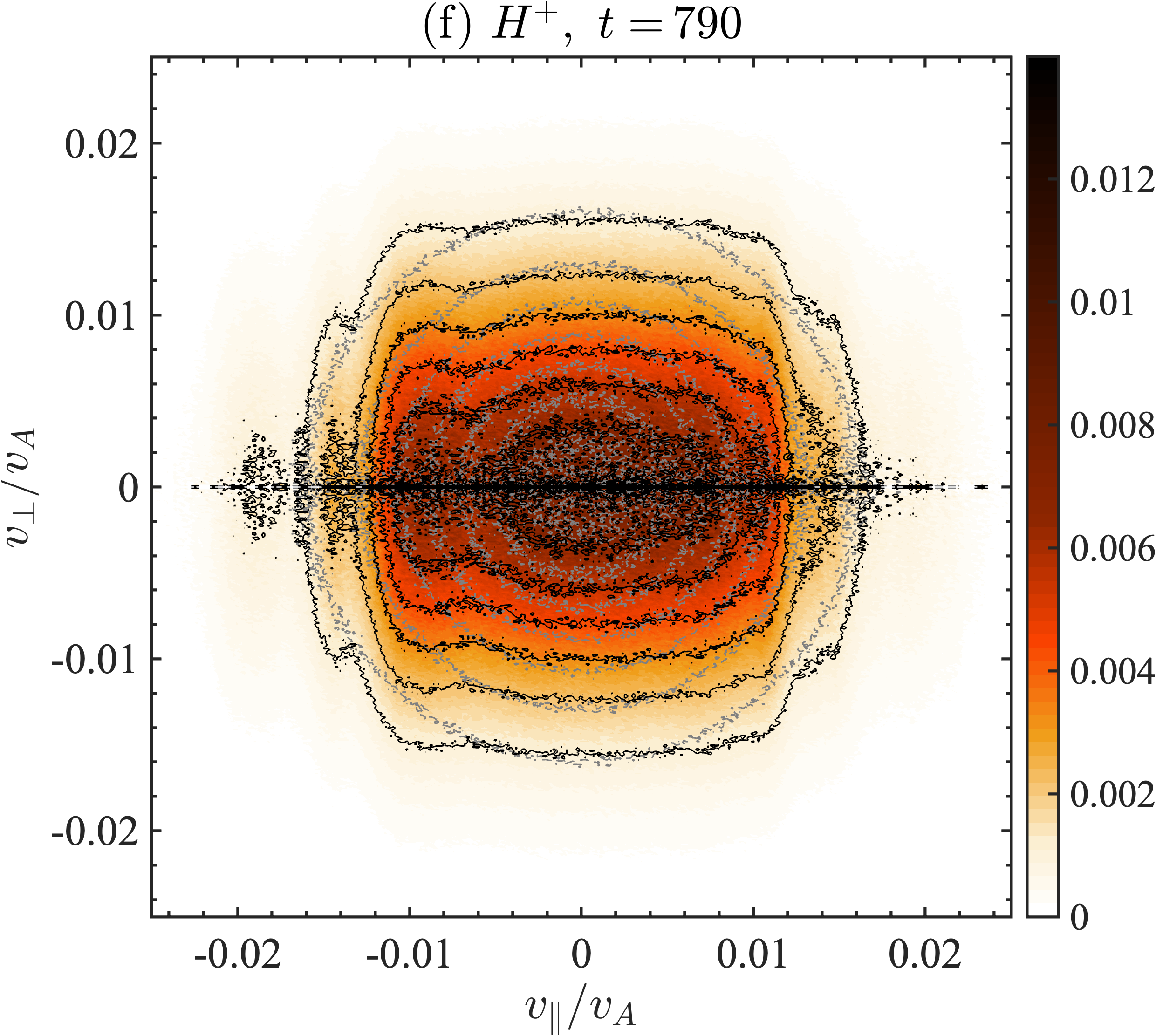}\\
\includegraphics[height=0.35\textwidth]{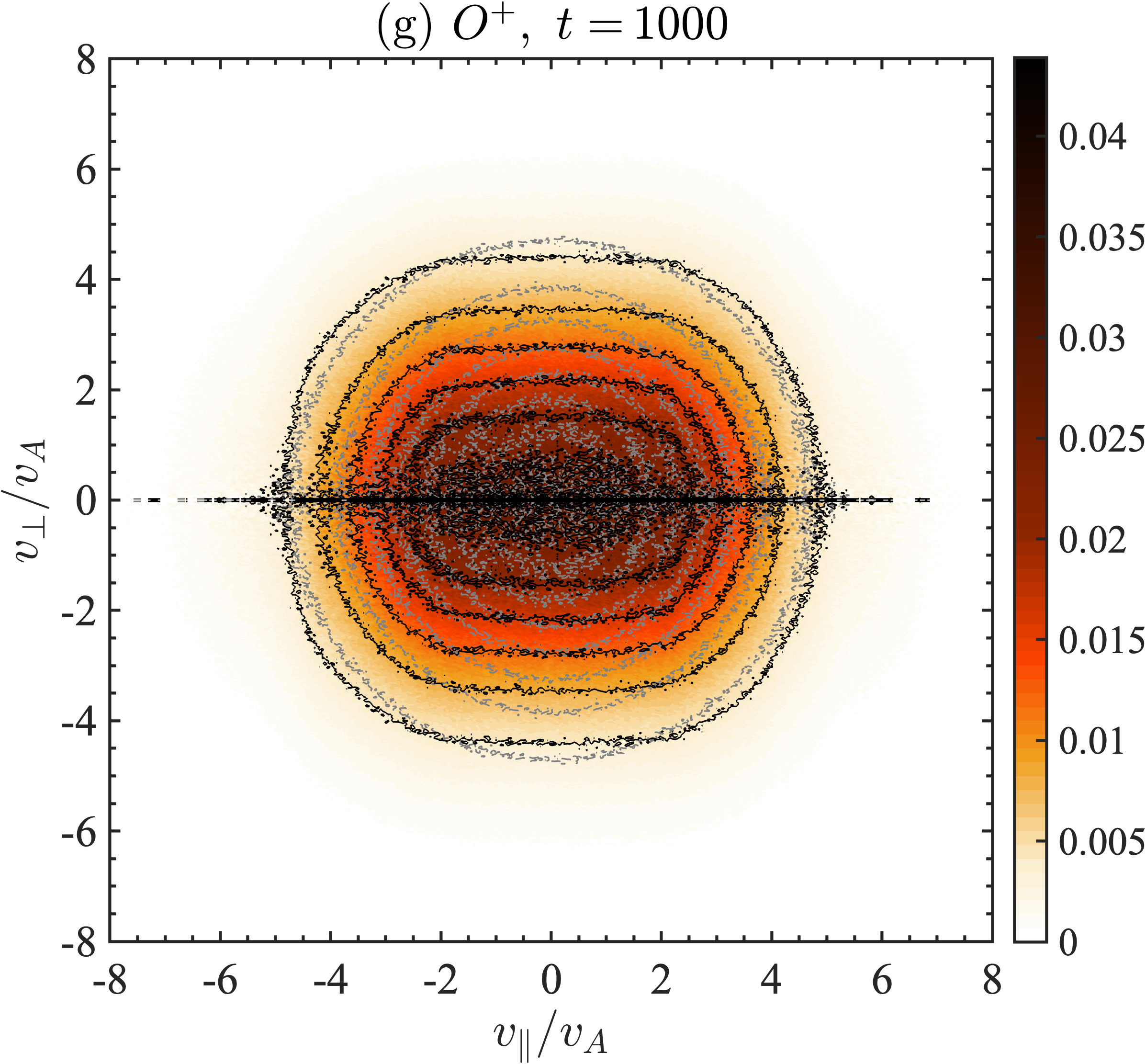}
\includegraphics[height=0.35\textwidth]{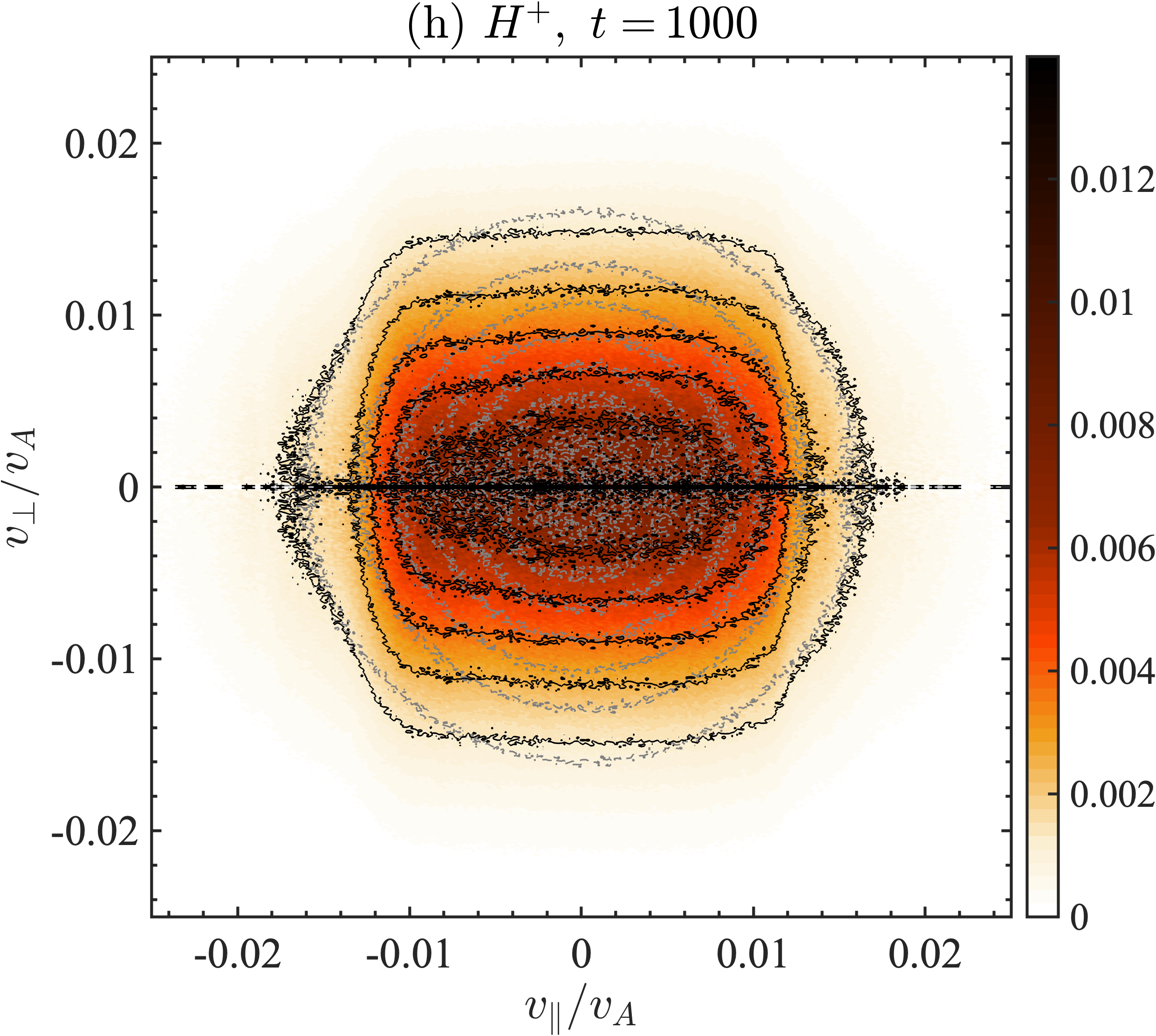}\\
   
\caption{Run K: Time evolution of the ion VDFs, $O^+$ in the left column and $H^+$ in the right column. In each panel pseudocolor plots and black contours are shown, with superimposed gray contours representing the initial VDFs for the respective species.} 
\label{fig:runKvdf}
\end{figure}

The temporal evolution of the ion VDFs is captured in a series of key snapshots presented in Figure \ref{fig:runKvdf} at different increasing times from top to bottom.  The $O^+$ VDFs are shown in the left panels, while the $H^+$ VDF are shown in right ones. The ion VDFs begin to evolve away from the initial state at $t=120$, (Panels a and b, coinciding with  the appearance of the spectral density peaks. Subsequently, a significant spreading of the VDFs is observed along both positive and negative parallel directions (Panels c-d, $t=280$). The distribution tails continue to fill, and at some point in the simulation ($t \gtrsim 400$), distinct ion beams start to appear. The $H^+$ VDF, in particular, exhibits three prominent pairs of beams at $v_\parallel/v_A \simeq \pm 0.01$, $v_\parallel/v_A \simeq \pm 0.015$, and $v_\parallel/v_A \simeq \pm 0.02$  (see Panel e-f, representing VDFs at $t=790$). At longer times, both ion VDF exhibit minor oscillations, confirming the new quasi-equilibrium phase (see Panels g-h, $t=1000$).

\section{Conclusion}\label{sec:conclusion}

While PDI is a well-known process in plasma physics, its precise effects on ion dynamics in ultra-low-beta, ionospheric-like plasmas remain unexplored. To address this gap, we performed a comprehensive study employing 1D hybrid simulations of a low-beta plasmas perturbed by small-amplitude Alfv\'en waves, a setup mimicking ionospheric conditions. Our primary diagnostic was the temporal evolution of the ion VDF, to reveal the microscopic kinetic changes induced by the instability. Our findings show that the PDI is the key driver shaping the VDF. The instability introduces pronounced nonthermal effects and drives the plasma away from its initial Maxwellian state. Specifically, the final velocity distributions show clear evidence of parallel ion heating, manifesting as either a plateau or a distinct ion beam, depending on the specific characteristics of the run.

To perform a realistic simulation campaign, we focused on three key aspects; (A) the modeling of an ultra-low-beta plasma; (B) the incorporation of a realistic ionospheric composition; (C) the consideration of a pump wave with a realistic amplitude. Regarding aspect (A), as detailed in Table \ref{tab:ionoParam}, ionospheric components exhibit a plasma beta of the order of $10^{-5}$ for electrons and $O^+$ ions, and $10^{-7}$ for $H^+$ ions. Runs J and K were designed with these precise beta values. Notably, to our knowledge, this represents the first hybrid simulation of a plasma in this ultra-low-beta regime. For aspect (B), a realistic ionospheric composition at 500 km was implemented, comprised of a mixture of $95\%$ heavier $O^+$ ions and $5\%$ lighter $H^+$ ions, a ratio that accurately represents the top-side ionosphere as shown in Table \ref{tab:ionoParam}. Concerning (C), given the ubiquitous Earth's magnetic field of approximately $10^4$ nT, realistic ionospheric perturbations constitute only a small fraction of the ambient field.

The distinctive characteristics of our setup provided an opportunity to investigate the PDI in a domain of parameters not yet explored by prior studies. Our investigation employed a methodical series of simulations. We first established a baseline using a well-studied plasma regime, then systematically varied parameters across a series of runs (Run A through Run K). This systematic tuning was essential for disentangling the complex influence of each parameter on the instability's evolution and enabled a controlled progression toward a more representative physical scenario. A summary of the key results from all runs is presented in Table \ref{tab:finale}.

\begin{table}
	\centering
	\begin{tabular}{p{0.05\textwidth}p{0.24\textwidth}p{0.7\textwidth}}
	\hline
	\textbf{Run} & \textbf{Summary} & \textbf{Results}  \\
	\hline
	\textbf{A}   & Test of the setup & Single parametric decay at $t\approx 200$, with the generation of a daughter wave at lower $k$. Correspondingly, a modification of the VDF is observed, with the generation of populations of faster ions. \\
	\hline
    \textbf{B} & Halved $k_m$ with respect to run A & Single decay, same as run A but slower evolution.\\
	\hline
	\textbf{C}   & Double $k_m$ w.r.t.  run A & Double decay, faster evolution. First decay $\rightarrow$ daughter wave at lower $k$ $\rightarrow$ fast ion beams (positive side of the parallel VDF). Second decay $\rightarrow$ granddaughter wave at even lower $k$ $\rightarrow$ ion beams (negative side).\\
	\hline
	\textbf{D} & Reduced beta w.r.t. run C & Faster evolution w.r.t. run C. Generation of a daughter wave at lower $k$. Rapid generation and reabsorption of several harmonics in density and magnetic field. Widening of the VDF in both side of the parallel direction.\\
	\hline
	\textbf{E} & Reduced beta w.r.t. run D &  Same behaviour of run D, but with an even faster evolution, more harmonics on the power spectrum and VDF significantly spread out along the parallel direction.\\
	\hline
	\textbf{F} & Reduced $\delta B/B_0$ w.r.t. run E & Density peak at $t=60$ and second harmonics at $t=120$ on the spectrum. No secondary daughter wave magnetic peak. Minor modification of the VDF, with the central peak slightly lower and barely visible beams on both side of the parallel direction at $t\geq 400$.\\
	\hline
	\textbf{G} & Spectrum of Alfv\'en waves &  Rising of a triple peak at $t=120$ and correspondingly the second harmonic on the magnetic field. Effects on the VDF slightly increased.\\
	\hline
	\textbf{H} & Reverse polarization w.r.t run G & Faster evolution w.r.t. G. Appearance of several harmonics. Significant evolution of the VDF, with generation of fast ion beams at $t > 400$.\\
	\hline
	\textbf{I} & As run H but with two species plasma  & Similar behaviour of run H for the $O^+$ ions. Generation of fast proton beams in both sides and filling of the tails of the $H^+$ VDF.\\
	\hline
	\textbf{J} & Lower beta w.r.t. run I (reaching realistic values) & Similar situation as run E with the fast appearance of several harmonics on the spectrum and a VDF significantly spread out along both sides of the parallel direction.\\
	\hline
	\textbf{K} & As run J but with lowered $\delta B/B_0$ &  Several harmonics in both the density and the magnetic field rise and decrease with a slow evolution. Remarkable deviation of both species VDF from its original configuration. For $H^+$, generation of tails in the VDF and of fast ion beams at $t\geq 300$.\\
	\hline
	\end{tabular}
	\caption{Summary of simulation results.}
    \label{tab:finale}
\end{table}

Specifically, in our simulation campaign, we investigated the impact of the following parameters:

\begin{itemize}

\item \textbf{Wave number ($k_m$)}: Runs A, B, and C  were conducted under identical initial conditions, with the exception of the mother wave's wavenumber $k_m$. A reduction in $k_m$ (Run B) resulted in a decrease in kinetic effects, due to a closer approximation to the MHD regime. Although Run B underwent a parametric decay akin to Run A, its ion VDF evolved at a reduced pace. Conversely, increasing $k_m$ (Run C) accelerated VDF evolution. A noteworthy finding in Run C was the observation of a double decay,  leading to the emergence of both a daughter and a granddaughter wave and the corresponding generation of ion beams on both sides of the VDF. The appearance of a granddaughter wave is consistent with previous findings on non-linear wave dynamics \citep[see e.g.][]{kojima1989nonlinear, del2001parametric, umeda2018decay}.

\item \textbf{Plasma beta ($\beta$)}: We investigated its influence on the PDI by comparing Runs C, D, and E, which shared identical initial parameters except for $\beta$. A significant  modification in PDI characteristics was observed when the plasma beta was reduced to $\beta \leq 10^{-3}$ (Runs D and E). As illustrated in Figure \ref{fig:parametrica}, at these low $\beta$ values, the reflected daughter wave's wavenumber became nearly equal to the mother wave wavenumber ($|k_r|\simeq |k_m|$). Consequently, a distinct magnetic field peak corresponding to the daughter wave was not discernible from the original pump wave in the power spectrum. However, the PDI still served as a trigger for the emergence of one or more density peaks (depending on $\beta$ and amplitude of the mother wave). The density fluctuations exhibited rapid growth, achieving significantly higher amplitudes compared to the plasma regimes investigated in previous studies. For context, \cite{gonzalez2023particle} observed a saturation of the instability and a constant value of $n_{\mathrm{RMS}}/\langle n \rangle \simeq 0.2$ after the saturation. In contrast, as shown in Figure \ref{fig:runErmsDensity}, our results show an initial, high-amplitude overshoot in $n_{\mathrm{RMS}}$ before a sharp decrease and eventual stabilization. Indeed, initially,  the simultaneous growth of density and magnetic field fluctuations aligns with the expected pressure equilibrium ($B^2 \propto nk_BT$). Subsequently, as energy is transferred from the wave to the plasma particles, ion heating occurs, evidenced by the spreading of the ion VDF shown in Figure \ref{fig:runEvdf}. Consequently, the system reaches a quasi-stationary state where the increased thermal pressure offsets the need for large-amplitude density fluctuations to sustain pressure balance. Further confirmation is provided by the observed inverse relationship between the plasma beta and the ratio of density to magnetic field fluctuations ($\delta n/\delta B^2$), which increases progressively from Run C to Run E. Furthermore, we find that PDI features (the decay timing, the emergence of the spectral density peak, and the temporal scale of ion VDF evolution) are all enhanced with decreasing $\beta$. These observations directly confirm the well-documented result that the instability growth rate increases with decreasing plasma beta \citep[see e.g.][]{gonzalez2023particle}. 

\item \textbf{Ratio of the total plasma beta and the energy density associated to the magnetic field fluctuations ($\beta^\star$)}: Our results demonstrate that a constant perturbation exerts a more pronounced effect on lower-beta plasmas. Specifically, a comparison of simulations with identical initial parameters (in particular the same perturbing wave amplitude) but lower beta values (Run E vs. Run D; Run J vs. Run I) reveals a faster and more significant evolution of the ion VDF. Conversely, reducing the perturbing wave amplitude (e.g., Run F vs. Run E) yields a slower and less pronounced modification of the ion VDF, indicating a combined effect of the pump wave amplitude and plasma beta. A critical transition is observed when $1/\beta^* > 1$. This regime is characterized by the concurrent, rapid generation of multiple harmonics in both the density and magnetic field spectra, along with rapid and complete VDF broadening. This behaviour indicates the development of intense electric fields driven by large pressure gradients at short-scale (Figure \ref{fig:runEdensityExProfile}), which accelerates particles in both sides of the parallel direction. To the best of our knowledge, this is the first report in the literature of such radical modification of the VDF. Indeed, while other works have marginally analyzed regimes where $1/\beta^* \gtrsim 1$ \citep{matteini2010kinetics,gonzalez2023particle}, the $1/\beta^*>10$ regime analyzed in our Run E has not been previously investigated.

\item \textbf{Initial power spectrum}: A narrowband spectrum of perturbing waves induces slightly more pronounced VDF modifications with respect to a monochromatic pump wave. This has been verified in Run G, where all the parameters are unchanged with respect to Run F but the monochromatic pump wave has been replaced by a localized (narrowband) wavepacket. This result confirms the fact that the energy gain of particles interacting with a varying-frequency (chirped) electromagnetic pulse increases with respect to a non-chirped electromagnetic pulse \citep{khachatryan2005effect}.

\end{itemize}

Although Runs A-I were necessary to understand the peculiar manifestation of PDI in the target regime, providing  insight into PDI characteristics in a very-low plasma beta, the final simulations (J and K) were key for reaching our research goal.
These two runs represent the most realistic modeling of the ionospheric environment within our simulation campaign, in terms of wave amplitudes and background parameters, enabling the investigation of two scenarios distinguished by the pump wave amplitude.
 The first scenario involves waves with amplitude $\delta B/B_0 \gtrsim 5 \cdot 10^{-3}$ which corresponds to ionospheric magnetic field perturbations of hundreds of $nT$. This configuration represents  extreme ionospheric perturbations, such as those associated with severe space weather events. Under these conditions, we observe a rapid and complete ion VDF  spreading in the parallel direction for both ion species,  strongly indicating the potential for particle precipitation events.  This mechanism is particularly relevant given the widely reported connection between severe space weather events (e.g. geomagnetic storms) and ionospheric particle precipitation \citep{longden2007driving,longden2008particle,song2025topside}. 

In contrast, the second scenario involves waves with amplitude $\delta B/B_0 \lesssim 2 \cdot 10^{-3}$, which corresponds to ionospheric magnetic field perturbations of tens of $nT$. This configuration represents a more common ionospheric state, as such perturbation levels occur even during geomagnetically quiet periods. In this case, we observe slower and less pronounced VDF changes, primarily affecting $H^+$ ions. Specifically, the formation of fast ion beams after approximately $t=300$. However, even in this case, the generated ion beams still possess the potential to induce particle precipitation, particularly for $H^+$ ions. This mechanism aligns with recent studies demonstrating that EMIC waves can cause proton precipitation \citep{tian2022effects,d2024case}.

Furthermore, numerous studies over the past decades have suggested a concurrent occurrence of energetic particle flux variations and electromagnetic activity in the proximity of, or even preceding, large earthquakes \citep{SGRIGNA20051448,anagnostopoulos2010temporal,fidani2010study,sidiropoulos2011comparative,pulinets2022earthquake}. In particular, whistlers have been proposed as seismic precursors \cite{liu2023pre,wang2024lightning}.
However, a critical limitation in these investigations is the neglect of the time delay between the arrival of the perturbing wave and the resultant modification of the ion VDF.  As demonstrated in Run K, the formation of ion beams is delayed until $t=300 \,\Omega_i^{-1}$, corresponding to approximately $10.5$ seconds. Therefore, this study provides a crucial first estimation of the expected time delay between electromagnetic wave impact and VDF modifications, establishing a fundamental step for future work linking electromagnetic anomalies and particle flux enhancements connected to natural events.

While this investigation constitutes a first step towards the description of wave-particle interactions from the PDI in a ionospheric-like plasma, there are certainly limitations in the approach and possible extensions should be considered in the future.
Charged particle trajectories in a plasma mainly follow magnetic field lines 
and this is in particular expected in a low-beta environment, with strong background magnetic field, like the ionosphere.
Consequently, the 1D simulations presented here are expected to  represent accurately enough particle dynamics, which is assumed to develop mostly along the underlying ionospheric field lines. However, this constraint could be violated through various additional mechanisms \citep[][and reference therein]{minnie2009particles} and it is also known that the PDI can evolve differently when starting with a pump wave which is not field-aligned \citep{matteini2010parametric, del2015parametric}. Furthermore, in the Earth's ionosphere, the neutral particle density is significantly higher than the ionized particle density throughout most altitudes \citep[e.g.][]{kelley2009earth,lavstovivcka2012trends}. Therefore, while more computationally demanding, the inclusion of neutral particles (and thus collision with neutrals) and the extension to 2D or 3D domains would ensure an even higher degree of realism and are planned for future studies.


\vspace{1 cm}
L.F. is supported by the Royal Society
University Research Fellowship No. URF/R1/231710 and
was also supported by the Science and Technology Facilities
council (STFC) grant ST/W001071/1. 
This work used the DiRAC Data Intensive service (CSD3) at the University of Cambridge, managed by the University of Cambridge University Information Services on behalf of the STFC DiRAC HPC Facility (www.dirac.ac.uk). The DiRAC component of CSD3 at Cambridge was funded by BEIS, UKRI and STFC capital funding and STFC operations grants. DiRAC is part of the UKRI Digital Research Infrastructure.
This work also used the DiRAC Memory Intensive service (Cosma8) at Durham University, managed by the Institute for Computational Cosmology on behalf of the STFC DiRAC HPC Facility (www.dirac.ac.uk). The DiRAC service at Durham was funded by BEIS, UKRI and STFC capital funding, Durham University and STFC operations grants.

\bibliographystyle{jpp}

\bibliography{jpp-instructions}

\end{document}